\documentclass[11pt,envcountsame,envcountsect,runningheads,fleqn]{llncs} 
\usepackage[a4paper,margin=2.9cm,textheight=19cm]{geometry}

\newcommand{\VersionA}[1]{#1} 
\newcommand{\VersionB}[1]{} 

\def\tuple#1{\langle#1\rangle} 
\usepackage{latexsym} 
\usepackage{xspace} 
\usepackage{amssymb} 
\usepackage{stmaryrd} 
\usepackage{url} 
\usepackage{hyperref} 
\usepackage{graphicx} 
\usepackage{tikz}
\usepackage{fancybox}
\usepackage{multirow}
\usetikzlibrary{matrix,arrows,decorations.pathmorphing}
\input{xy} 
\xyoption{all} 
\usepackage[ruled,vlined,linesnumbered]{algorithm2e}

\def\eqref#1{(\ref{#1})}

\newcommand{\mand}{\sqcap} 
\newcommand{\mor}{\sqcup} 
\newcommand{\V}{\forall} 
\newcommand{\E}{\exists}

\newcommand{\ndoteq}{\neq}

\newcommand{\mL}{\mathcal{L}} 
 
\newcommand{\mT}{\mathcal{T}} 
\newcommand{\mR}{\mathcal{R}} 
\newcommand{\mA}{\mathcal{A}} 
\newcommand{\mI}{\mathcal{I}}

\newcommand{\mZ}{\mathcal{Z}} 

\newcommand{\bP}{\mathbb{P}} 
\newcommand{\bL}{\mathbb{L}} 
\newcommand{\bX}{\mathbb{X}}

\newcommand{\ALC}{$\mathcal{ALC}$\xspace} 
\newcommand{\ALCQIO}{$\mathcal{ALCQIO}$\xspace} 
\newcommand{\SHIQ}{$\mathcal{SHIQ}$\xspace} 
 
\newcommand{\SHOIQ}{$\mathcal{SHOIQ}$\xspace} 
\newcommand{\SROIQ}{$\mathcal{SROIQ}$\xspace} 

\newcommand{\CN}{\Sigma_C} 
\newcommand{\RN}{\Sigma_R} 
\newcommand{\RNpm}{\Sigma_R^\pm} 
\newcommand{\IN}{\Sigma_I} 

\newcommand{\Self}{\mathtt{Self}} 

\newcommand{\QU}{\mathtt{Q}} 
\newcommand{\SE}{\mathtt{S}} 

\newcommand{\ECond}{\textit{ECond}_{\,\Phi}}

\newcommand{\mLP}{\mL_\Phi}
\newcommand{\simLP}{{\sim_{\,\Phi,\mI}}}
\newcommand{\equivLP}{{\equiv_{\,\Phi,\mI}}}

\newcommand{\ALCreg}{$\mathcal{ALC}_{reg}$\xspace}

\newcommand{\hasChild}{\mathit{hasChild}}
\newcommand{\hasParent}{\mathit{hasParent}}
\newcommand{\hasDescendant}{\mathit{hasDescendant}}
\newcommand{\Human}{\mathit{Human}}
\newcommand{\Male}{\mathit{Male}}
\newcommand{\Female}{\mathit{Female}}
\newcommand{\John}{\mathit{John}}
\newcommand{\Doctor}{\mathit{Doctor}}
\newcommand{\Lawyer}{\mathit{Lawyer}}

\SetKwInOut{Input}{input}
\SetKwInOut{Output}{output}

\newcommand{\myend}{\mbox{}\hfill{\footnotesize$\blacksquare$}} 
\newcommand{\myendVS}{
	\mbox{ }\hfill{\footnotesize$\blacksquare$}
	
	\medskip
	
} 

\newcommand{\comment}[1]{} 

\newcommand{\KB}{\mathit{KB}}

\newcommand{\Alice}{\mathit{Alice}}
\newcommand{\Bob}{\mathit{Bob}}
\newcommand{\Claudia}{\mathit{Claudia}}
\newcommand{\Dave}{\mathit{Dave}}
\newcommand{\Eva}{\mathit{Eva}}
\newcommand{\Frank}{\mathit{Frank}}
\newcommand{\George}{\mathit{George}}
\newcommand{\Helen}{\mathit{Helen}}

\newcommand{\Mother}{\mathit{Mother}}
\newcommand{\Father}{\mathit{Father}}

\newcommand{\HasChild}{\mathit{hasChild}}

\newcommand{\HasParent}{\mathit{hasParent}}

\def\mybox#1{\begin{center} 
\fbox{\parbox{14.85cm}{\begin{center}#1\end{center}}} 
\end{center}} 

\newcommand{\BeginDefinition}[1]{\begin{definition}[{\bf #1}]\begin{em}}
\newcommand{\EndDefinition}{\end{em}\end{definition}}

\newcommand{\bS}{\mathbf{S}} 

\newcommand{\mIqz}{{\mI/_\sim}}
\newcommand{\mIqzQS}{{\mI/_\sim^{QS}}}

\newcommand{\mIq}{{\mI/_\simLP}}
\newcommand{\mIqQS}{{\mI/_\simLP^{QS}}}

\newcommand{\degree}{\mathit{deg}}
\newcommand{\Count}{\mathit{count}}

\newcommand{\notequivLP}{{\not\equiv_{\,\Phi,\mI}}}

\newcommand{\RL}{\mbox{OWL$\,$2$\,$RL}\xspace}
\newcommand{\RLz}{\mbox{OWL$\,$2$\,$RL$_0$}\xspace}
\newcommand{\RLp}{\mbox{OWL$\,$2$\,$RL$^+$}\xspace}

\newcommand{\BBCLearn}{BBCL\xspace}

\newcommand{\BBCLearnS}{BBCL2\xspace}

\newcommand{\PTIME}{{\sc PTime}\xspace}

\newcommand{\mLSP}{\mL_{\Sigma,\Phi}}
\newcommand{\mLSPd}{\mL_{\Sigma,\Phi,d\,}}
\newcommand{\mLSPD}{\mL_{\Sigma^\dag,\Phi^\dag}}

\newcommand{\LargestContainer}{\mathit{LargestContainer}}

\newcommand{\SigmaDag}{\Sigma^\dag}
\newcommand{\SigmaDagI}{\Sigma^\dag_I}
\newcommand{\SigmaDagA}{\Sigma^\dag_A}
\newcommand{\SigmaDagC}{\Sigma^\dag_C}

\newcommand{\SdPdI}{{\SigmaDag,\Phi^\dag,\mI}}

\newcommand{\SigmaDagNA}{\Sigma^\dag_{nA}}
\newcommand{\SigmaDagOR}{\Sigma^\dag_{oR}}
\newcommand{\SigmaDagDR}{\Sigma^\dag_{dR}}

\newcommand{\Range}{\mathit{range}}

\title{On Bisimulations for Description Logics\thanks{$\,$This is a revised and extended version of~\cite{BSDLv1,BSDL-CSP,DivroodiN15}.}} 

\author{Ali Rezaei Divroodi \and Linh Anh Nguyen}
\institute{
Institute of Informatics, University of Warsaw\\
Banacha 2, 02-097 Warsaw, Poland\\
\email{\{rezaei,nguyen\}@mimuw.edu.pl}
}

\authorrunning{A.R. Divroodi and L.A. Nguyen}

\tocauthor{Ali Rezaei Divroodi (University of Warsaw), 
	   Linh Anh Nguyen (University of Warsaw)}

\begin{document} 
\maketitle  
\sloppy 

%----------------------------------------------------------------------------- 

\begin{abstract}
We study bisimulations for useful description logics. The simplest among the considered logics is \ALCreg (a variant of PDL). The others extend that logic with inverse roles, nominals, quantified number restrictions, the universal role, and/or the concept constructor for expressing the local reflexivity of a role. They also allow role axioms. We give results about invariance of concepts, TBoxes and ABoxes, preservation of RBoxes and knowledge bases, and the Hennessy-Milner property w.r.t.\ bisimulations in the considered description logics. Using the invariance results we compare the expressiveness of the considered description logics w.r.t.\ concepts, TBoxes and ABoxes. Our results about separating the expressiveness of description logics are naturally extended to the case when instead of \ALCreg we have any sublogic of \ALCreg that extends \ALC. We also provide results on the largest auto-bisimulations and quotient interpretations w.r.t.\ such equivalence relations. Such results are useful for minimizing interpretations and concept learning in description logics. To deal with minimizing interpretations for the case when the considered logic allows quantified number restrictions and/or the constructor for the local reflexivity of a role, we introduce a new notion called QS-interpretation, which is needed for obtaining expected results. By adapting Hopcroft's automaton minimization algorithm and the Paige-Tarjan algorithm, we give efficient algorithms for computing the partition corresponding to the largest auto-bisimulation of a finite interpretation.
\end{abstract}

%===============================================================================

\section{Introduction}

Description logics (DLs) are variants of modal logic. They are of particular importance in providing a logical formalism for ontologies and the Semantic Web.
DLs represent the domain of interest in terms of concepts, individuals, and roles. A concept is interpreted as a set of individuals, while a role is interpreted as a binary relation among individuals. A~DL is characterized by a set of concept constructors, a set of role constructors, and a set of allowed forms of role axioms and individual assertions. A~knowledge base in a DL usually has three parts: an RBox consisting of axioms about roles, a TBox consisting of terminology axioms, and an ABox consisting of assertions about individuals. 
The basic DL \ALC allows basic concept constructors listed in Table~\ref{table: constr-features}, but does not allow role constructors nor role axioms. The most common additional features for extending \ALC are also listed in Table~\ref{table: constr-features}.

\begin{table}
\begin{center}
\begin{tabular}{|l|c|c|}
\hline
\multicolumn{3}{|c|}{Concept constructors of \ALC} \\ \hline
Constructor & Syntax & Example \\ \hline
complement & $\lnot C$ & $\lnot \Male$ \\ \hline
intersection & $C \mand D$ & $\Human \mand \Male$ \\ \hline
union & $C \mor D$ & $\Doctor \mor \Lawyer$ \\ \hline 
existential restriction & $\E r.C$ & $\E \hasChild.\Male$ \\ \hline
universal restriction & $\V r.C$ & $\V \hasChild.\Female$ \\ \hline
\hline
\multicolumn{3}{|c|}{Some additional constructors/features of other DLs} \\ \hline
Constructor/Feature & Syntax & Example \\ \hline
\underline{i}nverse roles ($\mathcal{I}$) & $r^-$ & $\hasChild^-$ (i.e., $\hasParent$) \\ \hline
\underline{q}ualified number & $\geq\!n\,R.C$ & $\geq\!3\,\hasChild.\Male$ \\ 
restrictions ($\mathcal{Q}$) & $\leq\!n\,R.C$ & $\leq\!2\,\hasParent.\top$ \\ \hline
n\underline{o}minals ($\mathcal{O}$) & $\{a\}$ & $\{\John\}$ \\ \hline
\underline{h}ierarchies of roles ($\mathcal{H}$) & $R \sqsubseteq S$ & $\hasChild \sqsubseteq \hasDescendant$ \\ \hline
tran\underline{s}itive roles ($\mathcal{S}$) & $R \circ R \sqsubseteq R$ & $\hasDescendant \circ \hasDescendant \sqsubseteq \hasDescendant$ \\ \hline
\end{tabular}

\medskip

\caption{Concept constructors for \ALC and some additional constructors/features of other DLs.\label{table: constr-features}}
\end{center}
\end{table}

Given two individuals in an interpretation, sometimes we are interested in the question whether they are ``similar'' or not, i.e., whether they are indiscernible w.r.t.\ the considered description language. Indiscernibility is used, for example, in machine learning. In DLs, it is formally characterized by bisimulation. Roughly speaking, two individuals are indiscernible iff they are bisimilar. 

Bisimulations arose in modal logic~\cite{vBenthem76,vBenthem83,vBenthem84} and state transition systems~\cite{Park81,HennessyM85}. They were introduced by van Benthem under the name {\em p-relation} in~\cite{vBenthem76,vBenthem83} and the name {\em zigzag relation} in~\cite{vBenthem84}. Bisimulations reflect, in a particularly simple and direct way, the locality of the modal satisfaction definition. The famous Van Benthem Characterization Theorem states that modal logic is the bisimulation invariant fragment of first-order logic. Bisimulations have been used to analyze the expressivity of a wide range of extended modal logics (see, e.g., \cite{BRV2001} for details).
In state transition systems, bisimulation is viewed as a binary relation associating systems which behave in the same way in the sense that one system simulates the other and vice versa. Kripke models in modal logic are a special case of labeled state transition systems. Hennessy and Milner~\cite{HennessyM85} showed that weak modal languages could be used to classify various notions of process invariance.
In general, bisimulations are a very natural notion of equivalence for both mathematical and computational investigations.\footnote{This paragraph is based on~\cite{BRV2001}.}

Bisimilarity between two states is usually defined by three conditions (the states have the same label, each transition from one of the states can be simulated by a similar transition from the other, and vice versa). 
As shown in~\cite{BRV2001}, the four program constructors of PDL (propositional dynamic logic) are ``safe'' for these three conditions. That is, we need to specify the mentioned conditions only for atomic programs, and as a consequence, they hold also for complex programs. For bisimulation between two pointed-models, the initial states of the models are also required to be bisimilar. 
When converse is allowed (the case of CPDL), two additional conditions are required for bisimulation~\cite{BRV2001}. 
Bisimulation conditions for dealing with graded modalities were studied in~\cite{Rijke00,ConradieThesis,JaninL04}. In the field of hybrid logic, the bisimulation condition for dealing with nominals is well known (see, e.g., \cite{ArecesBM01}).

In this paper we study bisimulations for the family of DLs which extend \ALCreg (a variant of PDL) with an arbitrary combination of inverse roles, qualified number restrictions, nominals, the universal role, and the concept constructor $\E r.\Self$ for expressing the local reflexivity of a role. 
Inverse roles are like converse modal operators, qualified number restrictions are like graded modalities, and nominals are as in hybrid logic. 

The topic is worth studying due to the following reasons:
\begin{enumerate}
\item Despite that bisimulation conditions are known for PDL and for some features like converse modal operators, graded modal operators and nominals, we are not aware of previous work on bisimulation conditions for the universal role and the concept constructor $\E r.\Self$. More importantly, without proofs one cannot be sure that all the conditions can be combined together to guarantee standard properties like invariance and the Hennessy-Milner property. 

\smallskip

There are many papers on bisimulations, but just a few on bisimulations in DLs:
  \begin{itemize}
  \item In~\cite{KurtoninaR99} Kurtonina and de Rijke studied expressiveness of concept expressions in some DLs by using bisimulations. They considered a family of DLs that are sublogics of the DL $\mathcal{ALCNR}$, which extend \ALC with (unqualified) number restrictions and role conjunction. They did not consider individuals, nominals, qualified number restrictions, the concept constructor $\E r.\Self$, the universal role, and the role constructors like the program constructors of PDL. 

  \item In~\cite{LutzPW11} Lutz et al.\ characterized the expressiveness of TBoxes in the DL \ALCQIO and its sublogics, including the lightweight DLs such as DL-Lite and $\mathcal{EL}$. They also studied invariance of TBoxes and the problem of TBox rewritability. The logic \ALCQIO lacks the role constructors of PDL, the concept constructor $\E r.\Self$ and the universal role.

  \item In~\cite{DivroodiN13} we studied comparisons between interpretations in DLs with respect to ``logical consequences'' of the form of {\em semi-positive concepts} (like semi-positive concept assertions). Such comparisons are called {\em bisimulation-based comparisons} and characterized by conditions similar to the ones of bisimulations defined in the current paper. The problems studied in~\cite{DivroodiN13} are: preservation of semi-positive concepts with respect to comparisons, the Hennessy-Milner property for comparisons, and minimization of interpretations that preserves semi-positive concepts.

  \item Bisimulation-based concept learning in DLs was studied in~\cite{LbRoughification,KSE2012,SoICT2012,ICCCI2012a,TranHHNN2013,TranNH2014}. %All of these works are based on the notion of bisimulation and its properties investigated in the workshop version~\cite{BSDL-CSP} of the current paper. 
  A~survey on these papers is presented in Section~\ref{section: JHPSM}.
  \end{itemize}

The family of DLs studied in this work is large and contains useful DLs. Not only concept constructors and role constructors are allowed, but role axioms are also allowed. In particular, the DL \SROIQ, which is the logical basis of the Web Ontology Language OWL~2, belongs to this class.

\item DLs differ from other logics like modal logics and hybrid logics in the domain of applications and the settings. In DLs, there are special notions like named individual, RBox, TBox, ABox. Also, recall that a knowledge base in a DL usually consists of an RBox, a TBox and an ABox. 
Invariance of ABoxes and preservation of RBoxes and knowledge bases in DLs were not studied before. On the other hand, invariance of TBoxes was studied in the independent work~\cite{abs-1104-2844,LutzPW11} for the DL \ALCQIO and its sublogics. 
%Note that the first version~\cite{abs-1104-2844} of~\cite{LutzPW11} appeared to the public a few days later than the first version~\cite{BSDLv1} of the current paper. 
The works~\cite{abs-1104-2844,LutzPW11} use the notion of global bisimulation to characterize invariance of TBoxes, whose condition is the same as the bisimulation conditions introduced in~\cite{BSDLv1} and the current paper for the universal role.

\item Bisimulation is a very useful notion for DLs. Apart from analyzing expressiveness of DLs, it can be used for minimizing interpretations and concept learning in DLs:
  \begin{itemize}
  \item Roughly speaking, two objects that are bisimilar to each other can be merged. This is the basis for minimizing interpretations. In automated reasoning in DLs, sometimes we want to return a model of a knowledge base (e.g., as a counterexample for a subsumption problem or an instance checking problem). It is expected that the returned model is simple and as small as possible. One can just find some model and minimize it. As another example, given an information system specified by an acyclic knowledge base with a large ABox and a small TBox, one can compute that information system and minimize it to save space and increase efficiency of reasoning tasks. 

  \item Concept learning in DLs is similar to binary classification in traditional machine learning. The difference is that in DLs objects are described not only by attributes but also by relationships between the objects. As bisimulation is the notion for characterizing indiscernibility of objects in DLs,
  %it is very useful for concept learning in DLs~\cite{LbRoughification,KSE2012,SoICT2012,ICCCI2012a,TranHHNN2013,TranNH2014}.
  it is useful for concept learning in DLs.
  \end{itemize}
\end{enumerate}

In this paper we present conditions for bisimulation in a uniform way for the whole considered family of DLs. For this, we introduce bisimulation conditions for the universal role and the concept constructor $\E r.\Self$. 
A special point of our approach is that named individuals are treated as initial states, which requires an appropriate condition for bisimulation. 
Our bisimulation conditions for qualified number restrictions are relatively simpler than the ones given for graded modalities in~\cite{Rijke00,ConradieThesis}.
We prove the standard invariance property (Theorem~\ref{theorem: bs-inv-1}) and the Hennessy-Milner property (Theorem~\ref{theorem: H-M}) and address the following problems:
\begin{itemize}
\item When is a TBox invariant for bisimulation? (Corollary~\ref{cor: TBox-invar} and Theorem~\ref{theorem: TBox-inv-2})
\item When is an ABox invariant for bisimulation? (Theorem~\ref{theorem: ABox-inv})
\item What can be said about preservation of RBoxes w.r.t.~bisimulation? (Theorem~\ref{theorem: preserving RBox})
\item What can be said about invariance or preservation of knowledge bases w.r.t.~bisimulation? (Theorems~\ref{theorem: HJWKA} and~\ref{theorem: DFSHJ})
\end{itemize}

Furthermore, we give results (Theorems~\ref{theorem: quo1}, \ref{theorem: quo-x}, \ref{theorem: mz1}, \ref{theorem: quo-y} and \ref{theorem: mz2}) on the largest auto-bisimulation of an interpretation in a DL, the quotient interpretation w.r.t.~that equivalence relation, and minimality of such a quotient interpretation. To deal with minimizing interpretations for the case when the considered logic allows qualified number restrictions and/or the concept constructor $\E r.\Self$, we introduce a new notion called QS-interpretation, which is needed for obtaining expected results. 

Computing the largest auto-bisimulations in modal logics and state transition systems is standard like Hopcroft's automaton minimization algorithm~\cite{Hopcroft71} and the Paige-Tarjan algorithm~\cite{PaigeT87}. 
By adapting these algorithms, we give efficient algorithms for computing the partition corresponding to the largest auto-bisimulation of a finite interpretation in any DL of the considered family. The adaptation involves the allowed constructors of the considered DLs.

Using the invariance results we compare the expressiveness of the considered DLs w.r.t.\ concepts, TBoxes and ABoxes. Our results about separating the expressiveness of DLs are naturally extended to the case when instead of \ALCreg we have any sublogic of \ALCreg that extends \ALC. 

%In comparison with~\cite{DivroodiN13}, note that \cite{DivroodiN13} deals with semi-positive concepts, while the current paper deals with (general) concepts, RBoxes, TBoxes, ABoxes and knowledge bases. The overlap between \cite{DivroodiN13} and the current paper is small. 

The rest of this paper is structured as follows. In Section~\ref{section: defs} we present notation and semantics of the DLs considered in this paper. In Section~\ref{section: bis-inv} we define bisimulations in those DLs and give our results on invariance and preservation w.r.t. such bisimulations. 
In Section~\ref{section: H-M} we give our results on the Hennessy-Milner property of the considered DLs. Section~\ref{section: aut-bis min} is devoted to auto-bisimulation and minimization. Section~\ref{section: comp l-aut-bis} is devoted to computing the partition corresponding to the largest auto-bisimulation of a finite interpretation. Section~\ref{section: apps} is devoted to applications of bisimulations. In particular, in Section~\ref{section: OIDSJ} we present our results about separating the expressiveness of DLs w.r.t.\ concepts, TBoxes and ABoxes, in Section~\ref{section: IUEKA} we discuss applications of interpretation minimization, and in Section~\ref{section: JHPSM} we present a survey on bisimulation-based concept learning in DLs. 
Section~\ref{section: conc} concludes this work. \VersionB{All proofs of the results in Sections~\ref{section: bis-inv}--\ref{section: comp l-aut-bis} are presented in the appendix.}

%===============================================================================

\section{Preliminaries} 
\label{section: defs} 

\subsection{Notation of Description Logics} 

Our languages use a countable set $\CN$ of {\em concept names}\index{concept name} (atomic concepts)\index{atomic concept}, a countable set $\RN$ of {\em role names}\index{role name} (atomic roles)\index{atomic role}, and a countable set $\IN$ of {\em individual names}\index{individual name}. Let $\Sigma = \CN \cup \RN \cup \IN$. We denote concept names by letters like $A$ and $B$, denote role names by letters like $r$ and $s$, and denote individual names by letters like $a$ and~$b$. 
 
We consider some (additional) {\em DL-features}\index{DL-feature}\index{feature|see{DL-feature}} denoted by $I$ ({\em inverse})\index{inverse}\index{I@$I$|see{inverse}}, $O$ ({\em nominal})\index{nominal}\index{O@$O$|see{nominal}}, $Q$ ({\em qualified number restriction})\index{qualified number restriction}\index{Q@$Q$|see{qualified number restriction}}\index{number restriction}, $U$ ({\em universal role})\index{universal role}\index{U@$U$|see{universal role}}, $\Self$\index{Self@$\Self$}. A {\em set of DL-features} is a set consisting of some or zero of these names. We sometimes abbreviate sets of DL-features, writing e.g., $IOQ$ instead of $\{I,O,Q\}$.  

\BeginDefinition{Syntax of Concepts and Roles}\label{def:HGFDA}\index{LP@$\mLP$}\index{ALCreg@\ALCreg}\index{concept}\index{role}Let $\Phi$ be any set of DL-features and let $\mL$ stand for \ALCreg, which is the name of a DL corresponding to propositional dynamic logic (PDL). The DL language $\mLP$ allows {\em roles} and {\em concepts} defined inductively as follows:
\begin{itemize}
\item if $r \in \RN$ then $r$ is a role of $\mLP$
\item if $A \in \CN$ then $A$ is a concept of $\mLP$
\item if $R$ and $S$ are roles of $\mLP$ and $C$ is a concept of $\mLP$ then 
   \begin{itemize}
   \item $\varepsilon$, $R \circ S$ , $R \sqcup S$, $R^*$ and $C?$ are roles of $\mLP$
   \item $\top$, $\bot$, $\lnot C$, $C \mand D$, $C \mor D$, $\V R.C$ and $\E R.C$ are concepts of $\mLP$
   \item if $I \in \Phi$ then $R^-$ is a role of $\mLP$
   \item if $O \in \Phi$ and $a \in \Sigma_I$ then $\{a\}$ is a concept of $\mLP$
   \item if $Q \in \Phi$, $r \in \RN$ and $n$ is a natural number\\ then $\geq n\,r.C$ and $\leq n\,r.C$ are concepts of $\mLP$
   \item if $\{Q,I\} \subseteq \Phi$, $r \in \RN$ and $n$ is a natural number\\ then $\geq n\,r^-.C$ and $\leq n\,r^-.C$ are concepts of $\mLP$
   \item if $U \in \Phi$ then $U$ is a role of $\mLP$ (we assume $U \notin \RN$)
   \item if $\Self \in \Phi$ and $r \in \RN$ then $\E r.\Self$ is a concept of $\mLP$.
\myend
   \end{itemize}
\end{itemize}
\EndDefinition

We use letters like $R$ and $S$ to denote arbitrary roles, and use letters like $C$ and $D$ to denote arbitrary concepts. 
A role stands for a binary relation, while a concept stands for a unary relation. 

The intended meaning of the role constructors is the following:
\begin{itemize}
\item $R \circ S$ stands for the sequential composition of $R$ and $S$
\item $R \sqcup S$ stands for the set-theoretical union of $R$ and $S$
\item $R^*$ stands for the reflexive and transitive closure of $R$
\item $C?$ stands for the test operator (as of PDL)
\item $R^-$ stands for the {\em inverse} of $R$.
\end{itemize}

We say that a role $R$ is in the {\em converse normal form}\index{converse normal form} (CNF)\index{CNF|see{converse normal form}} if the inverse constructor is applied in $R$ only to role names and the role $U$ is not under the scope of any other role constructor. Since every role can be translated to an equivalent role in CNF,\footnote{For example, $((r \mor s^-)\circ r^*)^- = (r^-)^* \circ (r^- \mor s)$.} in this dissertation we assume that roles are presented in the CNF.

We refer to elements of $\RN$ also as {\em atomic roles}. Let $\RNpm = \RN \cup \{r^- \mid r \in \RN\}$. From now on, by {\em basic roles}\index{basic role} we refer to elements of $\RNpm$ if the considered language allows inverse roles, and refer to elements of $\RN$ otherwise. In general, the language decides whether inverse roles are allowed in the considered context. 

The concept constructors $\V R.C$ and $\E R.C$ correspond respectively to the modal operators $[R]C$ and $\tuple{R}C$ of PDL. 
The concept constructors $\geq n\,R.C$ and $\leq n\,R.C$ are called {\em qualified number restrictions}. They correspond to graded modal operators. 

%-------------------------------------------------------------------------------

\BeginDefinition{RBox -- Box of Role Axioms}\index{RBox}\index{role axiom}A {\em role (inclusion) axiom} in $\mLP$ is an expression of the form $\varepsilon \sqsubseteq r$ or $R_1 \circ \ldots \circ R_k \sqsubseteq r$, where $k \geq 1$ and $R_1,\ldots,R_k$ are basic roles of $\mLP$.\footnote{This definition depends only on whether $\mLP$ allows inverse roles, i.e., whether $I \in \Phi$.} 
An {\em RBox} in $\mLP$ is a finite set of role axioms in $\mLP$.
\myend
\EndDefinition

\BeginDefinition{TBox -- Box of Terminological Axioms}\index{TBox}\index{terminological axiom}\index{TBox axiom}A {\em terminological axiom} in $\mLP$, also called a {\em general concept inclusion} (GCI) in $\mLP$, is an expression of the form $C \sqsubseteq D$, where $C$ and $D$ are concepts in $\mLP$. 
A {\em TBox} in $\mLP$ is a finite set of terminological axioms in $\mLP$. 
\myend
\EndDefinition

\BeginDefinition{ABox -- Box of Individual Assertions}\index{ABox}\index{assertion!individual assertion}\index{assertion!concept assertion}\index{assertion!role assertion}An {\em individual assertion} in $\mLP$ is an expression of one of the forms $C(a)$ ({\em concept assertion}), $R(a,b)$ ({\em positive role assertion}), $\lnot R(a,b)$ ({\em negative role assertion}), $a \doteq b$, and $a \ndoteq b$, where $C$ is a concept and $R$ is a role in~$\mLP$.
An {\em ABox} in $\mLP$ is a finite set of individual assertions in $\mLP$. 
\myend
\EndDefinition

\BeginDefinition{Knowledge Base}\index{knowledge base}A {\em knowledge base} in $\mLP$ is a triple $\tuple{\mR,\mT,\mA}$, where $\mR$ (resp.\ $\mT$, $\mA$) is an RBox (resp.\ a TBox, an ABox) in $\mLP$. 
\myend
\EndDefinition

%-------------------------------------------------------------------------------

\subsection{Semantics of Description Logics} 

As usual, the semantics of a logic is specified by interpretations and the satisfaction relation.

\BeginDefinition{Interpretation}\index{interpretation}An {\em interpretation} $\mI = \langle \Delta^\mI, \cdot^\mI \rangle$ consists of a non-empty set $\Delta^\mI$, called the {\em domain}\index{domain} of $\mI$, and a function $\cdot^\mI$, called the {\em interpretation function} of $\mI$, which maps every concept name $A$ to a subset $A^\mI$ of $\Delta^\mI$, maps every role name $r$ to a binary relation $r^\mI$ on $\Delta^\mI$, and maps every individual name $a$ to an element $a^\mI$ of~$\Delta^\mI$. We say that $\mI$ is a {\em finite interpretation} if $\Delta^\mI$ and $\Sigma$ are finite. 
The interpretation function $\cdot^\mI$ is extended to complex roles and complex concepts as shown in Figure~\ref{fig: int-comp}, where $\#\Gamma$ stands for the cardinality of the set~$\Gamma$, $C^\mI(x)$ denotes $x \in C^\mI$, and $R^\mI(x,y)$ denotes $\tuple{x,y} \in R^\mI$. 
\myend
\EndDefinition

For a set $\Gamma$ of concepts, by $\Gamma^\mI$ we denote the set $\bigcap \{C^\mI \mid C \in \Gamma\}$. 
If $x \in \Gamma^\mI$ then we say that $x$ {\em satisfies}\index{satisfy}~$\Gamma$, $\mI$ {\em satisfies} $\Gamma$ (at $x$) and $\Gamma$ is {\em satisfied}\index{satisfied} (at~$x$) in~$\mI$.

If $R^\mI(x,y)$ holds then we call $y$ an {\em $R$-successor}\index{R-successor@$R$-successor}\index{successor|see{R-successor}} of~$x$.

\begin{figure}[t]
\mybox{ \(
\begin{array}{c}
\begin{array}{rcl}
(R \circ S)^\mI & = & R^\mI \circ S^\mI \\[0.5ex]
(R \sqcup S)^\mI & = & R^\mI \cup S^\mI \\[0.5ex]
(R^*)^\mI & = & (R^\mI)^* \\[0.5ex]
(C?)^\mI & = & \{ \tuple{x,x} \mid C^\mI(x) \} \\[0.5ex]
\varepsilon^\mI & = & \{\tuple{x,x} \mid x \in \Delta^\mI\} \\[0.5ex]
U^\mI & = & \Delta^\mI \times \Delta^\mI \\[0.5ex]
(R^-)^\mI & = & (R^\mI)^{-1} 
\end{array}
\quad\quad\quad
\begin{array}{rcl}
\top^\mI & = & \Delta^\mI \\[0.5ex]
\bot^\mI & = & \emptyset \\[0.5ex]
(\lnot C)^\mI & = & \Delta^\mI \setminus C^\mI \\[0.5ex]
(C \mand D)^\mI & = & C^\mI \cap D^\mI \\[0.5ex]
(C \mor D)^\mI & = & C^\mI \cup D^\mI \\[0.5ex]
\{a\}^\mI & = & \{a^\mI\} \\[0.5ex]
(\E r.\Self)^\mI & = & \{x \in \Delta^\mI \mid r^\mI(x,x)\}
\end{array} \\
\\[-1ex]
\begin{array}{rcl}
(\V R.C)^\mI & = & \{ x \in \Delta^\mI \mid \V y\,[R^\mI(x,y) \textrm{ implies } C^\mI(y)] \} \\[1ex]
(\E R.C)^\mI & = & \{ x \in \Delta^\mI \mid \E y\,[R^\mI(x,y) \textrm{ and } C^\mI(y)] \\[1ex]
(\geq n\,R.C)^\mI & = & \{x \in \Delta^\mI \mid \#\{y \mid R^\mI(x,y) \textrm{ and } C^\mI(y)\} \geq n \} \\[1ex]
(\leq n\,R.C)^\mI & = & \{x \in \Delta^\mI \mid \#\{y \mid R^\mI(x,y) \textrm{ and } C^\mI(y)\} \leq n \}
\end{array}
\end{array}
\) }%\vspace{-0.5em} 
\caption{Interpretation of complex roles and complex concepts.}
\label{fig: int-comp}
\end{figure}

%-------------------------------------------------------------------------------

\BeginDefinition{The Satisfaction Relation}
Given an interpretation $\mI$, define that:
\begin{tabbing}
mmmmmm \= $\mI \models R_1 \circ \ldots \circ R_k \sqsubseteq r$m \= mo \= \kill
\> $\mI \models C \sqsubseteq D$ \> if \> $C^\mI \subseteq D^\mI$ \\[0.5ex] 
\> $\mI \models R_1 \circ \ldots \circ R_k \sqsubseteq r$ \> if \> $R_1^\mI \circ \ldots \circ R_k^\mI \subseteq r^\mI$ \\[0.5ex]
\> $\mI \models \varepsilon \sqsubseteq r$ \> if \> $\varepsilon^\mI \subseteq r^\mI$ \\[0.5ex] 
\> $\mI \models a \doteq b$ \> if \> $a^\mI = b^\mI$ \\[0.5ex] 
\> $\mI \models a \ndoteq b$ \> if \> $a^\mI \neq b^\mI$ \\[0.5ex] 
\> $\mI \models C(a)$ \> if \> $C^\mI(a^\mI)$ holds \\[0.5ex] 
\> $\mI \models R(a,b)$ \> if \> $R^\mI(a^\mI,b^\mI)$ holds \\[0.5ex] 
\> $\mI \models \lnot R(a,b)$ \> if \> $R^\mI(a^\mI,b^\mI)$ does not hold, 
\end{tabbing} 
where the operator $\circ$ stands for the composition of binary relations. 
We say that $\mI$ {\em validates}\index{validate} an axiom (resp. {\em satisfies}\index{satisfy} an assertion) $\varphi$ if $\mI \models \varphi$. In that case, we also say that $\varphi$ is {\em validated}\index{validated} by (resp. {\em satisfied}\index{satisfied} in) $\mI$.
\myend
\EndDefinition

Note that reflexiveness and transitiveness of atomic roles are expressible by role axioms. When $I \in \Phi$, symmetry of an atomic role can also be expressed by a role axiom. 

\BeginDefinition{Semantics}
An interpretation $\mI$ is a {\em model}\index{model} of a ``box'' (RBox, TBox or ABox) if it validates all the axioms/assertions of that ``box''.
It is a {\em model} of a knowledge base $\tuple{\mR,\mT,\mA}$ if it is a model of $\mR$, $\mT$ and~$\mA$. A knowledge base is {\em satisfiable}\index{satisfiable} if it has a model. 
An individual $a$ is said to be an {\em instance} of a concept $C$ w.r.t.\ a knowledge base $\KB$, denoted by $\KB \models C(a)$, if, for every model $\mI$ of $\KB$, $a^\mI \in C^\mI$. 
\myend
\EndDefinition

%-------------------------------------------------------------------------------

\begin{example}\label{example: KJDSA}
Let 
\begin{itemize}
\item $\IN = \{\Alice,\Bob,\Claudia,\Dave,\Eva,\Frank,\George,\Helen\}$, 
\item $\CN = \{\Male,\Female,\Father,\Mother\}$, and 
\item $\RN = \{\HasChild,\HasParent\}$. 
\end{itemize}
Consider the interpretation $\mI$ specified by: 
\begin{itemize}
\item $\Delta^\mI = \{a,b,c,d,e,f,g,h,u,v\}$, 
\item $\Alice^\mI = a$, $\Bob^\mI = b$, \ldots, $\Helen^\mI = h$ ($u$ and $v$ are unnamed individuals),  
\item $\HasChild^\mI$ consists of elements illustrated by edges in the following graph:
\begin{center}
\begin{tabular}{c}
\xymatrix{
&
a:F
\ar@{->}[d] 
\ar@{->}[dr] 
&
b:M
\ar@{->}[dl] 
\ar@{->}[d] 
\\
c:F
\ar@{->}[d] 
\ar@{->}[dr] 
\ar@{->}[drr] 
&
d:M
\ar@{->}[dl] 
\ar@{->}[d] 
\ar@{->}[dr] 
&
e:F
\ar@{->}[dr] 
&
u:M
\ar@{->}[d] 
\\
f:M
&
g:M
&
h:F
&
v:F
}
\end{tabular}
\end{center}
(in this graph, the letter $M$ denotes $\Male$, and $F$ denotes $\Female$),
\item $\HasParent^\mI = (\HasChild^{-1})^\mI = (\HasChild^\mI)^{-1}$, 
\item $\Male^\mI = \{b,d,f,g,u\}$,\ \ $\Female^\mI = \Delta^\mI \setminus \Male^\mI = \{a,c,e,h,v\}`$,
\item $\Father^\mI = (\Male \mand \E\HasChild.\top)^\mI = \{b,d,u\}$,   
\item $\Mother^\mI = (\Female \mand \E\HasChild.\top)^\mI = \{a,c,e\}$.
\end{itemize}
As examples, we have that:
\begin{itemize}
\item $(\E\HasChild.\Self)^\mI = \emptyset$, 
\item $(\geq\!3\,\HasChild.\top)^\mI = \{c,d\}$,
\item $(\geq\!2\,\HasChild.\Male)^\mI = \{c,d\}$,
\item $(\Female\ \mand <\!2\,\HasChild.\top)^\mI = \{e,h,v\}$.
\myend
\end{itemize}
\end{example}

%-------------------------------------------------------------------------------

\begin{example} \label{example1}
Let $\IN = \{a,b,c\}$, $\CN = \{F,M\}$ and $\RN = \{r\}$. One can think of these names as {\em Alice}~($a$), {\em Bob}~($b$), {\em Claudia}~($c$), {\em female}~($F$), {\em male}~($M$), and {\em has\_child}~($r$). In Figure~\ref{fig: example1} we give three interpretations $\mI_1$, $\mI_2$ and $\mI_3$. 

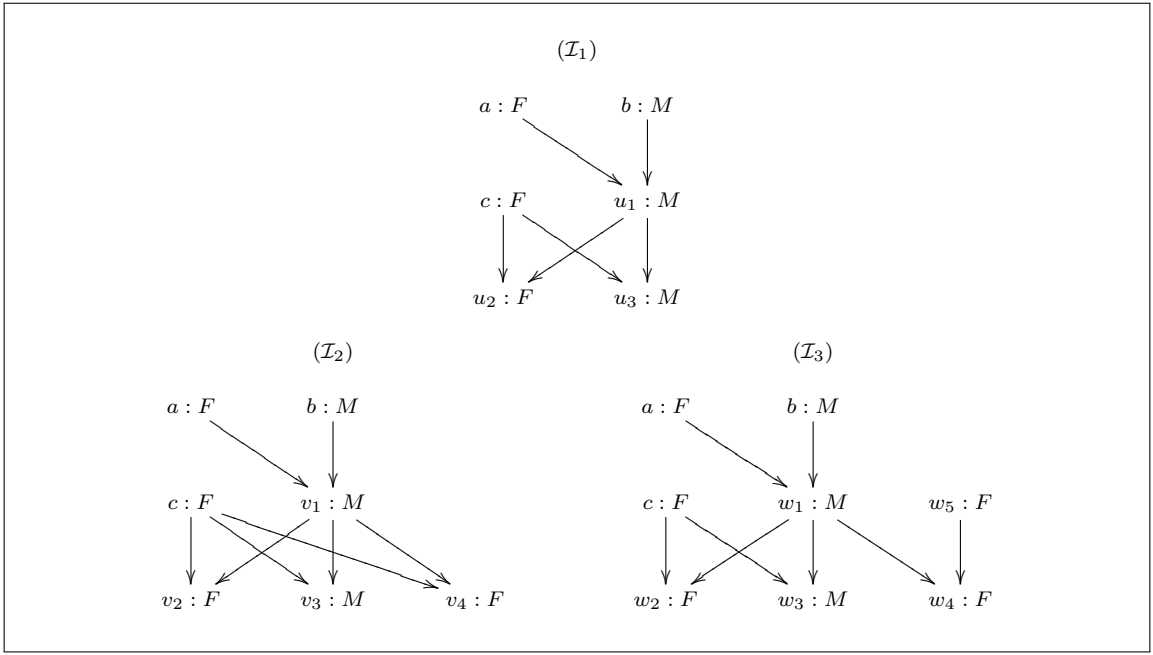
\begin{figure}[h!]
\mybox{
\begin{center}
\begin{tabular}{c}
\begin{scriptsize}
\begin{tabular}{c@{\extracolsep{5em}}c}
\multicolumn{2}{c}{($\mI_1$)}\\
\\
\multicolumn{2}{c}{
\xymatrix{
a:F
\ar@{->}[dr] 
&
b:M
\ar@{->}[d] 
\\
c:F
\ar@{->}[d] 
\ar@{->}[dr] 
&
u_1:M
\ar@{->}[dl] 
\ar@{->}[d] 
\\
u_2:F
&
u_3:M
}
} \\
\\
($\mI_2$) & ($\mI_3$) \\
\\
\xymatrix{
a:F
\ar@{->}[dr] 
&
b:M
\ar@{->}[d] 
\\
c:F
\ar@{->}[d] 
\ar@{->}[dr] 
\ar@{->}[drr] 
&
v_1:M
\ar@{->}[dl] 
\ar@{->}[d] 
\ar@{->}[dr] 
\\
v_2:F
&
v_3:M
&
v_4:F
}
&
\xymatrix{
a:F
\ar@{->}[dr] 
&
b:M
\ar@{->}[d] 
\\
c:F
\ar@{->}[d] 
\ar@{->}[dr] 
&
w_1:M
\ar@{->}[dl] 
\ar@{->}[d] 
\ar@{->}[dr] 
&
w_5:F
\ar@{->}[d] 
\\
w_2:F
&
w_3:M
&
w_4:F
}
\end{tabular}
\end{scriptsize}
\end{tabular}
\end{center}
}
\caption{\label{fig: example1}Interpretations used in Examples~\ref{example1} and \ref{example2}.} 
\end{figure}

The edges are instances of $r$. We have, for example, $\Delta^{\mI_1} = \{a^{\mI_1}, b^{\mI_1}, c^{\mI_1}, u_1, u_2, u_3\}$, where these six elements are pairwise different, $F^{\mI_1} = \{a^{\mI_1}, c^{\mI_1}, u_2\}$, and $M^{\mI_1} = \{b^{\mI_1}, u_1, u_3\}$.\footnote{The elements $u_i$, $v_j$, $w_k$ are unnamed objects. (The elements of $\IN$ can be called {\em named individuals}, while the elements $u_i$, $v_j$, $w_k$ can be called {\em unnamed individuals}.)}
All of these interpretations are models of the following ABox in $\mL_{IOQ}$, where $r^-$ can be read as {\em has\_parent}:
\[ \big\{\ F(a),\; M(b),\; F(c),\; \big(\E r.(\E r^-.\{b\}\ \mand \geq\!2\, r.\E r^-.\{c\})\big)(a)\ \big\}\]
Assuming that $r$ means $\textit{has\_child}$, then the last assertion of the above ABox means ``$a$ and $b$ have a child which in turn has at least two children with $c$''.

All the interpretations $\mI_1$, $\mI_2$ and $\mI_3$ validate the terminological axioms $\lnot F \sqsubseteq M$ and $\{a\} \sqsubseteq \V r^*.(\{a\}$ $\mor$ $\geq\!2\,r^-.\top)$ of $\mL_{IOQ}$. 
\myend
\end{example}

%-------------------------------------------------------------------------------

\section{Bisimulations and Invariance Results}
\label{section: bis-inv}

\BeginDefinition{Bisimulation}\index{bisimulation}\index{LP-bisimulation@$\mLP$-bisimulation}Let $\mI$ and $\mI'$ be interpretations. 
A non-empty binary relation $Z \subseteq \Delta^\mI \times \Delta^{\mI'}$ is called an {\em $\mLP$-bisimulation} between $\mI$ and $\mI'$ if the following conditions hold for every $a \in \Sigma_I$, $A \in \CN$, $r \in \RN$, $x,y \in \Delta^\mI$, $x',y' \in \Delta^{\mI'}$~:
\begin{eqnarray}
&& Z(a^\mI,a^{\mI'}) \label{bs:eqA} \\
&& Z(x,x') \Rightarrow [A^\mI(x) \Leftrightarrow A^{\mI'}(x')] \label{bs:eqB} \\
&& [Z(x,x') \land r^\mI(x,y)] \Rightarrow \E y' \in \Delta^{\mI'}[Z(y,y') \land r^{\mI'}(x',y')] \label{bs:eqC} \\
&& [Z(x,x') \land r^{\mI'}(x',y')] \Rightarrow \E y \in \Delta^\mI [Z(y,y') \land r^\mI(x,y)], \label{bs:eqD} 
\end{eqnarray}
if $I \in \Phi$ then
\begin{eqnarray}
&& [Z(x,x') \land r^\mI(y,x)] \Rightarrow \E y' \in \Delta^{\mI'}[Z(y,y') \land r^{\mI'}(y',x')] \label{bs:eqI1} \\
&& [Z(x,x') \land r^{\mI'}(y',x')] \Rightarrow \E y \in \Delta^\mI [Z(y,y') \land r^\mI(y,x)], \label{bs:eqI2} 
\end{eqnarray}
if $O \in \Phi$ then 
\begin{eqnarray}
&& Z(x,x') \Rightarrow [x = a^\mI \Leftrightarrow x' = a^{\mI'}], \label{bs:eqO0}
\end{eqnarray}
if $Q \in \Phi$ then
\begin{eqnarray}
\parbox{12cm}{if $Z(x,x')$ holds and $y_1,\ldots,y_n$ ($n \geq 1$) are pairwise different elements of $\Delta^\mI$ such that $r^\mI(x,y_i)$ holds for every $1 \leq i \leq n$ then there exist pairwise different elements $y'_1,\ldots,y'_n$ of $\Delta^{\mI'}$ such that $r^{\mI'}(x',y'_i)$ and $Z(y_i,y'_i)$ hold for every $1 \leq i \leq n$} \label{bs:eqQ1} \\[1.5ex]
\parbox{12cm}{if $Z(x,x')$ holds and $y'_1,\ldots,y'_n$ ($n \geq 1$) are pairwise different elements of $\Delta^{\mI'}$ such that $r^{\mI'}(x',y'_i)$ holds for every $1 \leq i \leq n$ then there exist pairwise different elements $y_1,\ldots,y_n$ of $\Delta^\mI$ such that $r^\mI(x,y_i)$ and $Z(y_i,y'_i)$ hold for every $1 \leq i \leq n$,} \label{bs:eqQ2}
\end{eqnarray}
if $\{Q,I\} \subseteq \Phi$ then (additionally)
\begin{eqnarray}
\parbox{12cm}{if $Z(x,x')$ holds and $y_1,\ldots,y_n$ ($n \geq 1$) are pairwise different elements of $\Delta^\mI$ such that $r^\mI(y_i,x)$ holds for every $1 \leq i \leq n$ then there exist pairwise different elements $y'_1,\ldots,y'_n$ of $\Delta^{\mI'}$ such that $r^{\mI'}(y'_i,x')$ and $Z(y_i,y'_i)$ hold for every $1 \leq i \leq n$} \label{bs:eqQI1} \\[1.5ex]
\parbox{12cm}{if $Z(x,x')$ holds and $y'_1,\ldots,y'_n$ ($n \geq 1$) are pairwise different elements of $\Delta^{\mI'}$ such that $r^{\mI'}(y'_i,x')$ holds for every $1 \leq i \leq n$ then there exist pairwise different elements $y_1,\ldots,y_n$ of $\Delta^\mI$ such that $r^\mI(y_i,x)$ and $Z(y_i,y'_i)$ hold for every $1 \leq i \leq n$,} \label{bs:eqQI2}
\end{eqnarray}
if $U \in \Phi$ then
\begin{eqnarray}
&& \V x \in \Delta^\mI\, \E x' \in \Delta^{\mI'}\, Z(x,x') \label{bs:eqU1} \\
&& \V x' \in \Delta^{\mI'}\, \E x \in \Delta^\mI\, Z(x,x'), \label{bs:eqU2} 
\end{eqnarray}
if $\Self \in \Phi$ then 
\begin{eqnarray}
&& Z(x,x') \Rightarrow [r^\mI(x,x) \Leftrightarrow r^{\mI'}(x',x')]. \label{bs:eqSelf}
\end{eqnarray}
For example, if $\Phi = \{I,Q\}$ then only the conditions \eqref{bs:eqA}-\eqref{bs:eqI2} and \eqref{bs:eqQ1}-\eqref{bs:eqQI2} (and all of them) are essential. 
\myend
\EndDefinition

Notice that our bisimulation conditions~\eqref{bs:eqQ1}-\eqref{bs:eqQI2} for qualified number restrictions are relatively simpler than the ones given for graded modalities in~\cite{Rijke00,ConradieThesis}. 

\BeginDefinition{Finitely Branching Interpretation}
An interpretation $\mI$ is {\em finitely branching}\index{finitely branching} (or {\em image-finite}) w.r.t.\ $\mLP$ if, for every $x \in \Delta^\mI$ and every basic role $R$ of $\mLP$, the set $\{y \in \Delta^\mI \mid R^\mI(x,y)\}$ is finite.
\myend
\EndDefinition

%\BeginDefinition{Countably Branching Interpretation}
%An interpretation $\mI$ is {\em countably branching}\index{countably branching} w.r.t.\ $\mLP$ if, for every $x \in \Delta^\mI$ and every basic role $R$ of $\mLP$, the set $\{y \in \Delta^\mI \mid R^\mI(x,y)\}$ is countable.
%\myend
%\EndDefinition

\noindent
Observe that, if $\mI$ and $\mI'$ are finitely branching interpretations, then:
\begin{itemize}
\item the combination of the conditions \eqref{bs:eqQ1} and \eqref{bs:eqQ2} is equivalent to: 
\begin{eqnarray*}
\parbox{13.5cm}{if $Z(x,x')$ holds then there exists a bijection $h: \{y \mid r^\mI(x,y)\} \to \{y' \mid r^{\mI'}(x',y')\}$ such that $h \subseteq Z$,} 
\end{eqnarray*}
\item the combination of the conditions \eqref{bs:eqQI1} and \eqref{bs:eqQI2} is equivalent to: 
\begin{eqnarray*}
\parbox{13.5cm}{if $Z(x,x')$ holds then there exists a bijection $h: \{y \mid r^\mI(y,x)\} \to \{y' \mid r^{\mI'}(y',x')\}$ such that $h \subseteq Z$.}
\end{eqnarray*}
\end{itemize} 

%-------------------------------------------------------------------------------

\begin{lemma} \label{lemma:pr-bis}
\ 
\begin{enumerate}
\item \label{pr-bis-1} The relation $\{\tuple{x,x} \mid x \in \Delta^\mI\}$ is an $\mLP$-bisimulation between $\mI$ and $\mI$.
\item \label{pr-bis-2} If $Z$ is an $\mLP$-bisimulation between $\mI$ and $\mI'$ then $Z^{-1}$ is an $\mLP$-bisimulation between $\mI'$ and $\mI$.
\item \label{pr-bis-3} If $Z_1$ is an $\mLP$-bisimulation between $\mI_0$ and $\mI_1$, and $Z_2$ is an $\mLP$-bisimulation between $\mI_1$ and $\mI_2$, then $Z_1 \circ Z_2$ is an $\mLP$-bisimulation between $\mI_0$ and $\mI_2$.
\item \label{pr-bis-4} If $\mZ$ is a set of $\mLP$-bisimulations between $\mI$ and $\mI'$ then $\bigcup\mZ$ is also an $\mLP$-bisimulation between $\mI$ and $\mI'$.
\end{enumerate}   
\end{lemma}

The proof of this lemma is straightforward. 

\BeginDefinition{$\mLP$-Bisimilarity}\index{bisimilar}\index{LP-bisimilarity@$\mLP$-bisimilarity}An interpretation $\mI$ is {\em $\mLP$-bisimilar} to $\mI'$ if there exists an $\mLP$-bisimulation between them. 
We say that $x \in \Delta^\mI$ is {\em $\mLP$-bisimilar} to $x' \in \Delta^{\mI'}$ if there exists an $\mLP$-bisimulation $Z$ between $\mI$ and $\mI'$ such that $Z(x,x')$ holds. 
\myend
\EndDefinition

By Lemma~\ref{lemma:pr-bis}, the former $\mLP$-bisimilarity relation is an equivalence relation between interpretations. 
The latter $\mLP$-bisimilarity relation is also an equivalence relation (between elements of interpretations' domains). 

\begin{algorithm}[t!]
\caption{checking $\mLP$-bisimilarity of two finite interpretations\label{alg: HGFDS}}
\Input{a set $\Phi$ of DL-features and finite interpretations $\mI$, $\mI'$}
\Output{an $\mLP$-bisimulation between $\mI$ and $\mI'$ if they are $\mLP$-bisimilar, or $\mathit{false}$ otherwise.} 
\BlankLine
$Z := \Delta^\mI \times \Delta^{\mI'}$\;
\BlankLine
\Repeat{$Z$ was not modified during the last iteration}{
  \ForEach{$x \in \Delta^\mI$ and $x' \in \Delta^{\mI'}$}{
     \lIf{some condition among \eqref{bs:eqB}-\eqref{bs:eqQI2}, \eqref{bs:eqSelf} is related to $\Phi$ but not satisfied for some $A, r, y, y', a$}{delete the pair $\tuple{x,x'}$ from $Z$}
  }
}
\BlankLine
\lIf{the condition~\eqref{bs:eqA} is not satisfied for some $a \in \IN$}{\Return $\mathit{false}$}\;
\lIf{$U \in \Phi$ and the condition~\eqref{bs:eqU1} or~\eqref{bs:eqU2} is not satisfied}{\Return $\mathit{false}$}\;
\BlankLine
\Return $Z$\;
\end{algorithm}

To check whether two finite interpretations $\mI$ and $\mI'$ are $\mLP$-bisimilar to each other, one can use Algorithm~\ref{alg: HGFDS} (on page~\pageref{alg: HGFDS}). It is straightforward to prove the following proposition.

\begin{proposition}\label{prop: JHDSS}
Algorithm~\ref{alg: HGFDS} is correct. Furthermore, if it returns $Z$ (but not ``$\mathit{false}$'') then $Z$ is a maximal $\mLP$-bisimulation between $\mI$ and $\mI'$.
\end{proposition}

\begin{example} \label{example2}
Consider the interpretations $\mI_1$, $\mI_2$ and $\mI_3$ given in Figure~\ref{fig: example1} (on page~\pageref{fig: example1}) and described in Example~\ref{example1}. 
\begin{itemize}
\item By using Algorithm~\ref{alg: HGFDS}, it can be checked that all the interpretations $\mI_1$, $\mI_2$ and $\mI_3$ are $\mL$-bisimilar. 
For example, running Algorithm~\ref{alg: HGFDS} for $\mI_1$ and $\mI_2$ with $\Phi = \emptyset$ results in $Z = \{\tuple{a^{\mI_1},a^{\mI_2}}$, $\tuple{b^{\mI_1},b^{\mI_2}}$, $\tuple{c^{\mI_1},c^{\mI_2}}$, $\tuple{u_1,v_1}$, $\tuple{u_2,v_2}$, $\tuple{u_2,v_4}$, $\tuple{u_3,v_3}\}$. By Proposition~\ref{prop: JHDSS}, this is a maximal $\mL$-bisimulation between $\mI_1$ and~$\mI_2$.
 
\item Let us construct a minimal $\mL$-bisimulation between $\mI_1$ and $\mI_2$. We try to construct a minimal relation $Z \subseteq \Delta^{\mI_1} \times \Delta^{\mI_2}$ that satisfies the conditions~\eqref{bs:eqA}-\eqref{bs:eqD}. 
Recall that $\Delta^{\mI_1} = \{a^{\mI_1}, b^{\mI_1}, c^{\mI_1}, u_1, u_2, u_3\}$ and $\Delta^{\mI_2} = \{a^{\mI_2}, b^{\mI_2}, c^{\mI_2}, v_1, v_2, v_3, v_4\}$. To satisfy the condition~\eqref{bs:eqA}, $Z(x^{\mI_1},x^{\mI_2})$ must hold for $x \in \{a,b,c\}$. To satisfy the condition~\eqref{bs:eqC} for the case when $x = a^{\mI_1}$, $x' = a^{\mI_2}$ and $y = u_1$, $Z(u_1,v_1)$ must hold. Observe that, due to the condition~\eqref{bs:eqB} with $A = M$, none of the pairs $\tuple{u_2,v_3}$, $\tuple{u_3,v_2}$, $\tuple{u_3,v_4}$ belongs to $Z$. Since $Z(u_1,v_1)$ holds, to satisfy the condition~\eqref{bs:eqC} for the case when $x = u_1$, $x' = v_1$ and $y = u_3$, $Z(u_3,v_3)$ must hold. Similarly, to satisfy the condition~\eqref{bs:eqD} for the case when $x = u_1$, $x' = v_1$ and $y' = v_2$ (resp.\ $y' = v_4$), we must have that $\tuple{u_2,v_2} \in Z$ (resp.\ $\tuple{u_2,v_4} \in Z$). Summing up, we must have 
$\{\tuple{a^{\mI_1},a^{\mI_2}},\tuple{b^{\mI_1},b^{\mI_2}},\tuple{c^{\mI_1},c^{\mI_2}},\tuple{u_1,v_1},\tuple{u_2,v_2},\tuple{u_2,v_4},\tuple{u_3,v_3}\} \subseteq Z$. Let $Z$ be the set in the left hand side of this inclusion. It is easy to check that $Z$ satisfies all the conditions~\eqref{bs:eqA}-\eqref{bs:eqD}. Hence, $Z$ is a minimal $\mL$-bisimulation between $\mI_1$ and $\mI_2$. Together with the above item, it follows that this is the unique $\mL$-bisimulation between $\mI_1$ and $\mI_2$. 

\item Running Algorithm~\ref{alg: HGFDS} for $\mI_1$ and $\mI_2$ with $\Phi = \{I,O\}$ results in the same set $Z = \{\tuple{a^{\mI_1},a^{\mI_2}}$, $\tuple{b^{\mI_1},b^{\mI_2}}$, $\tuple{c^{\mI_1},c^{\mI_2}}$, $\tuple{u_1,v_1}$, $\tuple{u_2,v_2}$, $\tuple{u_2,v_4}$, $\tuple{u_3,v_3}\}$ as in the case $\Phi = \emptyset$. By Proposition~\ref{prop: JHDSS}, $Z$ is a maximal $\mL_{IO}$-bisimulation between $\mI_1$ and $\mI_2$. It follows that the elements $u_2$ (of $\mI_1$) and $v_2$, $v_4$ (of $\mI_2$) are $\mL_{IO}$-bisimilar. 

\item Running Algorithm~\ref{alg: HGFDS} for $\mI_1$ and $\mI_2$ with $\Phi = \{Q\}$ results in $\mathit{false}$. Hence, $\mI_1$ and $\mI_2$ are not $\mL_Q$-bisimilar. Similarly, the interpretation $\mI_3$ is not $\mL_I$-bisimilar to $\mI_1$ nor $\mI_2$. 
\myend
\end{itemize} 
\end{example}
 
\begin{lemma} \label{lemma: bs-inv-1}
Let $\mI$ and $\mI'$ be interpretations and $Z$ be an $\mLP$-bisimulation between $\mI$ and $ \mI'$. Then the following properties hold for every concept $C$ in $\mLP$, every role $R$ in $\mLP$, every $x, y \in \Delta^\mI$, every $x', y' \in \Delta^{\mI'}$, and every $a \in \Sigma_I$:
\begin{eqnarray}
&& Z(x,x') \Rightarrow [C^\mI(x) \Leftrightarrow C^{\mI'}(x')] \label{bs:eqB-2} \\
&& [Z(x,x') \land R^\mI(x,y)] \Rightarrow \E y' \in \Delta^{\mI'}[Z(y,y') \land R^{\mI'}(x',y')] \label{bs:eqC-2} \\
&& [Z(x,x') \land R^{\mI'}(x',y')] \Rightarrow \E y \in \Delta^\mI [Z(y,y') \land R^\mI(x,y)] \label{bs:eqD-2}
\end{eqnarray}
if $O \in \Phi$ then:
\begin{eqnarray}
&& Z(x,x') \Rightarrow [R^\mI(x,a^\mI) \Leftrightarrow R^{\mI'}(x',a^{\mI'})]. \label{bs:eqO-2}
\end{eqnarray}
\end{lemma}

\newcommand{\LemmaBsInvF}{Let $\mI$ and $\mI'$ be interpretations and $Z$ be an $\mLP$-bisimulation between $\mI$ and $ \mI'$. Then the following properties hold for every concept $C$ in $\mLP$, every role $R$ in $\mLP$, every $x, y \in \Delta^\mI$, every $x', y' \in \Delta^{\mI'}$, and every $a \in \mI$:
\[
\begin{array}{ll}
\eqref{bs:eqB-2}\;\;\; & Z(x,x') \Rightarrow [C^\mI(x) \Leftrightarrow C^{\mI'}(x')]  \\[0.5ex]
\eqref{bs:eqC-2} & [Z(x,x') \land R^\mI(x,y)] \Rightarrow \E y' \in \Delta^{\mI'}[Z(y,y') \land R^{\mI'}(x',y')] \\[0.5ex]
\eqref{bs:eqD-2} & [Z(x,x') \land R^{\mI'}(x',y')] \Rightarrow \E y \in \Delta^\mI [Z(y,y') \land R^\mI(x,y)] 
\end{array}
\]
if $O \in \Phi$ then:
\[
\begin{array}{ll}
\eqref{bs:eqO-2}\;\;\; & Z(x,x') \Rightarrow [R^\mI(x,a^\mI) \Leftrightarrow R^{\mI'}(x',a^{\mI'})]. 
\end{array}
\]
} % \newcommand

\newcommand{\ProofBsInvF}{\begin{proof}
We prove this lemma by induction on the structures of $C$ and $R$.

Consider the assertion~\eqref{bs:eqC-2}. 
Suppose $Z(x,x')$ and $R^\mI(x,y)$ hold. By induction on the structure of~$R$ we prove that there exists $y' \in \Delta^{\mI'}$ such that $Z(y,y')$ and $R^{\mI'}(x',y')$ hold. The base case occurs when $R$ is a role name and the assertion for it follows from~\eqref{bs:eqC}. The induction steps are given below. 
\begin{itemize}
\item Case $R = S_1 \circ S_2$~: We have that $(S_1\circ S_2)^\mI(x,y)$ holds. Hence, there exists $z\in \Delta^\mI$ such that $S_1^\mI(x,z)$ and $S_2^\mI(z,y)$ hold. Since $Z(x,x')$ and $S_1^\mI(x,z)$ hold, by the inductive assumption of~\eqref{bs:eqC-2}, there exists $z' \in \Delta^{\mI'}$ such that $Z(z,z')$ and $S_1^{\mI'}(x',z')$ hold.
Since $Z(z,z')$ and $S_2^\mI(z,y)$ hold, by the inductive assumption of~\eqref{bs:eqC-2}, there exists $y' \in \Delta^{\mI'}$ such that $Z(y,y')$ and $S_2^{\mI'}(z',y')$ hold. Since $S_1^{\mI'}(x',z')$ and $S_2^{\mI'}(z',y')$ hold, we have that $(S_1 \circ S_2)^{\mI'}(x',y') $ holds, i.e.~$R^{\mI'}(x',y')$ holds. 

\item Case $R = S_1 \sqcup S_2$ is trivial.
 
\item Case $R = S^*$~: Since $R^\mI(x,y)$ holds, there exist $x_0, \ldots, x_k \in \Delta^\mI$ with $k \geq 0$ such that $x_0 = x$, $x_k = y$ and, for $1 \leq i \leq k$, $S^\mI(x_{i-1},x_i)$ holds. Let $x'_0 = x'$. For each $1 \leq i \leq k$, since $Z(x_{i-1},x'_{i-1})$ and $S^\mI(x_{i-1},x_i)$ hold, by the inductive assumption of~\eqref{bs:eqC-2}, there exists $x'_i \in \Delta^{\mI'}$ such that $Z(x_i,x'_i)$ and $S^{\mI'}(x'_{i-1},x'_i)$ hold. Hence, $Z(x_k,x'_k)$ and $(S^*)^{\mI'}(x'_0,x'_k)$ hold. Let $y' = x'_k$. Thus, $Z(y,y')$ and $R^{\mI'}(x',y')$ hold. 

\item Case $R = (D?)$~: Since $R^\mI(x,y)$ holds, we have that $D^\mI(x)$ holds and $x = y$. Since $Z(x,x')$ and $D^\mI(x)$ hold, by the inductive assumption of~\eqref{bs:eqB-2}, $D^{\mI'}(x')$ also holds, and hence $R^{\mI'}(x',x')$ holds. By choosing $y' = x'$, both $Z(y,y')$ and $R^{\mI'}(x',y')$ hold. 

\item Case $I \in \Phi$ and $R = r^-$~: The assertion for this case follows from~\eqref{bs:eqI1}.
\end{itemize}

By Lemma~\ref{lemma:pr-bis}(\ref{pr-bis-2}), the assertion~\eqref{bs:eqD-2} follows from the assertion~\eqref{bs:eqC-2}.

Consider the assertion~\eqref{bs:eqO-2} and suppose $O \in \Phi$. By Lemma~\ref{lemma:pr-bis}(\ref{pr-bis-2}), it suffices to show that if $Z(x,x')$ and $R^\mI(x,a^\mI)$ hold then $R^{\mI'}(x',a^{\mI'})$ also holds. We prove this by using similar argumentation as for~\eqref{bs:eqC-2}.
Suppose $Z(x,x')$ and $R^\mI(x,a^\mI)$ hold. We prove that $R^{\mI'}(x',a^{\mI'})$ also holds by induction on the structure of~$R$. 
The base case occurs when $R$ is a role name and the assertion for it follows from~\eqref{bs:eqC} and~\eqref{bs:eqO0}. The induction steps are given below. 
\begin{itemize}
\item Case $R = S_1 \circ S_2$~: We have that $(S_1\circ S_2)^\mI(x,a^\mI)$ holds. Hence, there exists $y\in \Delta^\mI$ such that $S_1^\mI(x,y)$ and $S_2^\mI(y,a^\mI)$ hold. Since $Z(x,x')$ and $S_1^\mI(x,y)$ hold, by the inductive assumption of~\eqref{bs:eqC-2}, there exists $y' \in \Delta^{\mI'}$ such that $Z(y,y')$ and $S_1^{\mI'}(x',y')$ hold. 
Since $Z(y,y')$ and $S_2^\mI(y,a^\mI)$ hold, by the inductive assumption of~\eqref{bs:eqO-2}, $S_2^{\mI'}(y',a^{\mI'})$ holds. Since $S_1^{\mI'}(x',y')$ and $S_2^{\mI'}(y',a^{\mI'})$ hold, we have that $(S_1 \circ S_2)^{\mI'}(x',a^{\mI'}) $ holds, i.e.~$R^{\mI'}(x',a^{\mI'})$ holds. 

\item Case $R = S_1 \sqcup S_2$ is trivial.
 
\item Case $R = S^*$~: Since $R^\mI(x,a^\mI)$ holds, there exist $x_0, \ldots, x_k \in \Delta^\mI$ with $k \geq 0$ such that $x_0 = x$, $x_k = a^\mI$ and, for $1 \leq i \leq k$, $S^\mI(x_{i-1},x_i)$ holds. 
  \begin{itemize}
  \item Case $k = 0$~: We have that $x = a^\mI$. Since $Z(x,x')$ holds, by~\eqref{bs:eqO0}, it follows that $x' = a^{\mI'}$. Hence $R^{\mI'}(x',a^{\mI'})$ holds. 

  \item Case $k > 0$~: Let $x'_0 = x'$. For each $1 \leq i < k$, since $Z(x_{i-1},x'_{i-1})$ and $S^\mI(x_{i-1},x_i)$ hold, by the inductive assumption of~\eqref{bs:eqC-2}, there exists $x'_i \in \Delta^{\mI'}$ such that $Z(x_i,x'_i)$ and $S^{\mI'}(x'_{i-1},x'_i)$ hold. Hence, $Z(x_{k-1},x'_{k-1})$ and $(S^*)^{\mI'}(x'_0,x'_{k-1})$ hold. Since $Z(x_{k-1},x'_{k-1})$ and $S^\mI(x_{k-1},a^\mI)$ hold, by the inductive assumption of~\eqref{bs:eqO-2}, we have that $S^{\mI'}(x'_{k-1},a^{\mI'})$ holds. Since $(S^*)^{\mI'}(x'_0,x'_{k-1})$ holds, it follows that $R^{\mI'}(x',a^{\mI'})$ holds. 
  \end{itemize}

\item Case $R = (D?)$~: Since $R^\mI(x,a^\mI)$ holds, we have that $x = a^\mI$ and $D^\mI(a^\mI)$ holds. Since $Z(x,x')$ holds, by~\eqref{bs:eqO0}, it follows that $x' = a^{\mI'}$. Since $Z(a^\mI,a^{\mI'})$ and $D^\mI(a^\mI)$ hold, by the inductive assumption of~\eqref{bs:eqB-2}, $D^{\mI'}(a^{\mI'})$ also holds. Since $x' = a^{\mI'}$, it follows that $R^{\mI'}(x',a^{\mI'})$ holds. 

\item Case $I \in \Phi$ and $R = r^-$~: The assertion for this case follows from~\eqref{bs:eqI1} and~\eqref{bs:eqO0}.
\end{itemize}

Consider the assertion~\eqref{bs:eqB-2}.
By Lemma~\ref{lemma:pr-bis}(\ref{pr-bis-2}), it suffices to show that if $Z(x,x')$ and $C^\mI(x)$ hold then $C^{\mI'}(x')$ also holds. Suppose $Z(x,x')$ and $C^\mI(x)$ hold.
The cases when $C$ is of the form $\top$, $\bot$, $A$, $\lnot D$, $D \mor D'$ or $D \mand D'$ are trivial.
\begin{itemize}
\item Case $C = \E R.D$~: Since $C^\mI(x)$ holds, there exists $y \in \Delta^\mI$ such that $R^\mI(x,y)$ and $D^\mI(y)$ hold. Since $Z(x,x')$ and $R^\mI(x,y)$ hold, by the assertion~\eqref{bs:eqC-2} (proved earlier), there exists $y' \in \Delta^{\mI'}$ such that $Z(y,y')$ and $R^{\mI'}(x',y')$ hold. Since $Z(y,y')$ and $D^\mI(y)$ hold, by the inductive assumption of~\eqref{bs:eqB-2}, it follows that $D^{\mI'}(y')$ holds. Therefore, $C^{\mI'}(x')$ holds.

\item Case $C = \V R.D$ is reduced to the above case, treating $\V R.D$ as $\lnot \E R.\lnot D$.

\item Case $O \in \Phi$ and $C = \{a\}$~: Since $C^\mI(x)$ holds, we have that $x = a^\mI$. Since $Z(x,x')$ holds, by~\eqref{bs:eqO0}, it follows that $x' = a^{\mI'}$. Hence $C^{\mI'}(x')$ holds. 

\item Case $Q \in \Phi$ and $C = (\geq\!n\,R.D)$, where $R$ is a basic role: Since $C^\mI(x)$ holds, there exist pairwise different $y_1$, \ldots, $y_n \in \Delta^\mI$ such that $R^\mI(x,y_i)$ and $D^\mI(y_i)$ hold for all $1 \leq i \leq n$.
Since $Z(x,x')$ holds, by the conditions~\eqref{bs:eqQ1} and~\eqref{bs:eqQI1}, there exist pairwise different $y'_1$, \ldots, $y'_n \in \Delta^{\mI'}$ such that $R^{\mI'}(x',y'_i)$ and $Z(y_i,y'_i)$ hold for all $1 \leq i \leq n$. 
Since $Z(y_i,y'_i)$ and $D^\mI(y_i)$ hold, by the inductive assumption of~\eqref{bs:eqB-2}, it follows that $D^{\mI'}(y'_i)$ holds. Since $R^{\mI'}(x',y'_i)$ and $D^{\mI'}(y'_i)$ hold for all $1 \leq i \leq n$, it follows that $C^{\mI'}(x')$ holds. 

\item Case $Q \in \Phi$ and $C = (\leq\!n\,R.D)$, where $R$ is a basic role: This case is reduced to the above case, treating $\leq n\,R.D$ as $\lnot(\geq (n+1)\,R.D)$.

\item Case $\Self \in \Phi$ and $C = \E r.\Self$~: Since $C^\mI(x)$ holds, we have that $r^\mI(x,x)$ holds. By~\eqref{bs:eqSelf}, it follows that $r^{\mI'}(x',x')$ holds. Hence $C^{\mI'}(x')$ holds.
\myend
\end{itemize}
\end{proof}
} % \newcommand

\VersionA{\ProofBsInvF}

%-------------------------------------------------------------------------------

\BeginDefinition{Invariance of a Concept}\index{invariance}A~concept $C$ in $\mLP$ is said to be {\em invariant for $\mLP$-bisimulation} if, for any interpretations $\mI$, $\mI'$ and any $\mLP$-bisimulation $Z$ between $\mI$ and $\mI'$, if $Z(x,x')$ holds then $x \in C^\mI$ iff $x' \in C^{\mI'}$. 
\myend
\EndDefinition

\begin{theorem} \label{theorem: bs-inv-1}
All concepts in $\mLP$ are invariant for $\mLP$-bisimulation.
\end{theorem}

This theorem follows immediately from the assertion~\eqref{bs:eqB-2} of Lemma~\ref{lemma: bs-inv-1}.

\BeginDefinition{Invariance of a TBox, an ABox or a Knowledge Base}\index{invariance}A TBox $\mT$ in $\mLP$ is said to be {\em invariant for $\mLP$-bisimulation} if, for every interpretations $\mI$ and $\mI'$, if there exists an $\mLP$-bisimulation between $\mI$ and $\mI'$ then $\mI$ is a model of $\mT$ iff $\mI'$ is a model of $\mT$. 
The notions of whether an ABox or a knowledge base in $\mLP$ is invariant for $\mLP$-bisimulation are defined similarly. 
\myend
\EndDefinition

\newcommand{\CorTBoxInvar}{If $U \in \Phi$ then all TBoxes in $\mLP$ are invariant for $\mLP$-bisimulation.}

\begin{corollary} \label{cor: TBox-invar}
\CorTBoxInvar
\end{corollary}

\newcommand{\ProofCorTBoxInvar}{\begin{proof}
Suppose $U \in \Phi$ and let $\mT$ be a TBox in $\mLP$ and $\mI$, $\mI'$ be interpretations. Suppose that $\mI$ is a model of $\mT$, and $Z$ is an $\mLP$-bisimulation between $\mI$ and $\mI'$. We show that $\mI'$ is a model of $\mT$. 
Let $C \sqsubseteq D$ be an axiom from $\mT$ and let $x' \in \Delta^{\mI'}$. We need to show that $x' \in (\lnot C \mor D)^{\mI'}$. 
By~\eqref{bs:eqU2}, there exists $x \in \Delta^\mI$ such that $Z(x,x')$ holds. Since $\mI$ is a model of $\mT$, we have that $x \in (\lnot C \mor D)^\mI$, which, by Theorem~\ref{theorem: bs-inv-1}, implies that $x' \in (\lnot C \mor D)^{\mI'}$. 
\myendVS
\end{proof}
} % \newcommand

\VersionA{\ProofCorTBoxInvar}

\BeginDefinition{Unreachable-Objects-Free Interpretation}\index{unreachable-objects-free}An interpretation $\mI$ is said to be {\em unreachable-objects-free} (w.r.t.\ the considered language) if every element of $\Delta^\mI$ is reachable from some $a^\mI$, where $a \in \IN$, via a path consisting of edges being instances of basic roles.
\myend
\EndDefinition

%An element of the domain of $\mI$ is called a named object if it is $a^\mI$ for some $a \in \IN$. From the point of view of users, named objects are the most important ones, and unnamed objects are less important. 

It is clear that, if $U \notin \Phi$, $C$ is a concept of $\mLP$ and $a \in \IN$, then $\mI \models C(a)$ iff $\mI' \models C(a)$, where $\mI'$ is the unreachable-objects-free interpretation obtained from $\mI$ by deleting from the domain unreachable objects. That is, when $U \notin \Phi$, unreachable objects are redundant for the instance checking problem. Therefore, it is worth considering the class of unreachable-objects-free interpretations. 

Like Corollary~\ref{cor: TBox-invar}, the following theorem concerns invariance of TBoxes w.r.t.~$\mLP$-bisimulation.

\newcommand{\TheoremTBoxInvS}{Let $\mT$ be a TBox in $\mLP$ and $\mI$, $\mI'$ be unreachable-objects-free interpretations (w.r.t.~$\mLP$) such that there exists an $\mLP$-bisimulation between $\mI$ and $\mI'$. Then $\mI$ is a model of $\mT$ iff $\mI'$ is a model of~$\mT$.}

\begin{theorem} \label{theorem: TBox-inv-2}
\TheoremTBoxInvS
\end{theorem}

\newcommand{\ProofThmTBoxInvS}{\begin{proof}
Let $Z$ be an $\mLP$-bisimulation between $\mI$ and $\mI'$. 
By Lemma~\ref{lemma:pr-bis}(\ref{pr-bis-2}), it suffices to show that if $\mI$ is a model of $\mT$ then $\mI'$ is also a model of $\mT$. Suppose $\mI$ is a model of $\mT$. Let $C \sqsubseteq D$ be an axiom from $\mT$. We need to show that $C^{\mI'} \subseteq D^{\mI'}$. Let $x' \in C^{\mI'}$. We show that $x' \in D^{\mI'}$. 

Since $\mI'$ is an unreachable-objects-free interpretation, there exist elements $x'_0$, \ldots, $x'_k$ of $\Delta^{\mI'}$ and basic roles $R_1$, \ldots, $R_k$ with $k \geq 0$ such that $x'_0 = a^{\mI'}$ for some $a \in \IN$, $x'_k = x'$ and, for $1 \leq i \leq k$, $R_i^{\mI'}(x'_{i-1},x'_i)$ holds.

By~\eqref{bs:eqA}, $Z(a^\mI,a^{\mI'})$ holds. Let $x_0 = a^\mI$. For each $1 \leq i \leq k$, since $Z(x_{i-1},x'_{i-1})$ and $R_i^{\mI'}(x'_{i-1},x'_i)$ hold, by~\eqref{bs:eqD-2}, there exists $x_i \in \Delta^\mI$ such that $Z(x_i,x'_i)$ and $R_i^\mI(x_{i-1},x_i)$ hold. Let $x = x_k$. Thus, $Z(x,x')$ holds. 
Since $x' \in C^{\mI'}$, by Theorem~\ref{theorem: bs-inv-1}, we have that $x \in C^\mI$. Since $\mI$ is a model of $\mT$, it follows that $x \in D^\mI$. By Theorem~\ref{theorem: bs-inv-1}, we derive that $x' \in D^{\mI'}$, which completes the proof. 
\myendVS
\end{proof}
} % \newcommand

\VersionA{\ProofThmTBoxInvS}

To justify that Corollary~\ref{cor: TBox-invar} and Theorem~\ref{theorem: TBox-inv-2} are as strong as possible, we present here a simple example with $U \notin \Phi$ and one of $\mI$, $\mI'$ being not unreachable-objects-free such that $\mI$ and $\mI'$ are $\mLP$-bisimilar but there exists a TBox $\mT$ such that $\mI \models \mT$ and $\mI' \nvDash \mT$:
\begin{example}
Assume that $U \notin \Phi$ and let $\CN = \{A\}$, $\RN = \emptyset$, $\IN = \{a\}$ (i.e., the signature consists of only concept name $A$ and individual name $a$). Let $\mI$ and $\mI'$ be the interpretations specified by: $\Delta^\mI = \{a\}$, $\Delta^{\mI'} = \{a,u\}$, $a^\mI = a^{\mI'} = a$, $A^\mI = A^{\mI'} = \{a\}$. It is easy to see that $Z = \{\tuple{a,a}\}$ is an $\mLP$-bisimulation between $\mI$ and $\mI'$ (it satisfies all of the conditions \eqref{bs:eqA}-\eqref{bs:eqQI2} and \eqref{bs:eqSelf}). However, $\mI$ is a model of the TBox $\{\top \sqsubseteq A\}$, while $\mI'$ is not.
\myend
\end{example}

%-------------------------------------------------------------------------------

As mentioned in the introduction, in the independent work~\cite{LutzPW11} Lutz et al.\ use the notion of global bisimulation to characterize invariance of TBoxes, whose condition is the same as our bisimulation conditions \eqref{bs:eqU1} and \eqref{bs:eqU2} for the universal role. 
Their result on invariance of TBoxes is not stronger than our Corollary~\ref{cor: TBox-invar}: one can just add $U$ to $\Phi$, and the considered TBox, which may not use $U$, is invariant w.r.t.\ the corresponding bisimulation satisfying the conditions \eqref{bs:eqU1} and \eqref{bs:eqU2}. 
Furthermore, the family of DLs considered in this dissertation contains other logics than the DL \ALCQIO considered in~\cite{LutzPW11}. 
On the matter of originality of our Corollary~\ref{cor: TBox-invar} and Theorem~\ref{theorem: TBox-inv-2}, note that they appeared to the public early in~\cite{BSDLv1}. 

%-------------------------------------------------------------------------------

The following theorem concerns invariance of ABoxes w.r.t.~$\mLP$-bisimulation.

\newcommand{\TheoremABoxInv}{Let $\mA$ be an ABox in $\mLP$. 
If $O \in \Phi$ or $\mA$ contains only assertions of the form $C(a)$ then $\mA$ is invariant for $\mLP$-bisimulation.}

\begin{theorem} \label{theorem: ABox-inv}
\TheoremABoxInv
\end{theorem}

\newcommand{\ProofThmABoxInv}{\begin{proof}
Suppose that $O \in \Phi$ or $\mA$ contains only assertions of the form $C(a)$. 
Let $\mI$ and $\mI'$ be interpretations and let $Z$ be an $\mLP$-bisimulation between $\mI$ and $\mI'$. By Lemma~\ref{lemma:pr-bis}(\ref{pr-bis-2}), it suffices to show that if $\mI$ is a model of $\mA$ then $\mI'$ is also a model of $\mA$. Suppose $\mI$ is an model of $\mA$. Let $\varphi$ be an assertion from $\mA$. We need to show that $\mI' \models \varphi$.
\begin{itemize}
\item Case $\varphi = (a \doteq b)$~: Since $\mI \models \varphi$, we have that $a^\mI = b^\mI$. By~\eqref{bs:eqA}, $Z(a^\mI,a^{\mI'})$ and $Z(b^\mI,b^{\mI'})$ hold. Since $a^\mI = b^\mI$, by~\eqref{bs:eqO0}, it follows that $a^{\mI'} = b^{\mI'}$. Hence $\mI' \models \varphi$.

\item Case $\varphi = (a \ndoteq b)$ is reduced to the above case, by using Lemma~\ref{lemma:pr-bis}(\ref{pr-bis-2}).

\item Case $\varphi = C(a)$~: By~\eqref{bs:eqA}, $Z(a^\mI,a^{\mI'})$ holds. Since $\mI \models \varphi$, $C^\mI(a^\mI)$ holds. By~\eqref{bs:eqB-2}, it follows that $C^{\mI'}(a^{\mI'})$ holds. Thus $\mI' \models \varphi$.  

\item Case $\varphi = R(a,b)$~: By~\eqref{bs:eqA}, $Z(a^\mI,a^{\mI'})$ holds. Since $\mI \models \varphi$, $R^\mI(a^\mI,b^\mI)$ holds. By~\eqref{bs:eqC-2}, there exists $y' \in \Delta^{\mI'}$ such that $Z(b^\mI,y')$ and $R^{\mI'}(a^{\mI'},y')$ hold. Consider $C = \{b\}$ (the assumption $O \in \Phi$ is used here). Since $Z(b^\mI,y')$ and $C^\mI(b^\mI)$ hold, by~\eqref{bs:eqB-2}, $C^{\mI'}(y')$ holds, which means $y' = b^{\mI'}$. Thus $R^{\mI'}(a^{\mI'},b^{\mI'})$ holds, i.e., $\mI' \models \varphi$.

\item Case $\varphi = \lnot R(a,b)$ is reduced to the above case, by using Lemma~\ref{lemma:pr-bis}(\ref{pr-bis-2}). 
\myend
\end{itemize}
\end{proof}
} % \newcommand

\VersionA{\ProofThmABoxInv}

Clearly, the condition ``$O \in \Phi$ or $\mA$ contains only assertions of the form $C(a)$'' of the above theorem covers many useful cases. 
The following example justifies that this theorem is as strong as possible. 

\begin{example}
We show that if $O \notin \Phi$ then none of the ABoxes $\mA_1 = \{a \doteq b\}$, $\mA_2 = \{a \ndoteq b\}$, $\mA_3 = \{r(a,b)\}$, $\mA_4 = \{\lnot r(a,b)\}$ is invariant for $\mLP$-bisimulation.  
Assume that $O \notin \Phi$ and let $\CN = \emptyset$, $\IN = \{a,b\}$, $\RN = \{r\}$. Let $\mI$ and $\mI'$ be the interpretations specified by: 
\begin{center}
\includegraphics{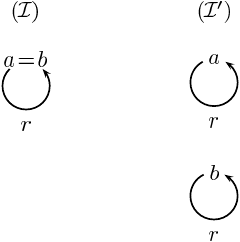}
\end{center}
$\Delta^\mI = \Delta^{\mI'} = \{u,v\}$ with $u \neq v$, $a^\mI = b^\mI = a^{\mI'} = u$, $b^{\mI'} = v$, and $r^\mI = r^{\mI'} = \{\tuple{u,u},\tuple{v,v}\}$. 
It can be checked that $Z = \Delta^\mI \times \Delta^{\mI'}$ is an $\mLP$-bisimulation between $\mI$ and $\mI'$. However:
  \begin{itemize}
  \item $\mI$ is a model of $\mA_1$, while $\mI'$ is not
  \item $\mI'$ is a model of $\mA_2$, while $\mI$ is not
  \item $\mI$ is a model of $\mA_3$, while $\mI'$ is not
  \item $\mI'$ is a model of $\mA_4$, while $\mI$ is not.
\myend
  \end{itemize} 
\end{example}

%-------------------------------------------------------------------------------

In general, RBoxes are not invariant for $\mLP$-bisimulations. (The Van~Benthem Characterization Theorem states that a first-order formula is invariant for bisimulations iff it is equivalent to the translation of a modal formula (see, e.g., \cite{BRV2001}).) We give below a simple example about this:
\begin{example}
Let $\CN = \emptyset$, $\RN = \{r\}$, $\IN = \{a\}$ (i.e., the signature consists of only role name $r$ and individual name $a$) and $\Phi = \emptyset$. Let $\mI$ and $\mI'$ be the interpretations specified~by: 
\begin{center}
\includegraphics{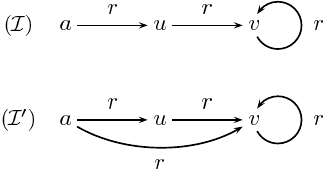}
\end{center}
$\Delta^\mI = \Delta^{\mI'} = \{a,u,v\}$, $a^\mI = a^{\mI'} = a$, $r^\mI = \{\tuple{a,u},\tuple{u,v},\tuple{v,v}\}$ and $r^{\mI'} = r^\mI \cup \{\tuple{a,v}\}$. It can be checked that $Z = \Delta^\mI \times \Delta^{\mI'}$ is an $\mLP$-bisimulation between $\mI$ and $\mI'$. However, $\mI'$ is a model of the RBox $\{r \circ r \sqsubseteq r\}$, while $\mI$ is not.
\myend
\end{example}

\BeginDefinition{Least R-Extension of an Interpretation}\index{least r-extension}An interpretation $\mI'$ is an r-extension of an interpretation $\mI$ if $\Delta^{\mI'} = \Delta^\mI$, $\cdot^{\mI'}$~differs from $\cdot^\mI$ only in interpreting role names, and for all $r \in \RN$, $r^{\mI'} \supseteq r^\mI$. 

Given an interpretation $\mI$ and an RBox $\mR$, the {\em least r-extension of $\mI$ validating $\mR$} is the r-extension $\mI'$ of $\mI$ such that $\mI'$ is a model of $\mR$ and, for every r-extension $\mI''$ of $\mI$, if $\mI''$ is a model of $\mR$ then $r^{\mI'} \subseteq r^{\mI''}$ for all $r \in \RN$.
\myend
\EndDefinition

The least r-extension exists and is unique because the axioms of $\mR$ correspond to non-negative Horn clauses of first-order logic. 
Namely, a role axiom $\varepsilon \sqsubseteq r$ corresponds to the following Horn clause
\[ \V x\ r(x,x), \]
and a role axiom $R_1 \circ \ldots \circ R_k \sqsubseteq r$ corresponds to the following Horn clause 
\[ \V x_0\ldots\V x_k\ [R_1(x_0,x_1) \land \ldots \land R_k(x_{k-1},x_k) \to r(x_0,x_k)], \]
where $s^-(x,y)$ stands for $s(y,x)$. It is clear that one can extend the relations standing for the roles in a minimal way to satisfy all of the Horn clauses corresponding to the axioms of~$\mR$.

\newcommand{\TheoremPreservingRBox}{Suppose $\Phi \subseteq \{I,O,U\}$ and let $\mR$ be an RBox in $\mLP$. Let $\mI_0$ be a model of $\mR$, $Z$ be an $\mLP$-bisimulation between $\mI_0$ and an interpretation $\mI_1$, and $\mI'_1$ be the least r-extension of $\mI_1$ validating $\mR$. Then $Z$ is an $\mLP$-bisimulation between $\mI_0$ and $\mI'_1$.}
 
\begin{theorem} \label{theorem: preserving RBox}
\TheoremPreservingRBox
\end{theorem}

This theorem states that, even in the case when interpretations $\mI_0$ and $\mI_1$ are $\mLP$-bisimilar but $\mI_0 \models \mR$ while $\mI_1 \nvDash \mR$, we can modify $\mI_1$ slightly by adding some edges (i.e. instances of roles) to obtain a model $\mI'_1$ of $\mR$ that is $\mLP$-bisimilar with $\mI_0$ (and hence also with $\mI_1$). This theorem is thus natural. 

\medskip

\newcommand{\ProofThmPreservingRBox}{\begin{proof}
We only need to prove that, for every $r \in \RN$, $x \in \Delta^\mI_0$, $x',y' \in \Delta^{\mI'_1}$~:
\begin{enumerate}
\item $[Z(x,x') \land r^{\mI'_1}(x',y')] \Rightarrow \E y \in \Delta^{\mI_0} [Z(y,y') \land r^{\mI_0}(x,y)]$

\item if $I \in \Phi$ then 
\( 
	[ Z(x,x') \land r^{\mI'_1}(y',x')] \Rightarrow 
		\E y \in \Delta^{\mI_0} [Z(y,y') \land r^{\mI_0}(y,x)]. 
\)
\end{enumerate}

We prove these assertions by induction on the timestamps of the steps that extend relations $r^{\mI_1}$ to $r^{\mI'_1}$, for $r \in \RN$. 

Consider the first assertion. Suppose $Z(x,x')$ and $r^{\mI'_1}(x',y')$ hold. We need to show there exists $y \in \Delta^{\mI_0}$ such that $Z(y,y')$ and $r^{\mI_0}(x,y)$ hold. There are the following three cases:
\begin{itemize}
\item Case $r^{\mI'_1}(x',y')$ holds because $r^{\mI_1}(x',y')$ holds: The assertion holds because $Z$ is an $\mLP$-bisimulation between $\mI_0$ and $\mI_1$. 

\item Case $r^{\mI'_1}(x',y')$ holds because $(\varepsilon \sqsubseteq r) \in \mR$ and $y' = x'$~: Take $y = x$. Thus, $Z(y,y')$ holds. Since $\mI_0$ is a model of $\mR$, it validates $\varepsilon \sqsubseteq r$, and hence $r^{\mI_0}(x,y)$ also holds. 

\item Case $r^{\mI'_1}(x',y')$ holds because $R_1 \circ\ldots\circ R_k \sqsubseteq r$ is an axiom of $\mR$ and there exist $x'_0 = x, x'_1, \ldots, x'_{k-1}, x'_k = y'$ such that $R_i^{\mI'_1}(x'_{i-1},x'_i)$ holds for all $1 \leq i \leq k$~: 
Let $x_0 = x$. 
For each $1 \leq i \leq k$, since $Z(x_{i-1},x'_{i-1})$ and $R_i^{\mI'_1}(x'_{i-1},x'_i)$ hold, by the inductive assumptions of both the assertions, there exists $x_i \in \Delta^{\mI_0}$ such that $Z(x_i,x'_i)$ and $R_i^{\mI_0}(x_{i-1},x_i)$ hold. Thus, $Z(x_k,x'_k)$ holds. Since $\mI_0$ validates the axiom $R_1 \circ\ldots\circ R_k \sqsubseteq r$ of $\mR$, we also have that $r^{\mI_0}(x_0,x_k)$ holds. We choose $y = x_k$ and finish with the proof of the first assertion.
\end{itemize}

The proof of the second assertion is similar to the proof of the first one. 
\myendVS
\end{proof}
} % \newcommand

\VersionA{\ProofThmPreservingRBox}

%-------------------------------------------------------------------------------

\begin{example}
To justify that the form of the above theorem is as strong as possible, we show that allowing either $Q$ or $\Self$ in $\Phi$ can make the theorem wrong. In the following: $u_i \neq u_j$ if $i \neq j$; $v_i \neq v_j$ if $i \neq j$; and $u_i \neq v_j$ for all $i$, $j$. Here are examples:
\begin{enumerate}
\item 
Assume that $\Self \in \Phi$ and $\Phi \subseteq \{\Self,O,U\}$. Let $\CN = \emptyset$, $\IN = \{a\}$ and $\RN = \{r\}$. 
Let $\mI_0$ and $\mI_1$ be the interpretations specified by: 
\begin{center}
\includegraphics{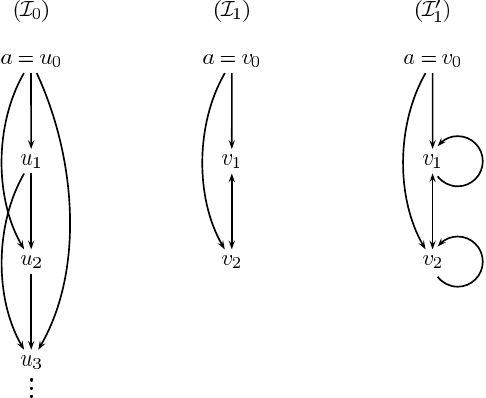}
\end{center}
  \begin{itemize}
  \item $\Delta^{\mI_0} = \{u_i \mid i \geq 0\}$, $a^{\mI_0} = u_0$, 
	$r^{\mI_0} = \{\tuple{u_i,u_j} \mid i < j\}$
  \item $\Delta^{\mI_1} = \{v_0,v_1,v_2\}$, $a^{\mI_1} = v_0$,
	$r^{\mI_1} = \{\tuple{v_0,v_1},\tuple{v_1,v_2},\tuple{v_2,v_1},\tuple{v_0,v_2}\}$.
  \end{itemize}

Let $Z = \{\tuple{u_0,v_0}\} \cup \{\tuple{u_i,v_j} \mid i \geq 1\ \textrm{and}\ (j = 1\ \textrm{or}\ j = 2)\}$. It is easy to check that $Z$ is an $\mLP$-bisimulation between $\mI_0$ and $\mI_1$, $\mI_0$ is a model of the RBox $\mR = \{r \circ r \sqsubseteq r\}$, but $\mI_1$ is not. 
Let $\mI'_1$ be the least r-extension of $\mI_1$ validating $\mR$. We have that $\{\tuple{v_1,v_1},\tuple{v_2,v_2}\} \subseteq r^{\mI'_1}$, while $\tuple{u_i,u_i} \notin r^{\mI_0}$ for all $i \geq 0$. Hence $\{v_1,v_2\} \subseteq (\E\Self.r)^{\mI'_1}$, while $u_i \notin (\E\Self.r)^{\mI_0}$ for all $i \geq 0$, Thus, it is easy to check that $Z$ is not an $\mLP$-bisimulation between $\mI_0$ and $\mI'_1$. 

\item Assume that $Q \in \Phi$ and $\Self \notin \Phi$. Let $\CN = \emptyset$, $\IN = \{a\}$ and $\RN = \{r\}$. Let $\mI_0$ and $\mI_1$ be the interpretations specified by: 
\begin{center}
\includegraphics{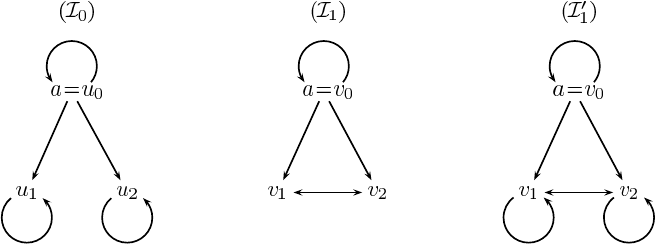}
\end{center}
  \begin{itemize}
  \item $\Delta^{\mI_0} = \{u_0,u_1,u_2\}$, $a^{\mI_0} = u_0$,\\ 
	$r^{\mI_0} = \{\tuple{u_0,u_0},\tuple{u_0,u_1},\tuple{u_0,u_2},\tuple{u_1,u_1},\tuple{u_2,u_2}\}$
  \item $\Delta^{\mI_1} = \{v_0,v_1,v_2\}$, $a^{\mI_1} = v_0$,\\
	$r^{\mI_1} = \{\tuple{v_0,v_0},\tuple{v_0,v_1},\tuple{v_0,v_2},\tuple{v_1,v_2},\tuple{v_2,v_1}\}$.
  \end{itemize}

Let $Z = \{\tuple{u_0,v_0}\} \cup (\{u_1,u_2\} \times \{v_1,v_2\})$. It is easy to check that $Z$ is an $\mLP$-bisimulation between $\mI_0$ and $\mI_1$, $\mI_0$ is a model of the RBox $\mR = \{\varepsilon \sqsubseteq r\}$, but $\mI_1$ is not. 
Let $\mI'_1$ be the least r-extension of $\mI_1$ validating $\mR$. We have that
$\{\tuple{v_1,v_1},\tuple{v_2,v_2}\} \subseteq r^{\mI'_1}$. Hence $\{v_1,v_2\} \subseteq (\geq\!2\,r.\top)^{\mI'_1}$, while $u_i \notin (\geq\!2\,r.\top)^{\mI_0}$ for both $i \in \{1,2\}$. Thus, it is easy to check that $Z$ is not an $\mLP$-bisimulation between $\mI_0$ and $\mI'_1$. 

\item 
Assume that $Q \in \Phi$. 
Let $\CN = \emptyset$, $\IN = \{a\}$, $\RN = \{r,s\}$ and let $\mI_0$, $\mI_1$ be the interpretations specified by: 
\begin{center}
\includegraphics{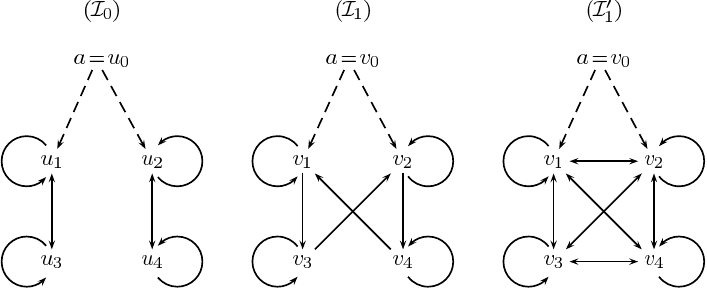}
\end{center}
  \begin{itemize}
  \item $\Delta^{\mI_0} = \{u_0,\ldots,u_4\}$, $a^{\mI_0} = u_0$, 
	$r^{\mI_0} = \{\tuple{u_0,u_1},\tuple{u_0,u_2}\}$,\\
	$s^{\mI_0} = \{\tuple{u_i,u_j} \mid \{i,j\} \subseteq \{1,3\}
			\textrm{ or } \{i,j\} \subseteq \{2,4\}\}$
  \item $\Delta^{\mI_1} = \{v_0,\ldots,v_4\}$, $a^{\mI_1} = v_0$, 
	$r^{\mI_1} = \{\tuple{v_0,v_1},\tuple{v_0,v_2}\}$,\\
	$s^{\mI_1} = \{\tuple{v_i,v_i} \mid 1 \leq i \leq 4\} \cup \{\tuple{v_1,v_3},\tuple{v_3,v_2},\tuple{v_2,v_4},\tuple{v_4,v_1}\}$. 
  \end{itemize}

Let $Z = \{\tuple{u_0,v_0}\} \cup (\{u_1,u_2\} \times \{v_1,v_2\}) \cup (\{u_3,u_4\} \times \{v_3,v_4\})$. It is easy to check that $Z$ is an $\mLP$-bisimulation between $\mI_0$ and $\mI_1$, $\mI_0$ is a model of the RBox \mbox{$\mR = \{s \circ s \sqsubseteq s\}$}, but $\mI_1$ is not. 
Let $\mI'_1$ be the least r-extension of $\mI_1$ validating $\mR$. We have that
$\{\tuple{v_3,v_4},\tuple{v_3,v_1}\} \subseteq s^{\mI'_1}$. Hence $v_3 \in (\geq\!4\,s.\top)^{\mI'_1}$, while $u_i \notin (\geq\!4\,s.\top)^{\mI_0}$ for all $0 \leq i \leq 4$. Thus, it is easy to check that $Z$ is not an $\mLP$-bisimulation between $\mI_0$ and~$\mI'_1$. 
\myend
\end{enumerate}
\end{example}

%-------------------------------------------------------------------------------

The following theorem concerns invariance of knowledge bases w.r.t.~$\mLP$-bisimulation. 
As stated before, in general, RBoxes are not invariant for $\mLP$-bisimulations. Thus, it is natural to consider the case when the considered RBox is empty. Restricting to this case, generality of the below theorem follows from the generality of Theorems~\ref{theorem: TBox-inv-2} and \ref{theorem: ABox-inv}. The case when the considered RBox is not empty is addressed in Theorem~\ref{theorem: DFSHJ}. 

\begin{theorem} \label{theorem: HJWKA}
Let $\tuple{\mR,\mT,\mA}$ be a knowledge base in $\mLP$ such that $\mR = \emptyset$ and either $O \in \Phi$ or $\mA$ contains only assertions of the form $C(a)$. Let $\mI$ and $\mI'$ be unreachable-objects-free interpretations (w.r.t.~$\mLP$) such that there exists an $\mLP$-bisimulation between $\mI$ and $\mI'$. 
Then $\mI$ is a model of $\tuple{\mR,\mT,\mA}$ iff $\mI'$ is a model of $\tuple{\mR,\mT,\mA}$.
\end{theorem}

This theorem follows immediately from Theorems~\ref{theorem: TBox-inv-2} and \ref{theorem: ABox-inv}.

The following theorem concerns preservation of knowledge bases under $\mLP$-bisimulation. Its generality follows from the generality of Theorems~\ref{theorem: TBox-inv-2}, \ref{theorem: ABox-inv} and~\ref{theorem: preserving RBox}. Clearly, it covers many useful cases. 

\begin{theorem} \label{theorem: DFSHJ}
Suppose $\Phi \subseteq \{I,O,U\}$ and let $\tuple{\mR,\mT,\mA}$ be a knowledge base in $\mLP$ such that if $O \notin \Phi$ then $\mA$ contains only assertions of the form $C(a)$. 
Let $\mI_0$ and $\mI_1$ be interpretations such that: $\mI_0$ is a model of $\mR$, there is an $\mLP$-bisimulation $Z$ between $\mI_0$ and $\mI_1$, and if $U \notin \Phi$ then $\mI_0$ and $\mI_1$ are unreachable-objects-free (w.r.t.~$\mLP$). 
Let $\mI'_1$ be the least r-extension of $\mI_1$ validating $\mR$. Then:
\begin{enumerate}
\item $Z$ is an $\mLP$-bisimulation between $\mI_0$ and $\mI'_1$,
\item $\mI'_1$ is a model of $\tuple{\mR,\mT,\mA}$ iff $\mI_0$ is a model of $\tuple{\mR,\mT,\mA}$.
\end{enumerate}
\end{theorem}

This theorem follows immediately from Corollary~\ref{cor: TBox-invar} and Theorems~\ref{theorem: TBox-inv-2}, \ref{theorem: ABox-inv}, \ref{theorem: preserving RBox}.

%-------------------------------------------------------------------------------

\section{The Hennessy-Milner Property}
\label{section: H-M}

\BeginDefinition{Modally Saturated Interpretation}\index{modally saturated}An interpretation $\mI$ is said to be {\em modally saturated} w.r.t.~$\mLP$ if the following conditions hold:
\begin{itemize}
\item for every $x \in \Delta^\mI$, every basic role $R$ of $\mLP$ and every infinite set $\Gamma$ of concepts in~$\mLP$, if for every finite subset $\Lambda$ of $\Gamma$ there exists an $R$-successor of $x$ that satisfies $\Lambda$, then there exists an $R$-successor of $x$ that satisfies $\Gamma$;

\item if $Q \in \Phi$ then, for every $x \in \Delta^\mI$, every basic role $R$ of $\mLP$, every infinite set $\Gamma$ of concepts in~$\mLP$ and every natural number $n$, if for every finite subset $\Lambda$ of $\Gamma$ there exist $n$ pairwise different $R$-successors of $x$ that satisfy $\Lambda$, then there exist $n$ pairwise different $R$-successors of $x$ that satisfy $\Gamma$;

\item if $U \in \Phi$ and $\mI$ is not unreachable-objects-free then, for every infinite set $\Gamma$ of concepts in~$\mLP$, if every finite subset $\Lambda$ of $\Gamma$ is satisfied in $\mI$ (i.e.~$\Lambda^\mI \neq \emptyset$) then $\Gamma$ is also satisfied in $\mI$ (i.e.~$\Gamma^\mI \neq \emptyset$). 
\myend
\end{itemize}
\EndDefinition

%\noindent
Observe that {\em $\omega$-saturated} interpretations (defined, e.g., as in~\cite{Rijke00}) are modally saturated. 

\begin{proposition}
Every finite interpretation is modally saturated. 
Every finitely branching and unreachable-objects-free interpretation is modally saturated. If $U \notin \Phi$ then every finitely branching interpretation is modally saturated. 
\end{proposition}

The proof of this proposition is straightforward. 

\BeginDefinition{$\mLP$-Equivalence}\index{LP-equivalence@$\mLP$-equivalence}Let $\mI$ and $\mI'$ be interpretations, and let $x \in \Delta^\mI$ and $x' \in \Delta^{\mI'}$. We say that $x$ is {\em $\mLP$-equivalent} to $x'$ if, for every concept $C$ in $\mLP$, $x \in C^\mI$ iff $x' \in C^{\mI'}$.
\myend
\EndDefinition

\newcommand{\TheoremHM}{Let $\mI$ and $\mI'$ be modally saturated interpretations (w.r.t.~$\mLP$) such that, for every $a \in \IN$, $a^\mI$ is $\mLP$-equivalent to $a^{\mI'}$. Suppose that if $U \in \Phi$ then either both $\mI$ and $\mI'$ are unreachable-objects-free or both of them are not unreachable-objects-free. Then $x \in \Delta^\mI$ is $\mLP$-equivalent to $x' \in \Delta^{\mI'}$ iff there exists an $\mLP$-bisimulation $Z$ between $\mI$ and $\mI'$ such that $Z(x,x')$ holds. In particular, the relation $\{\tuple{x,x'} \in \Delta^ \mI\times \Delta^{\mI'} \mid x$ is $\mLP$-equivalent to $x'\}$ is an $\mLP$-bisimulation between $\mI$ and $\mI'$ when it is not empty.}

\begin{theorem}[The Hennessy-Milner Property\index{Hennessy-Milner property}] \label{theorem: H-M}
\TheoremHM
\end{theorem}

\newcommand{\ProofThmHM}{\begin{proof}
Consider the ``$\Leftarrow$'' direction. 
Suppose $Z$ is an $\mLP$-bisimulation between $\mI$ and $\mI'$ such that $Z(x,x')$ holds. By~\eqref{bs:eqB-2}, for every concept $C$ in $\mLP$, $C^\mI(x)$ holds iff $C^{\mI'}(x')$ holds. Therefore, $x$ is $\mLP$-equivalent to $x'$.

Now consider the ``$\Rightarrow$'' direction. 
Define $Z = \{\tuple{x,x'} \in \Delta^\mI \times \Delta^{\mI'}  \mid x$ is $\mLP$-equivalent to $x'\}$ and assume that $Z$ is not empty. We show that $Z$ is an $\mLP$-bisimulation between $\mI$ and $\mI'$.
\begin{itemize} 
\item The assertion~\eqref{bs:eqA} follows from the assumption of the theorem.

\item Consider the assertion~\eqref{bs:eqB} and suppose $Z(x,x')$ holds. By the definitions of $Z$ and $\mLP$-equivalence, it follows that, for every concept name~$A$, $A^\mI(x)$ holds iff $A^{\mI'}(x')$ holds.

\item Consider the assertion \eqref{bs:eqC} and suppose that $Z(x,x')$ and $r^\mI(x,y)$ hold. Let $\bS = \{y' \mid r^{\mI'}(x',y')\}$. We want to show there exists $y' \in \bS$ such that $Z(y,y')$ holds. 
For the sake of contradiction, suppose that, for every $y' \in \bS$, $Z(y,y')$ does not hold, which means $y$ is not $\mLP$-equivalent to $y'$. Thus, for every  $y' \in \bS$, there exists a concept $C_{y'}$ such that $y \in C_{y'}^\mI$ but $y' \notin C_{y'}^{\mI'}$. Let $\Gamma = \{C_{y'} \mid y' \in \bS\}$. Thus, no $y' \in \bS$ satisfies $\Gamma$ (i.e.~$\bS \cap \Gamma^{\mI'} = \emptyset$). Since $\mI'$ is modally saturated, it follows that there exists a finite set $\Lambda$ of $\Gamma$ such that, for every $y' \in \bS$, $y' \notin \Lambda^{\mI'}$. Let $C = \E r.\bigsqcap\Lambda$, where $\bigsqcap\{C_1,\ldots,C_n\} = C_1 \mand\ldots\mand C_n$ and $\bigsqcap\emptyset = \top$.  
Thus, $x \in C^\mI$ but $x' \notin C^{\mI'}$, which contradicts the fact that $x$ is $\mLP$-equivalent to $x'$. Therefore, there exists $y' \in \bS$ such that $Z(y,y')$ holds.

\item The assertion \eqref{bs:eqD} can be proved analogously as for \eqref{bs:eqC}. 

\item Consider the assertions~\eqref{bs:eqI1} and~\eqref{bs:eqI2} and the case $I \in \Phi$. Observe that the argumentation used for proving \eqref{bs:eqC} are still applicable when replacing $r$ by $r^-$. Hence the assertion~\eqref{bs:eqI1} holds. Similarly, the assertion~\eqref{bs:eqI2} also holds. 

\item Consider the assertion \eqref{bs:eqO0} and the case $O \in \Phi$. Suppose $Z(x,x')$ holds. Take $C = \{a\}$. Since $x$ is $\mLP$-equivalent to $x'$, $x \in C^\mI$ iff $x' \in C^{\mI'}$. Hence, $x = a^\mI$ iff $x' = a^{\mI'}$. 

\item Consider the assertion~\eqref{bs:eqQ1} and the case $Q \in \Phi$. Suppose $Z(x,x')$ holds. Let $\bS = \{y \in \Delta^\mI \mid r^\mI(x,y)\}$ and $\bS' = \{y' \in \Delta^{\mI'} \mid r^{\mI'}(x',y')\}$. Let $y_1,\ldots,y_n$ be pairwise different elements of $\bS$. We need to show that there exist pairwise different $y'_1,\ldots,y'_n \in \bS'$ such that $y'_i$ is $\mLP$-equivalent to $y_i$ for every $1 \leq i \leq n$. Without loss of generality, assume that $y_1,\ldots,y_n$ are $\mLP$-equivalent to each other. Let $\bS'' = \{y' \in \bS' \mid y'$ is not $\mLP$-equivalent to $y_1\}$. 
Thus, for every  $y' \in \bS''$, there exists a concept $C_{y'}$ such that $y_1 \in C_{y'}^\mI$ but $y' \notin C_{y'}^{\mI'}$. Let $\Gamma = \{C_{y'} \mid y' \in \bS''\}$. Note that every element of $\Gamma^{\mI'}$ is $\mLP$-equivalent to $y_1$. 
For every finite subset $\Lambda$ of $\Gamma$, since $y_1,\ldots,y_n \in \Lambda^\mI$, we have $x \in (\geq\!n\,r.\bigsqcap\Lambda)^\mI$, and since $Z(x,x')$ holds, we also have that $x' \in (\geq\!n\,r.\bigsqcap\Lambda)^{\mI'}$, which means there are at least $n$ pairwise different $y'_1,\ldots,y'_n \in \bS'$ that belong to $\Lambda^{\mI'}$. Since $\mI'$ is modally saturated, it follows that there are at least $n$ pairwise different $y'_1,\ldots,y'_n \in \bS'$ that belong to $\Gamma^{\mI'}$ and are thus $\mLP$-equivalent to $y_1$ and any $y_i$ with $2 \leq i \leq n$. 

\item The assertion~\eqref{bs:eqQ2} for the case $Q \in \Phi$ and the assertions~\eqref{bs:eqQI1} and~\eqref{bs:eqQI2} for the case $\{Q,I\} \subseteq \Phi$ can be proved analogously as for~\eqref{bs:eqQ1}. 

\item Consider the assertion~\eqref{bs:eqU1} and the case $U \in \Phi$. If $\mI$ is unreachable-objects-free then the assertion~\eqref{bs:eqU1} follows from the assertions~\eqref{bs:eqA}, \eqref{bs:eqC} and \eqref{bs:eqI1}. So, assume that $\mI$ is not unreachable-objects-free. Thus, $\mI'$ is also not unreachable-objects-free. Since $Z$ is not empty, there exists $\tuple{y,y'} \in Z$. Let $x \in \Delta^\mI$. For the sake of contradiction suppose that no $x' \in \Delta^{\mI'}$ is $\mLP$-equivalent to~$x$. Thus, for every $x' \in \Delta^{\mI'}$, there exists a concept $C_{x'}$ such that $x \in C_{x'}^\mI$ but $x' \notin C_{x'}^{\mI'}$. Let $\Gamma = \{C_{x'} \mid x' \in \Delta^{\mI'}\}$. For any finite subset $\Lambda$ of $\Gamma$, since $x \in \Lambda^\mI$, we have that $y \in (\E U.\bigsqcap\Lambda)^\mI$, which implies that $y' \in (\E U.\bigsqcap\Lambda)^{\mI'}$, which means $\Lambda$ is satisfied in $\mI'$. Since $\mI'$ is modally saturated and not unreachable-objects-free, it follows that $\Gamma$ is satisfied in $\mI'$, which is a contradiction.  

\item The assertion \eqref{bs:eqU2} can be proved analogously as for \eqref{bs:eqU1}. 

\item Consider the assertion~\eqref{bs:eqSelf} and the case $\Self \in \Phi$. Suppose $Z(x,x')$ holds. Thus, $x \in (\E r.\Self)^\mI$ iff $x' \in (\E r.\Self)^{\mI'}$. Hence, $r^{\mI}(x,x)$ holds iff $r^{\mI'}(x',x')$ holds. 
\myend
\end{itemize}
\end{proof}
} % \newcommand

\VersionA{\ProofThmHM}

%-------------------------------------------------------------------------------

\section{Auto-Bisimulation and Minimization}
\label{section: aut-bis min}

\BeginDefinition{$\mLP$-Auto-Bisimulation}\index{LP-auto-bisimulation@$\mLP$-auto-bisimulation}\index{auto-bisimulation}An $\mLP$-bisimulation between $\mI$ and itself is called an {\em $\mLP$-auto-bisimulation of $\mI$}. An $\mLP$-auto-bisimulation of $\mI$ is said to be the {\em largest} if it is larger than or equal to ($\supseteq$) any other $\mLP$-auto-bisimulation of~$\mI$. 
\myend
\EndDefinition

\begin{proposition}\label{prop: JFDSP}
For every interpretation $\mI$, the largest $\mLP$-auto-bisimulation of $\mI$ exists and is an equivalence relation.
\end{proposition}

This proposition follows from Lemma~\ref{lemma:pr-bis}. 

Given an interpretation $\mI$, by $\simLP$\index{0simLP@$\simLP$} we denote the largest $\mLP$-auto-bisimulation of $\mI$, and by $\equivLP$\index{0equivLP@$\equivLP$} we denote the binary relation on $\Delta^\mI$ with the property that \mbox{$x\ \equivLP\ x'$} iff $x$ is $\mLP$-equivalent to $x'$. 

\newcommand{\PropLABS}{For every modally saturated interpretation $\mI$, $\equivLP$ is the largest $\mLP$-auto-bisimulation of $\mI$ (i.e.\ the relations $\equivLP$ and $\simLP$ coincide).}
 
\begin{proposition}\label{prop: LABS}
\PropLABS
\end{proposition}

\newcommand{\ProofPropLABS}{\begin{proof}
By Theorem~\ref{theorem: H-M}, $\equivLP$ is an $\mLP$-auto-bisimulation of $\mI$. We now show that it is the largest one. Suppose $Z$ is another $\mLP$-auto-bisimulation of $\mI$. If $Z(x,x')$ holds then, by~\eqref{bs:eqB-2}, for every concept $C$ of $\mLP$, $C^\mI(x)$ holds iff $C^\mI(x')$ holds, and hence $x\ \equivLP\ x'$. Therefore, $Z \subseteq\, \equivLP$.
\myendVS
\end{proof}
} % \newcommand

\VersionA{\ProofPropLABS}

\subsection{The Case without $Q$ and $\Self$}

\BeginDefinition{Quotient Interpretation}\index{quotient interpretation}\index{I0q@$\mIqz$|see{quotient interpretation}}Given an interpretation $\mI$, the {\em quotient interpretation} $\mIqz$ of $\mI$ w.r.t.\ an equivalence relation $\sim\ \subseteq \Delta^\mI \times \Delta^\mI$ is defined as usual: 
\begin{itemize}
\item $\Delta^\mIqz = \{[x]_\sim \mid x \in \Delta^\mI\}$, where $[x]_\sim$ is the equivalence class of $x$ w.r.t.~$\sim$, 
\item $a^\mIqz = [a^\mI]_\sim$, for $a \in \IN$, 
\item $A^\mIqz = \{[x]_\sim \mid x \in A^\mI\}$, for $A \in \CN$, 
\item $r^\mIqz = \{\tuple{[x]_\sim,[y]_\sim} \mid \tuple{x,y} \in r^\mI\}$, for $r \in \RN$.
\myend
\end{itemize}
\EndDefinition

\newcommand{\TheoremQuoF}{If $\Phi \subseteq \{I,O,U\}$ then, for every interpretation $\mI$, the relation $Z = \{\tuple{x,[x]_\simLP} \mid x \in \Delta^\mI\}$ is an $\mLP$-bisimulation between $\mI$ and $\mIq$.}

\begin{theorem} \label{theorem: quo1}
\TheoremQuoF
\end{theorem}

\newcommand{\ProofThmQuoF}{\begin{proof}
Suppose $\Phi \subseteq \{I,O,U\}$. 
We have to consider the assertions~\eqref{bs:eqA}-\eqref{bs:eqO0}, \eqref{bs:eqU1}, \eqref{bs:eqU2} for $\mI' = \mIq$.
By the definition of $\mIq$, the assertions~\eqref{bs:eqA} and~\eqref{bs:eqB} clearly hold. 
Similarly, the assertion~\eqref{bs:eqO0} for the case $O \in \Phi$ and the assertions~\eqref{bs:eqU1}, \eqref{bs:eqU2} for the case $U \in \Phi$ also hold.  

Consider the assertion~\eqref{bs:eqC}. Suppose $Z(x,x')$ and $r^\mI(x,y)$ hold. We need to show there exists $y' \in \Delta^\mIq$ such that $Z(y,y')$ and $r^\mIq(x',y')$ hold. We must have that $x' = [x]_\simLP$. Take $y' = [y]_\simLP$. Clearly, the goals are satisfied. 

For a similar reason, the assertion~\eqref{bs:eqI1} for the case $I \in \Phi$ holds. 

Consider the assertion~\eqref{bs:eqD}. Suppose $Z(x,x')$ and $r^\mIq(x',y')$ hold. We need to show there exists $y \in \Delta^\mI$ such that $Z(y,y')$ and $r^\mI(x,y)$ hold. We must have that $x' = [x]_\simLP$. Since $r^\mIq(x',y')$ holds, there exists $y \in y'$ such that $r^\mI(x,y)$ holds. Clearly, $y' = [y]_\simLP$ and $Z(y,y')$ holds.

For a similar reason, the assertion~\eqref{bs:eqI2} for the case $I \in \Phi$ holds. 
\myendVS
\end{proof}
} % \newcommand

\VersionA{\ProofThmQuoF}

The following theorem concerns invariance of terminological axioms and concept assertions, as well as preservation of role axioms and other individual assertion under the transformation of an interpretation to its quotient using the largest $\mLP$-auto-bisimulation. 

\newcommand{\TheoremQuoX}{Suppose $\Phi \subseteq \{I,O,U\}$ and let $\mI$ be an interpretation. Then:
\begin{enumerate}
\item For every expression $\varphi$ which is either a terminological axiom in $\mLP$ or a concept assertion (of the form $C(a)$) in $\mLP$, $\mI \models \varphi$ iff $\mIq \models \varphi$.
\item For every expression $\varphi$ which is either a role inclusion axiom or an individual assertion of the form $R(a,b)$ or $a \doteq b$, if $\mI \models \varphi$ then $\mIq \models \varphi$.
\end{enumerate}}

\begin{theorem} \label{theorem: quo-x}
\TheoremQuoX
\end{theorem}

\newcommand{\ProofThmQuoX}{\begin{proof}
The first assertion follows from Theorems~\ref{theorem: quo1}, \ref{theorem: bs-inv-1} and the definition of $\mIq$. 
Consider the second assertion. 
This assertion for the cases when $\varphi$ is of the form $\varepsilon \sqsubseteq r$, $R(a,b)$ or $a \doteq b$ follows immediately from the definition of $\mIq$. Let $\varphi = (R_1 \circ\ldots\circ R_k \sqsubseteq r)$ and suppose $\mI \models \varphi$. We show that $\mIq \models \varphi$. 
Let $v_0, \ldots, v_k$ be elements of $\Delta^\mIq$ such that, for $1 \leq i \leq k$, $R_i^\mIq(v_{i-1},v_i)$ holds. We need to show that $r^\mIq(v_0,v_k)$ holds. 

For $1 \leq i \leq k$, since $R_i^\mIq(v_{i-1},v_i)$ holds, there exist $y_{i-1} \in v_{i-1}$ and $z_i \in v_i$ such that $R_i^\mI(y_{i-1},z_i)$ holds. 
Let $x_0 = y_0$. 
For $1 \leq i \leq k$, since $x_{i-1}\, \simLP\, y_{i-1}$ and $R_i^\mI(y_{i-1}, z_i)$ hold, by~\eqref{bs:eqC-2}, there exists $x_i$ such that $x_i\, \simLP\, z_i$ and $R_i^\mI(x_{i-1},x_i)$ hold, which implies $x_i \in v_i$ and $x_i\, \simLP\, y_i$ (when $i < k$). Since $\mI \models (R_1 \circ\ldots\circ R_k \sqsubseteq r)$, it follows that $r^\mI(x_0,x_k)$ holds. Therefore, by definition, $r^\mIq(v_0,v_k)$ holds. 
\myendVS
\end{proof}
} % \newcommand

\VersionA{\ProofThmQuoX}

An interpretation $\mI$ is said to be {\em minimal}\index{minimal} among a class of interpretations if $\mI$ belongs to that class and, for every other interpretation $\mI'$ of that class, $\#\Delta^\mI \leq \#\Delta^{\mI'}$ (the cardinality of $\Delta^\mI$ is less than or equal to the cardinality of~$\Delta^{\mI'}$).
The following theorem concerns minimality of quotient interpretations generated by using the largest $\mLP$-auto-bisimulations. 

\newcommand{\TheoremMZF}{Suppose $\Phi \subseteq \{I,O,U\}$ and let $\mI$ be an interpretation. 
\begin{enumerate}
\item If $\mI$ is unreachable-objects-free or $U \in \Phi$ then $\mIq$ is a minimal interpretation $\mLP$-bisimilar to~$\mI$.
\item If $\mIq$ is finite then it is a minimal interpretation that validates the same set of terminological axioms in $\mLP$ as~$\mI$.
\item If $\mIq$ is unreachable-objects-free and finitely branching then it is a minimal interpretation that satisfies the same set of concept assertions in $\mLP$ as~$\mI$.
\end{enumerate}}

\begin{theorem} \label{theorem: mz1}
\TheoremMZF
\end{theorem}

\newcommand{\ProofThmMZF}{\begin{proof}
By Theorems~\ref{theorem: quo1} and~\ref{theorem: quo-x}, $\mIq$ is $\mLP$-bisimilar to~$\mI$, validates the same set of terminological axioms in $\mLP$ as $\mI$, and satisfies the same set of concept assertions in $\mLP$ as~$\mI$. 

Since $\simLP$ is the largest $\mLP$-auto-bisimulation of $\mI$, by Lemma~\ref{lemma:pr-bis}(\ref{pr-bis-4}), for $u,v \in \Delta^\mIq$, if $u \neq v$ then $u$ is not $\mLP$-bisimilar to $v$. 
Let $Z = \{\tuple{[x]_\simLP,x} \mid x \in \Delta^\mI\}$.
By Theorem~\ref{theorem: quo1} and Lemma~\ref{lemma:pr-bis}(\ref{pr-bis-2}), $Z$ is an $\mLP$-bisimulation between $\mIq$ and~$\mI$.

Consider the first assertion and suppose that either $\mI$ is unreachable-objects-free or $U \in \Phi$. Let $\mI'$ be any interpretation $\mLP$-bisimilar to $\mI$. We show that $\#\Delta^\mIq \leq \#\Delta^{\mI'}$. 
Let $Z'$ be an $\mLP$-bisimulation between $\mI$ and $\mI'$, and let $Z'' = Z \circ Z'$. By Lemma~\ref{lemma:pr-bis}(\ref{pr-bis-3}), $Z''$ is an $\mLP$-bisimulation between $\mIq$ and $\mI'$. 
If $\mI$ is unreachable-objects-free, then $\mIq$ is also unreachable-objects-free, and by \eqref{bs:eqA}, \eqref{bs:eqC} and \eqref{bs:eqI1}, for every $u \in \Delta^\mIq$, there exists $x_u \in \Delta^{\mI'}$ such that $Z''(u,x_u)$ holds.
If $U \in \Phi$ then, by~\eqref{bs:eqU1}, we also have that, for every $u \in \Delta^\mIq$, there exists $x_u \in \Delta^{\mI'}$ such that $Z''(u,x_u)$ holds. 
Let $u,v \in \Delta^\mIq$ and $u \neq v$. If $x_u = x_v$ then, since $u$ is $\mLP$-bisimilar to $x_u$ and $x_v$ is $\mLP$-bisimilar to $v$, we would have that $u$ is $\mLP$-bisimilar to $v$, which is a contradiction. Therefore $x_u \neq x_v$ and we conclude that $\#\Delta^\mIq \leq \#\Delta^{\mI'}$.

Consider the second assertion and suppose $\mIq$ is finite. 
Let $\Delta^\mIq = \{v_1,\ldots,v_n\}$. 
Since $\simLP$ is the largest $\mLP$-auto-bisimulation of $\mI$, by Theorem~\ref{theorem: H-M} and Lemma~\ref{lemma:pr-bis}, if $1 \leq i < j \leq n$ then $v_i$ is not $\mLP$-equivalent to $v_j$. 
For $1 \leq i,j \leq n$ with $i \neq j$, let $C_{i,j}$ be a concept in $\mLP$ such that $v_i \in C_{i,j}^\mIq$ and $v_j \notin C_{i,j}^\mIq$. For $1 \leq i \leq n$, let $C_i = (C_{i,1} \mand\ldots\mand C_{i,i-1} \mand C_{i,i+1} \mand\ldots\mand C_{i,n})$. We have that $v_i \in C_i^\mIq$ and $v_j \notin C_i^\mIq$ if $j \neq i$. Let $C = (C_1 \mor\ldots\mor C_n)$ and, for $1 \leq i \leq n$, let $D_i = (C_1 \mor\ldots\mor C_{i-1} \mor C_{i+1} \mor\ldots\mor C_n)$. Thus, $\mIq$ validates $\top \sqsubseteq C$ but does not validate any $\top \sqsubseteq D_i$ for $1 \leq i \leq n$. Any other interpretation with such properties must have at least $n$ elements in the domain. That is, $\mIq$ is a minimal interpretation that validates the same set of terminological axioms in $\mLP$ as~$\mI$. 

Consider the third assertion and suppose $\mIq$ is unreachable-objects-free and finitely branching.  
Let $\mI'$ be any interpretation that satisfies the same set of concept assertions in $\mLP$ as $\mI$. We show that $\#\Delta^\mIq \leq \#\Delta^{\mI'}$. 
By Theorem~\ref{theorem: quo-x}, $\mIq$ satisfies the same set of concept assertions in $\mLP$ as $\mI$ and $\mI'$. Thus, for every individual name $a$, $a^\mIq$ is $\mLP$-equivalent to $a^{\mI'}$. 
Since $\IN$ is countable and $\mIq$ is unreachable-objects-free and finitely branching, $\Delta^\mIq$ is countable. 
If $\mI'$ is not finitely branching then it is infinite and the assertion clearly holds. So, assume that $\mI'$ is finitely branching. 
Let $Z = \{\tuple{x,x'} \in \Delta^\mIq \times \Delta^{\mI'} \mid x$ is $\mLP$-equivalent to $x'\}$. Like the proof of Theorem~\ref{theorem: H-M}, the conditions~\eqref{bs:eqA}, \eqref{bs:eqC} and \eqref{bs:eqI1} hold, and since $\mIq$ is unreachable-objects-free, the condition~\eqref{bs:eqU1} also holds. Analogously to the proof of the first assertion, it follows that $\#\Delta^\mIq \leq \#\Delta^{\mI'}$.
\myendVS
\end{proof}
} % \newcommand

\VersionA{\ProofThmMZF}

\subsection{The Case with $Q$ and/or $\Self$}

The following two examples show that we cannot make Theorems~\ref{theorem: quo1} and~\ref{theorem: quo-x} stronger by allowing $\Self \in \Phi$ or $Q \in \Phi$. 

\begin{example}\label{example: JQPDY}
Let $\CN = \emptyset$, $\IN = \{a_1,a_2\}$ and $\RN = \{r\}$, where $a_1 \neq a_2$. Consider the interpretation $\mI$ specified by: 
\begin{center}
\includegraphics{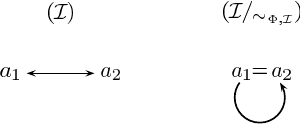}
\end{center}
$\Delta^\mI = \{a_1,a_2\}$, $a_1^\mI = a_1$, $a_2^\mI = a_2$ and $r^\mI = \{\tuple{a_1,a_2},\tuple{a_2,a_1}\}$. For any $\Phi$, we have that $a_1 \simLP\ a_2$. Denote $a = [a_1]_\simLP$ ($= \{a_1,a_2\}$). The quotient interpretation $\mIq$ is thus specified by: $\Delta^\mIq = \{a\}$, $a_1^\mIq = a_2^\mIq = a$ and $r^\mIq = \{\tuple{a,a}\}$. Observe that if $\Self \in \Phi$ then:
\begin{itemize}
\item $\mIq$ is not $\mLP$-bisimilar to $\mI$,  
\item for $\varphi$ being any of the axioms/assertions $\top \sqsubseteq \E r.\Self$, $\varepsilon \sqsubseteq r$, $(\E r.\Self)(a_1)$, $a_1 = a_2$, $r(a_1,a_1)$, we have that $\mIq \models \varphi$, but $\mI \not\models \varphi$.
\myend
\end{itemize}
\end{example}

\begin{example}\label{example: HGWQU}
Let $\CN = \emptyset$, $\IN = \{a,b_1,b_2\}$ and $\RN = \{r\}$, where $a,b_1,b_2$ are pairwise disjoint. Assume that $Q \in \Phi$ and consider the interpretation $\mI$ specified by: 
\begin{center}
\includegraphics{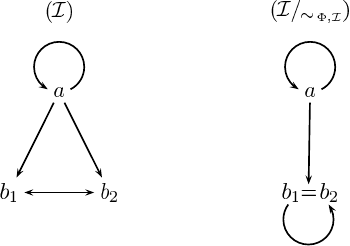}
\end{center}
$\Delta^\mI = \{a,b_1,b_2\}$, $a^\mI = a$, $b_1^\mI = b_1$, $b_2^\mI = b_2$ and $r^\mI = \{\tuple{a,a}$, $\tuple{a,b_1}$, $\tuple{a,b_2}$, $\tuple{b_1,b_2}$, $\tuple{b_2,b_1}\}$. Note that $b_1$ is $\mLP$-bisimilar to $b_2$ and is not $\mLP$-bisimilar to $a$. Denote $a' = [a]_\simLP$ and $b' = [b_1]_\simLP$ ($= \{b_1,b_2\}$). The quotient interpretation $\mIq$ is thus specified by: $\Delta^\mIq = \{a',b'\}$, $a^\mIq = a'$, $b_1^\mIq = b_2^\mIq = b'$ and $r^\mIq = \{\tuple{a',a'}$, $\tuple{a',b'}$, $\tuple{b',b'}\}$. Observe that:
\begin{itemize}
\item $\mIq$ is not $\mLP$-bisimilar to $\mI$, 
\item for $\varphi$ being any of the axioms/assertions $\geq\!2\,r.\top \sqsubseteq\ \geq\!3\,r.\top$, $\varepsilon \sqsubseteq r$, $(\geq\!3\,r.\top)(a)$, $b_1 = b_2$, $r(b_1,b_1)$, we have that $\mI \models \varphi$ iff $\mIq \not\models \varphi$.
\myend
\end{itemize}
\end{example}

For the case when $Q \in \Phi$ or $\Self \in \Phi$, in order to obtain results similar to Theorems~\ref{theorem: quo-x} and~\ref{theorem: mz1}, we introduce QS-interpretations as follows. 

\BeginDefinition{QS-Interpretation}\\
A {\em QS-interpretation} is a tuple $\mI = \langle \Delta^\mI, \cdot^\mI, \QU^\mI, \SE^\mI \rangle$, where 
\begin{itemize}
\item $\langle \Delta^\mI, \cdot^\mI\rangle$ is an interpretation, 
\item $\QU^\mI$ is a function that maps every basic role to a function $\Delta^\mI \times \Delta^\mI \to \mathbb{N}$ such that $\QU^\mI(R)(x,y) > 0$ iff $\tuple{x,y} \in R^\mI$, where $\mathbb{N}$ is the set of natural numbers, 
\item $\SE^\mI$ is a function that maps every role name to a subset of $\Delta^\mI$.
\end{itemize}
If $\mI$ is a QS-interpretation then we redefine
\begin{eqnarray*}
(\E r.\Self)^\mI & = & \{x \in \Delta^\mI \mid x \in \SE^\mI(r)\}\\
(\geq n\,R.C)^\mI & = & \{x \in \Delta^\mI \mid \Sigma\{\QU^\mI(R)(x,y) \mid C^\mI(y)\} \geq n \} \\
(\leq n\,R.C)^\mI & = & \{x \in \Delta^\mI \mid \Sigma\{\QU^\mI(R)(x,y) \mid C^\mI(y)\} \leq n \},
\end{eqnarray*}
where the sum of a set of natural numbers is assumed to be $+\infty$ if it is not finitely bounded.
Other notions for interpretations remain unchanged for QS-interpretations. 
\ \myend
\EndDefinition

\BeginDefinition{Quotient QS-Interpretation}\index{quotient QS-interpretation}\index{I0qQS@$\mIqzQS$|see{quotient QS-interpretation}}Given a finitely branching interpretation $\mI$, the {\em quotient QS-interpretation} of $\mI$ w.r.t.\ an equivalence relation $\sim\ \subseteq \Delta^\mI \times \Delta^\mI$, denoted by $\mIqzQS$, is the QS-interpretation $\mI' = \langle \Delta^{\mI'}, \cdot^{\mI'}, \QU^{\mI'}, \SE^{\mI'}\rangle$ such that: 
\begin{itemize}
\item $\langle \Delta^{\mI'}, \cdot^{\mI'}\rangle$ is the quotient interpretation of $\mI$ w.r.t.\ $\sim$, 
\item for every basic role $R$ and every $x,y \in \Delta^\mI$, \[ \QU^{\mI'}(R)([x]_\sim,[y]_\sim) = \min_{x' \in [x]_\sim} \#\{y' \in [y]_\sim \mid \tuple{x',y'} \in R^\mI\}, \]
\item for every role name $r$, 
\[ \SE^{\mI'}(r) = \{ [x]_\sim \mid \tuple{x,x} \in r^\mI \}. \]
\myend
\end{itemize}
\EndDefinition

Note that, in the case when $Q \in \Phi$, we have 
\[ \QU^{\mI'}(R)([x]_\simLP,[y]_\simLP) = \#\{y' \in [y]_\simLP \mid \tuple{x,y'} \in R^\mI\}. \]

\newcommand{\LemmaHGEAO}{Let $\mI$ be a finitely branching interpretation and let $\mI' = \mIqQS$. Then $Z = \{\tuple{x,[x]_\simLP} \in \Delta^\mI \times \Delta^{\mI'}\}$ satisfies all the properties \eqref{bs:eqA}-\eqref{bs:eqO0}, \eqref{bs:eqU1}, \eqref{bs:eqU2}, \eqref{bs:eqB-2}-\eqref{bs:eqO-2}. In particular, the assertion~\eqref{bs:eqB-2} states that, for every concept $C$ in $\mLP$ and every $x \in \Delta^\mI$, $x \in C^\mI$ iff $[x]_\simLP \in C^{\mI'}$.}

\begin{lemma}\label{lemma: HGEAO}
\LemmaHGEAO
\end{lemma}

\newcommand{\ProofLemmaHGEAO}{\begin{proof}
The properties \eqref{bs:eqA}-\eqref{bs:eqO0}, \eqref{bs:eqU1} and \eqref{bs:eqU2} can be shown as in the proof of Theorem~\ref{theorem: quo1}. The properties \eqref{bs:eqB-2}-\eqref{bs:eqO-2} can be shown as in Lemma~\ref{lemma: bs-inv-1} except that the case when $Q \in \Phi$ and $C = (\geq\!n\,R.D)$ and the case when $\Self \in \Phi$ and $C = \E r.\Self$ in the proof of the assertion~\eqref{bs:eqB-2} are changed to the following: 
\begin{itemize}
\item Case $Q \in \Phi$ and $C = (\geq n\,R.D)$, where $R$ is a basic role: Since $Z(x,x')$ holds, we have that $x' = [x]_\simLP$. Since $C^\mI(x)$ holds, there exist pairwise different $y_1$, \ldots, $y_n \in \Delta^\mI$ such that $R^\mI(x,y_i)$ and $D^\mI(y_i)$ hold for all $1 \leq i \leq n$. Let the partition of $\{y_1,\ldots,y_n\}$ that corresponds to the equivalence relation $\simLP$ consist of pairwise different blocks $Y_{i_1},\ldots,Y_{i_k}$, where $\{i_1,\ldots,i_k\} \subseteq \{1,\ldots,n\}$ and $y_{i_j} \in Y_{i_j}$ for all $1 \leq j \leq k$. By the inductive assumption, $D^{\mI'}([y_{i_j}]_\simLP)$ holds for all $1 \leq j \leq k$. By the definition of $\mI'$, $\QU^{\mI'}(R)([x]_\simLP,[y_{i_j}]_\simLP) \geq \#Y_{i_j}$ for all $1 \leq j \leq k$. Hence $C^{\mI'}([x]_\simLP)$ holds, which means $C^{\mI'}(x')$ holds. 

\item Case $\Self \in \Phi$ and $C = \E r.\Self$~: Since $Z(x,x')$ holds, we have that $x' = [x]_\simLP$. Since $C^\mI(x)$ holds, we have that $r^\mI(x,x)$ holds. Hence $[x]_\simLP \in \SE^{\mI'}(r)$ and consequently $[x]_\simLP \in (\E r.\Self)^{\mI'}$, which means $C^{\mI'}(x')$ holds. 
\myend
\end{itemize}
\end{proof}
} % \newcommand

\VersionA{\ProofLemmaHGEAO}

The following theorem is a counterpart of Theorem~\ref{theorem: quo-x}, with no restrictions on~$\Phi$.

\newcommand{\TheoremQuoY}{Let $\mI$ be a finitely branching interpretation. Then:
\begin{enumerate}
\item For every expression $\varphi$ which is either a terminological axiom in $\mLP$ or a concept assertion (of the form $C(a)$) in $\mLP$, $\mI \models \varphi$ iff $\mIqQS \models \varphi$.
\item For every expression $\varphi$ which is either a role inclusion axiom or an individual assertion of the form $R(a,b)$ or $a \doteq b$, if $\mI \models \varphi$ then $\mIqQS \models \varphi$.
\end{enumerate}}

\begin{theorem} \label{theorem: quo-y}
\TheoremQuoY
\end{theorem}

\newcommand{\ProofThmQuoY}{\begin{proof}
Denote $\mI' = \mIqQS$ and let $Z = \{\tuple{x,[x]_\simLP} \in \Delta^\mI \times \Delta^{\mI'}\}$. By Lemma~\ref{lemma: HGEAO}, for every concept $C$ in $\mLP$, $x \in C^\mI$ iff $[x]_\simLP \in C^{\mI'}$. The first assertion follows immediately from this property. The second assertion can be proved as for Theorem~\ref{theorem: quo-x}.
\myendVS
\end{proof}
} % \newcommand

\VersionA{\ProofThmQuoY}

The following theorem is a counterpart of Theorem~\ref{theorem: mz1}, with no restrictions on~$\Phi$.

\newcommand{\TheoremMZS}{Let $\mI$ be a finitely branching interpretation.
\begin{enumerate}
\item If $\mIqQS$ is finite then it is a minimal QS-interpretation that validates the same set of terminological axioms in $\mLP$ as $\mI$.
\item If $\mIqQS$ is unreachable-objects-free then it is a minimal QS-interpretation that satisfies the same set of concept assertions in $\mLP$ as $\mI$.
\end{enumerate}}

\begin{theorem} \label{theorem: mz2}
\TheoremMZS
\end{theorem}

\newcommand{\ProofThmMZS}{\begin{proof}
By Lemma~\ref{lemma: HGEAO}, every $x \in \Delta^\mI$ is $\mLP$-equivalent to $[x]_\simLP$. Since $\equivLP$ and $\simLP$ coincide, if $[x]_\simLP \neq [x']_\simLP$ then $[x]_\simLP$ and $[x']_\simLP$ are not $\mLP$-equivalent to each other. 
Denote $\mI' = \mIqQS$. 

Consider the first assertion and suppose $\mI'$ is finite. 
Let $\Delta^{\mI'} = \{v_1,\ldots,v_n\}$, where $v_1,\ldots,v_n$ are pairwise different and each $v_i$ is some $[x_i]_\simLP$. For $1 \leq i,j \leq n$ with $i \neq j$, let $C_{i,j}$ be a concept in $\mLP$ such that $v_i \in C_{i,j}^{\mI'}$ and $v_j \notin C_{i,j}^{\mI'}$. For $1 \leq i \leq n$, let $C_i = (C_{i,1} \mand\ldots\mand C_{i,i-1} \mand C_{i,i+1} \mand\ldots\mand C_{i,n})$. We have that $v_i \in C_i^{\mI'}$ and $v_j \notin C_i^{\mI'}$ if $j \neq i$. Let $C = (C_1 \mor\ldots\mor C_n)$ and, for $1 \leq i \leq n$, let $D_i = (C_1 \mor\ldots\mor C_{i-1} \mor C_{i+1} \mor\ldots\mor C_n)$. Thus, $\mI'$ validates $\top \sqsubseteq C$ but does not validate any $\top \sqsubseteq D_i$ for $1 \leq i \leq n$. Any other QS-interpretation with such properties must have at least $n$ elements in the domain. That is, $\mI'$ is a minimal QS-interpretation that validates the same set of terminological axioms in $\mLP$ as~$\mI$.

Consider the second assertion and suppose $\mI'$ is unreachable-objects-free. Since $\mI$ is finitely branching, $\mI'$ is also finitely branching. 
Let $\mI''$ be any QS-interpretation that satisfies the same set of concept assertions in $\mLP$ as $\mI$. We show that $\#\Delta^{\mI'} \leq \#\Delta^{\mI''}$. 
Since $\IN$ is countable and $\mI'$ is unreachable-objects-free and finitely branching, $\Delta^{\mI'}$ is countable. 
If $\mI''$ is not finitely branching then it is infinite and the assertion clearly holds. So, assume that $\mI''$ is finitely branching. 
Since $\mI'$ is unreachable-objects-free and $\mI''$ is a finitely branching QS-interpretation that satisfies the same set of concept assertions in $\mLP$ as $\mI$ and $\mI'$, it can be shown that, for every $x' \in \Delta^{\mI'}$, there exists $x'' \in \Delta^{\mI''}$ that is $\mLP$-equivalent to $x'$. Recall that if $x'_1$ and $x'_2$ are different elements of $\Delta^{\mI'}$ then they are not $\mLP$-equivalent to each other. This implies that $\#\Delta^{\mI'} \leq \#\Delta^{\mI''}$.
\myendVS
\end{proof}
} % \newcommand

\VersionA{\ProofThmMZS}

%-------------------------------------------------------------------------------

\section{Minimizing Interpretations}
\label{section: comp l-aut-bis}

In this section, we adapt Hopcroft's automaton minimization algorithm~\cite{Hopcroft71} and the Paige-Tarjan algorithm~\cite{PaigeT87} to obtain efficient algorithms for computing the partition corresponding to the equivalence relation $\simLP$ for the case when $\mI$ is finite. The partition is used to minimize $\mI$ to obtain $\mIq$ for the case $\{Q,\Self\} \cap \Phi = \emptyset$, or $\mIqQS$ for the other case. We do not require any restrictions on~$\Phi$. 

For $x \in \Delta^\mI$, $Y \subseteq \Delta^\mI$ and a basic role $R$ of $\mLP$, define 
\[ \degree_R(x,Y) = \#\{y \in Y \mid \tuple{x,y} \in R^\mI\}\; \textrm{and}\; \degree_R(x) = \degree_R(x,\Delta^\mI). \] 

The similarity between minimizing automata and minimizing interpretations relies on that equivalence between two states in a finite deterministic automaton is similar to $\mLP$-equivalence between two objects (i.e.\ elements of the domain) of an interpretation. The alphabet $\Sigma$ of an automaton corresponds to $\RN$ in the case $I \notin \Phi$, and $\RNpm$ in the other case. 
Note that the end conditions for equivalence are different; in the case of automata, it is required that the two considered states are either both accepting states or both unaccepting states; in the case of interpretations, it is required that the two considered objects $x$ and $x'$ satisfy the conjunction of the following conditions:   
\begin{itemize}
\item for every $A \in \CN$, $x \in A^\mI$ iff $x' \in A^\mI$,
\item if $Q \notin \Phi$ then, for every basic role $R$ of $\mLP$, 
$\degree_R(x) = 0$ iff $\degree_R(x') = 0$, 
\item if $Q \in \Phi$ then, for every basic role $R$ of $\mLP$, 
$\degree_R(x) = \degree_R(x')$, 
\item if $O \in \Phi$ then, for every $a \in \IN$, $x = a^\mI$ iff $x' = a^\mI$,
\item if $\Self \in \Phi$ then, for every $r \in \RN$, $\tuple{x,x} \in r^\mI$ iff $\tuple{x',x'} \in r^\mI$.
\end{itemize}

Denote the conjunction of the above conditions by $\ECond(x,x')$.\footnote{This is a correction for~\cite{DivroodiN15}.} 

Interpretations are like nondeterministic automata, while Hopcroft's algorithm~\cite{Hopcroft71} works only for deterministic automata. The Paige-Tarjan algorithm~\cite{PaigeT87} for the relational coarsest partition problem works on a graph and exploits the idea ``process the smaller half'' of Hopcroft's algorithm. We adapt it for computing the partition corresponding to $\simLP$ for the case $Q \notin \Phi$. We directly adapt Hopcroft's algorithm for the case $Q \in \Phi$. The idea for a similar problem related with number restrictions was formulated for graphs in~\cite{PaigeT87} and efficient algorithms for other similar problems were proposed even earlier (see~\cite{PaigeT87}). 

%-------------------------------------------------------------------------------

\subsection{The Case $Q \in \Phi$}

Algorithm~\ref{alg: CGSB} (given on page~\pageref{alg: CGSB}) computes the partition corresponding to $\simLP$ for the case when $Q \in \Phi$ and $\mI$ is finite. It starts by splitting $\Delta^\mI$ into blocks using the equivalence relation $\ECond$ and after that follows the idea of Hopcroft's algorithm~\cite{Hopcroft71} to refine that partition. Like Hopcroft's algorithm, Algorithm~\ref{alg: CGSB} keeps the current partition $\bP$ and a collection $\bL$ of pairs $\tuple{Y,R}$ for refining the partition, where $Y \in \bP$ and $R$ is a basic role. Splitting a block $X \in \bP$ by a pair $\tuple{Y,R}$ is done so that $x,x' \in X$ are separated when $\degree_R(x,Y) \neq \degree_R(x',Y)$, and may result in more than two blocks. Technically, it is done as follows: for each $y \in Y$ and for each edge coming to $y$ via $R$ from some $x$ (i.e.\ for each $\tuple{x,y} \in R^\mI$), do 
\begin{enumerate}
\item if it is the first time $x$ is considered for this task then set $\Count(x) := 1$, remove $x$ from its current block $X$ and put $x$ into the block $X_{\Count=1}$,
\item else if $\Count(x) = k$ then increase $\Count(x)$ by 1, remove $x$ from its current block $X_{\Count=k}$ and put it into the block $X_{\Count=k+1}$. 
\end{enumerate}
The non-empty blocks created from $X$ together with the modified block $X$, if not empty, form the set $\bX$ mentioned in the algorithm.

\begin{algorithm}[t!]
\caption{computing the partition corresponding to $\simLP$ for the case $Q \in \Phi$\label{alg: CGSB}}
\Input{a set $\Phi$ of DL-features with $Q \in \Phi$ and a finite interpretation $\mI$}
\Output{the partition $\bP$ corresponding to the largest $\mLP$-auto-bisimulation of $\mI$} 
%\DontPrintSemicolon
\BlankLine

\lIf{$I \notin \Phi$}{let $\RN^\dag = \RN$}
\lElse{let $\RN^\dag = \RNpm$}\;

\BlankLine

set $\bP$ to the partition corresponding to the equivalence relation $\ECond$\label{CGSB: Step2}\;

set $Z$ to a maximal block of $\bP$\;

\BlankLine
set $\bL$ to the empty collection\label{CGSB: Step4}\;
\ForEach{$X \in \bP \setminus \{Z\}$ and $R \in \RN^\dag$}{add $\tuple{X,R}$ to $\bL$\label{CGSB: Step3}\;}

\BlankLine
\While{$\bL \neq \emptyset$}
{
  extract a pair $\tuple{Y,R}$ from $\bL$\;
  \ForEach{$X \in \bP$ split by $\tuple{Y,R}$}{
      split $X$ by $\tuple{Y,R}$ into a set $\bX$ of blocks\;
      replace $X$ in $\bP$ by all the blocks of $\bX$\;
      set $Z$ to a maximal block of $\bX$\;
      \ForEach{$S \in \RN^\dag$}{
	\uIf{$\tuple{X,S} \in \bL$}{replace $\tuple{X,S}$ in $\bL$ by all the pairs $\tuple{X',S}$ with $X' \in \bX$\;}
	\Else{add all the pairs $\tuple{X',S}$ with $X' \in \bX \setminus \{Z\}$ to $\bL$\label{CGSB: Step14}\;}
      }
  }
}
\end{algorithm}

%-------------------------------------------------------------------------------

\begin{lemma}\label{lemma: UYCFM}
Consider an execution of Algorithm~\ref{alg: CGSB}. The resulting partition $\bP$ corresponds to an $\mLP$-auto-bisimulation of $\mI$.
\end{lemma}

\begin{proof}
Let $Z$ be the equivalence relation corresponding to the partition $\bP$. Consider the conditions \eqref{bs:eqA}--\eqref{bs:eqSelf} with $\mI' = \mI$. Clearly, \eqref{bs:eqA}, \eqref{bs:eqB}, \eqref{bs:eqO0}, \eqref{bs:eqU1}, \eqref{bs:eqU2}, \eqref{bs:eqSelf} hold. As \eqref{bs:eqC}--\eqref{bs:eqI2} are instances of \eqref{bs:eqQ1}--\eqref{bs:eqQI2}, respectively, we need to prove only \eqref{bs:eqQ1}--\eqref{bs:eqQI2}. It is sufficient to show that, for every $x,x' \in \Delta^\mI$, every basic role $R$ of $\mLP$ and every block $Y \in \bP$, if $\degree_R(x,Y) \neq \degree_R(x',Y)$ then $x$ and $x'$ belong to different blocks of $\bP$. This is clear for the case $\degree_R(x) \neq \degree_R(x')$. So, assume that $\degree_R(x) = \degree_R(x')$. Let $Y'$ be the smallest block appeared during the execution of the algorithm such that $Y'$ is a superset of $Y$ and $\degree_R(x,Y') = \degree_R(x',Y')$ (the biggest one is $\Delta^\mI$). Let $Y_1,\ldots,Y_k$ be the blocks obtained from the splitting of $Y'$. There exist $1 \leq i,j \leq k$ such that $i \neq j$, $\degree_R(x,Y_i) \neq \degree_R(x',Y_i)$ and $\degree_R(x,Y_j) \neq \degree_R(x',Y_j)$. Hence, $\tuple{Y_i,R}$ or $\tuple{Y_j,R}$ is inserted into $\bL$ when $Y'$ is split. It follows that, at some step, a pair $\tuple{Y'',R}$ such that $\degree_R(x,Y'') \neq \degree_R(x',Y'')$ is extracted from $\bL$ ($Y''$ is some subset of $Y_i$ or $Y_j$). This pair separates $x$ and $x'$.
\myend
\end{proof}

\begin{lemma}\label{lemma: YFDLS}
Consider an execution of Algorithm~\ref{alg: CGSB}. If $x,x' \in \Delta^\mI$ are separated (i.e., belong to different blocks of the partition $\bP$) then $x\ \notequivLP\ x'$.
\end{lemma}

\begin{proof}
Assume that $x,x' \in \Delta^\mI$ are separated. We prove that $x\ \notequivLP\ x'$ by induction on the iteration $k$ of the main loop at which $x$ and $x'$ are separated. 

Consider the base case $k = 0$ when $x$ and $x'$ belong to different equivalence classes of the equivalence relation $\ECond$. There are the following subcases:
\begin{itemize}
\item there exists $A \in \CN$ such that $x \in A^\mI$ and $x' \notin A^\mI$ or vice versa (i.e., $x \notin A^\mI$ and $x' \in A^\mI$); 
\item there exists a basic role $R$ of $\mLP$ such that $\degree_R(x) \neq \degree_R(x')$; without loss of generality, assume that $\degree_R(x) = l > \degree_R(x')$;
\item $O \in \Phi$ and there exists $a \in \IN$ such that $x = a^\mI$ and $x' \neq a^\mI$ or vice versa (i.e., $x \neq a^\mI$ and $x' = a^\mI$);
\item $\Self \in \Phi$ and there exists $r \in \RN$ such that $\tuple{x,x} \in r^\mI$ and $\tuple{x',x'} \notin r^\mI$ or vice versa (i.e., $\tuple{x,x} \notin r^\mI$ and $\tuple{x',x'} \in r^\mI$).
\end{itemize}
The concept $A$, $\geq\!l\,R.\top$, $\{a\}$ or $\E r.\Self$ of $\mLP$, respectively for these subcases, distinguishes $x$ and $x'$. Hence $x\ \notequivLP\ x'$. 

Now consider the induction step and assume that $x$ and $x'$ are separated by a pair $\tuple{Y,R}$ at the iteration $k+1$ of the main loop. Thus, $\degree_R(x,Y) \neq \degree_R(x',Y)$. Without loss of generality, assume that $\degree_R(x,Y) > \degree_R(x',Y)$ and let $h = \degree_R(x,Y)$. Let the partition $\bP$ before the iteration $k+1$ be $\{Y_0,\ldots,Y_s\}$ with $Y_0 = Y$ and let $Y_i = \{y_{i,1},\ldots,y_{i,t_i}\}$ for $0 \leq i \leq s$. By the induction assumption, for each $1 \leq i \leq s$, $1 \leq j \leq t_0$ and $1 \leq j' \leq t_i$, there exists a concept $C_{i,j,j'}$ such that $y_{0,j} \in C_{i,j,j'}^\mI$ and $y_{i,j'} \notin C_{i,j,j'}^\mI$. 
For $1 \leq i \leq s$ and $1 \leq j \leq t_0$, let $C_{i,j} = C_{i,j,1} \mand\ldots\mand C_{i,j,t_i}$, then $y_{0,j} \in C_{i,j}^\mI$ and $y_{i,j'} \notin C_{i,j}^\mI$ for all $1 \leq j' \leq t_i$. 
For $1 \leq i \leq s$, let $C_i = C_{i,1} \mor\ldots\mor C_{i,t_0}$, then $y_{0,j} \in C_i^\mI$ for all $1 \leq j \leq t_0$, and $y_{i,j'} \notin C_i^\mI$ for all $1 \leq j' \leq t_i$. 
Let $C = C_1 \mand\ldots\mand C_s$. Thus, $Y_0 \subseteq C^\mI$ and $Y_i \cap C^\mI = \emptyset$ for all $1 \leq i \leq s$, which means that $Y_0 = C^\mI$. Therefore, $x \in (\geq\!h\,R.C)^\mI$ and $x' \notin (\geq\!h\,R.C)^\mI$, which implies $x\ \notequivLP\ x'$.
\myend
\end{proof}

\newcommand{\TheoremCGSB}{Algorithm~\ref{alg: CGSB} is correct and can be implemented to have time complexity $O(|\Sigma|(m+n)\log{n})$, where $m = \sum_{r \in \RN} |r^\mI|$ and $n = |\Delta^\mI|$. A tighter bound for the complexity is $O(|\IN| + |\CN|n + |\RN|(m+n)\log{n})$.}

\begin{proposition} \label{theorem: CGSB}
\TheoremCGSB
\end{proposition}

\newcommand{\ProofThmCGSB}{\begin{proof} (Sketch)
The contrapositive of Lemma~\ref{lemma: YFDLS} states that if $x\ \equivLP\ x'$ then $x$ and $x'$ are not separated. That is, $\equivLP$ is a subset of the equivalence relation corresponding to the partition $\bP$. As $\equivLP$ and $\simLP$ coincide (by Proposition~\ref{prop: LABS}), it follows that $\simLP$ is a subset of the equivalence relation corresponding to the partition $\bP$. By Lemma~\ref{lemma: UYCFM}, the latter is also an $\mLP$-auto-bisimulation of $\mI$, hence it is the same as~$\simLP$ (the largest $\mLP$-auto-bisimulation of $\mI$). That is, Algorithm~\ref{alg: CGSB} is correct. 

To estimate complexity, notice that the steps~\ref{CGSB: Step4}-\ref{CGSB: Step14} of Algorithm~\ref{alg: CGSB} are essentially the same as the skeleton of Hopcroft's automaton minimization algorithm~\cite{Hopcroft71,Parys2005} used for refining the partition. The only difference is the way of splitting $X$ by a pair $\tuple{Y,R}$. The technique for this has been mentioned earlier. 
The complexity analysis of~\cite{Parys2005} can be applied to the steps~\ref{CGSB: Step4}-\ref{CGSB: Step14} of Algorithm~\ref{alg: CGSB}. 
The first change is that instead of the occurrences of $|\{p \in Q:\delta(p,a)\in Y\}|$ we have $|\{\tuple{x,y} \in R^\mI : y \in Y\}|$.
The second change is that, as Hopcroft's automaton minimization algorithm is for deterministic automata but here we have nondeterminism (i.e., for each $R \in \RN^\dag$, $R^\mI$ is a binary relation but not a function), the last two lines of \cite{Parys2005} are modified so that $O(n)$ is replaced by $O(m)$ and $O(|\Sigma|n\log{n})$ is replaced by $O(|\RN|m\log{n})$. Thus, we can conclude that the steps~\ref{CGSB: Step4}-\ref{CGSB: Step14} can be implemented to have time complexity $O(|\RN^\dag|(m+n)\log{n})$, which is the same as $O(|\RN|(m+n)\log{n})$.

Consider complexity of the step~\ref{CGSB: Step2} of Algorithm~\ref{alg: CGSB}. To compute equivalence classes of the equivalence relation $\ECond$, we start from the partition $\{\Delta^\mI\}$ and then:
\begin{enumerate}
\item Refine the current partition by using the condition that, when $O \in \Phi$, $x$ and $x'$ should be in the same block only if, for every $a \in \IN$, $x = a^\mI$ iff $x' = a^\mI$. This can be done in $O(|\IN|)$ steps.  

\item Refine the current partition by using the condition that $x$ and $x'$ should be in the same block only if, for every $A \in \CN$, $x \in A^\mI$ iff $x' \in A^\mI$. This can be done in $O(|\CN|n)$ steps. 

\item Refine the current partition by using the condition that, when $\Self \in \Phi$, $x$ and $x'$ should be in the same block only if, for every $r \in \RN$, $\tuple{x,x} \in r^\mI$ iff $\tuple{x',x'} \in r^\mI$. This can be done in $O(|\RN|n)$ steps. 

%\item Refine the current partition by using the condition that, when $Q \notin \Phi$, $x$ and $x'$ should be in the same block only when: $\degree_R(x) = 0$ iff $\degree_R(x') = 0$. This can be done in $O(|\RN|n)$ steps. 

\item Refine the current partition by using the condition that $x$ and $x'$ should be in the same block only when $\degree_R(x) = \degree_R(x')$ (since when $Q \in \Phi$). This can be done in $O(|\RN|n)$ steps. 
\end{enumerate}
Summing up, the time complexity of the step~\ref{CGSB: Step2} of Algorithm~\ref{alg: CGSB} is of rank $O(|\IN| + |\CN|n + |\RN|n)$. Therefore, Algorithm~\ref{alg: CGSB} can be implemented to have time complexity $O(|\IN| + |\CN|n + |\RN|(m+n)\log{n})$. 
\myend
\end{proof}
} % \newcommand

\VersionA{\ProofThmCGSB}

%-------------------------------------------------------------------------------

\subsection{The Case $Q \notin \Phi$}

Computing the partition corresponding to $\simLP$ for the case $Q \notin \Phi$ differs from the relational coarsest partition problem studied in~\cite{PaigeT87}, among others, in that the ``edges'' are labeled by basic roles of $\mLP$. 
Algorithm~\ref{alg: CGSB2} (on page~\pageref{alg: CGSB2}) is our adaptation of the Paige-Tarjan algorithm~\cite{PaigeT87} for computing the partition corresponding to $\simLP$ for the case $Q \notin \Phi$.\footnote{It is a correction for~\cite{DivroodiN15}.} It is formulated in a way to reflect the traditional presentation of Hopcroft's automaton minimization algorithm.

Roughly speaking, the main problem is that the relation $R^\mI$ for a basic role $R$ of $\mLP$ need not to be a function. Elements $x$ and $x'$ of a block $X$ should be separated by a pair $\tuple{V,R}$, where $V$ is a block, not only when $\deg_R(x,V) > 0$ and $\deg_R(x',V) = 0$ or vice versa (i.e., $\deg_R(x,V) = 0$ and $\deg_R(x',V) > 0$), but also when $\deg_R(x,V') > 0$ and $\deg_R(x',V') = 0$ or vice versa (i.e., $\deg_R(x,V') = 0$ and $\deg_R(x',V') > 0$), where $V'$ is the complement of $V$ w.r.t.\ an appropriate block $Y$ containing~$V$. Such a $Y$ is the union of some blocks of the current partition $\bP$. In~\cite{PaigeT87}, it is called a compound block.

Suppose that a block $X$ cannot be split by a pair $\tuple{Y,R}$ (in the sense that for every $x,x' \in X$, $\deg_R(x,Y) > 0$ iff $\deg_R(x',Y) > 0$). Let $V \subset Y$ and $V' = Y \setminus V$. Then splitting $X$ by a tuple $\tuple{V,V',R}$ is done as follows:
\begin{itemize}
\item Split $X$ by $\tuple{V,R}$ to obtain 
$X_1 = \{x \in X \mid \deg_R(x,V) > 0\}$ and  
$X_2 = X \setminus X_1$.
\item Split $X_1$ by $\tuple{V',R}$ to obtain 
$X_{1,1} = \{x \in X_1 \mid \deg_R(x,V') > 0\}$ and $X_{1,2} = X_1 \setminus X_{1,1}$.
\item Then return the set $\{X_{1,1},X_{1,2},X_2\}$ after deleting empty sets.
\end{itemize}
If the result contains more than one block then we say that $X$ is split by $\tuple{V,V',R}$. 
Note that $X_2$ cannot be split by $\tuple{V',R}$. 
Denote $R^{-1}(U) = \{x \in \Delta^\mI \mid \deg_R(x,U) > 0\}$. Then, also observe that:
\begin{itemize}
\item $X_1 = X \cap R^{-1}(V)$ and $X_2 = X \setminus X_1$, 
\item $X_{1,2} = X_1 \cap (R^{-1}(V) \setminus R^{-1}(V'))$ and $X_{1,1} = X_1 \setminus X_{1,2}$.
\end{itemize}
This gives a good way for computing the split of $X$ via $R^{-1}(V)$ and $R^{-1}(V) \setminus R^{-1}(V')$, as the computation can ``start'' from $V$ and ``look back'' via $R$. This is a crucial observation of~\cite{PaigeT87}. 

Regarding the idea ``process the smaller half'', observe that if $Y$ is the union of at least two blocks from the current partition $\bP$ and $V$ is a minimal block of $\bP$ such that $V \subset Y$ then $\# V \leq \# Y / 2$.  

\begin{algorithm}[t!]
\caption{computing the partition corresponding to $\simLP$ for the case $Q \notin \Phi$\label{alg: CGSB2}}
\Input{a set $\Phi$ of DL-features without $Q$, and a finite interpretation $\mI$}
\Output{the partition $\bP$ corresponding to the largest $\mLP$-auto-bisimulation of $\mI$} 
%\DontPrintSemicolon
\BlankLine

\lIf{$I \notin \Phi$}{let $\RN^\dag = \RN$}
\lElse{let $\RN^\dag = \RNpm$}\;

\BlankLine

set $\bP$ to the partition corresponding to the equivalence relation $\ECond$\label{CGSB2: Step2}\;
$\bL := \{\tuple{\Delta^\mI,R} \mid R \in \RN^\dag\}$\label{CGSB2: Step4}\;

\BlankLine
\While{$\bL \neq \emptyset$}
{
  extract a pair $\tuple{Y,R}$ from $\bL$\;
  let $V$ be a minimal block of $\bP$ such that $V \subset Y$\;
  $V' := Y \setminus V$\;
  \If{more than one block of $\bP$ is a subset of $V'$}{add $\tuple{V',R}$ to $\bL$}

  \ForEach{$X \in \bP$ split by $\tuple{V,V',R}$}{
      split $X$ by $\tuple{V,V',R}$ into a set $\bX$ of blocks\;
      replace $X$ in $\bP$ by all the blocks of $\bX$\;
      \ForEach{$S \in \RN^\dag$}{
	\If{$\bL$ does not contain any pair $\tuple{U,S}$ such that $X \subset U$}{add $\tuple{X,S}$ to $\bL$\label{CGSB2: Step14}}
      }
  }
}
\end{algorithm}

%-------------------------------------------------------------------------------

\begin{lemma}\label{lemma: UYCFM2}
Consider an execution of Algorithm~\ref{alg: CGSB2}. The resulting partition $\bP$ corresponds to an $\mLP$-auto-bisimulation of $\mI$.
\end{lemma}

\begin{proof}
Let $Z$ be the equivalence relation corresponding to the partition $\bP$. Consider the conditions \eqref{bs:eqA}--\eqref{bs:eqO0} and \eqref{bs:eqU1}--\eqref{bs:eqSelf} with $\mI' = \mI$. Clearly, \eqref{bs:eqA}, \eqref{bs:eqB}, \eqref{bs:eqO0}, \eqref{bs:eqU1}--\eqref{bs:eqSelf} hold. We need to prove only \eqref{bs:eqC}--\eqref{bs:eqI2}. It is sufficient to show that, for every $x,x' \in \Delta^\mI$, every basic role $R$ of $\mLP$ and every block $V \in \bP$, if ($\degree_R(x,V) > 0$ and $\degree_R(x',V) = 0$) or ($\degree_R(x,V) = 0$ and $\degree_R(x',V) > 0$) then $x$ and $x'$ belong to different blocks of $\bP$. This is clear for the case when ($\degree_R(x) > 0$ and $\degree_R(x') = 0$) or ($\degree_R(x) = 0$ and $\degree_R(x') > 0$). So, assume that either ($\degree_R(x) > 0$ and $\degree_R(x') > 0$) or ($\degree_R(x) = 0$ and $\degree_R(x') = 0$). Let $Y$ be the smallest block such that $V \subset Y$, $\tuple{Y,R}$ appeared in $\bL$ at some step and either ($\degree_R(x,Y) > 0$ and $\degree_R(x',Y) > 0$) or ($\degree_R(x,Y) = 0$ and $\degree_R(x',Y) = 0$). Such a set exists due to the candidate $\Delta^\mI$. 

Consider the moment when $\tuple{Y,R}$ is extracted from $\bL$. We have $Y = V_1 \cup \ldots \cup V_k$, where $k \geq 2$ and $V_1,\ldots,V_k$ are blocks of $\bP$. Let $V_1$ be the minimal block among $V_1,\ldots,V_k$ that is taken for processing $\tuple{Y,R}$ and let $V'_1 = Y \setminus V_1$. 
If $x$ and $x'$ are still in the same block of $\bP$ and they are not separated when splitting that block using $\tuple{V_1,V'_1,R}$ then the following conditions hold:
\begin{itemize}
\item ($\degree_R(x,V_1) > 0$ and $\degree_R(x',V_1) > 0$) or ($\degree_R(x,V_1) = 0$ and $\degree_R(x',V_1) = 0$),
\item ($\degree_R(x,V'_1) > 0$ and $\degree_R(x',V'_1) > 0$) or ($\degree_R(x,V'_1) = 0$ and $\degree_R(x',V'_1) = 0$),
\item $k = 2$ and either $V \subset V_1$ or $V \subset V'_1$.
\end{itemize} 
This implies that $V_1$ or $V'_1$ is split at some step and that operation adds $\tuple{V_1,R}$ or $\tuple{V'_1,R}$ to $\bL$. This contradicts the minimality of $Y$. Therefore, $x$ are $x'$ must be separated by using $\tuple{V_1,V'_1,R}$.
\myend
\end{proof}

\begin{lemma}\label{lemma: YFDLS2}
Consider an execution of Algorithm~\ref{alg: CGSB2}. If $x,x' \in \Delta^\mI$ are separated (i.e., belong to different blocks of the partition $\bP$) then $x\ \notequivLP\ x'$.
\end{lemma}

\begin{proof}
Assume that $x,x' \in \Delta^\mI$ are separated. We prove that $x\ \notequivLP\ x'$ by induction on the iteration $k$ of the main loop at which $x$ and $x'$ are separated. 

Consider the base case $k = 0$ when $x$ and $x'$ belong to different equivalence classes of the equivalence relation $\ECond$. There are the following subcases:
\begin{itemize}
\item there exists $A \in \CN$ such that $x \in A^\mI$ and $x' \notin A^\mI$ or vice versa (i.e., $x \notin A^\mI$ and $x' \in A^\mI$); 
\item there exists a basic role $R$ of $\mLP$ such that either $\degree_R(x) > 0$ and $\degree_R(x') = 0$ or $\degree_R(x) = 0$ and $\degree_R(x') > 0$; 
\item $O \in \Phi$ and there exists $a \in \IN$ such that $x = a^\mI$ and $x' \neq a^\mI$ or vice versa (i.e., $x \neq a^\mI$ and $x' = a^\mI$);
\item $\Self \in \Phi$ and there exists $r \in \RN$ such that $\tuple{x,x} \in r^\mI$ and $\tuple{x',x'} \notin r^\mI$ or vice versa (i.e., $\tuple{x,x} \notin r^\mI$ and $\tuple{x',x'} \in r^\mI$).
\end{itemize}
The concept $A$, $\E R.\top$, $\{a\}$ or $\E r.\Self$ of $\mLP$, respectively for these subcases, distinguishes $x$ and $x'$. Hence $x\ \notequivLP\ x'$. 

Now consider the induction step and assume that $x$ and $x'$ are separated by a tuple $\tuple{V,V',R}$ at the iteration $k+1$ of the main loop. There are the following cases:
\begin{enumerate}
\item $\deg_R(x,V) > 0$ and $\deg_R(x',V) = 0$;
\item $\deg_R(x,V) = 0$ and $\deg_R(x',V) > 0$;
\item $\deg_R(x,V) > 0$, $\deg_R(x',V) > 0$, $\deg_R(x,V') > 0$ and $\deg_R(x',V') = 0$;
\item $\deg_R(x,V) > 0$, $\deg_R(x',V) > 0$, $\deg_R(x,V') = 0$ and $\deg_R(x',V') > 0$. 
\end{enumerate}
Using the induction assumption, in a similar way as in the proof of Lemma~\ref{lemma: YFDLS}, it can be shown that there exist concepts $C$ and $C'$ such that $V = C^\mI$ and $V' = C'^\mI$. It can be seen that either $\E R.C$ or $\E R.C'$ distinguishes $x$ and $x'$.
\myend
\end{proof}

\newcommand{\TheoremCGSBt}{Algorithm~\ref{alg: CGSB2} is correct and can be implemented to have time complexity $O(|\Sigma|(m+n)\log{n})$, where $m = \sum_{r \in \RN} |r^\mI|$ and $n = |\Delta^\mI|$. A tighter bound for the complexity is $O(|\IN| + |\CN|n + |\RN|(m+n)\log{n})$.}

\begin{proposition} \label{theorem: CGSB2}
\TheoremCGSBt
\end{proposition}

\newcommand{\ProofThmCGSBt}{\begin{proof} (Sketch)
The contrapositive of Lemma~\ref{lemma: YFDLS2} states that if $x\ \equivLP\ x'$ then $x$ and $x'$ are not separated. That is, $\equivLP$ is a subset of the equivalence relation corresponding to the partition $\bP$. As $\equivLP$ and $\simLP$ coincide (by Proposition~\ref{prop: LABS}), it follows that $\simLP$ is a subset of the equivalence relation corresponding to the partition $\bP$. By Lemma~\ref{lemma: UYCFM2}, the latter is also an $\mLP$-auto-bisimulation of $\mI$, hence it is the same as~$\simLP$ (the largest $\mLP$-auto-bisimulation of $\mI$). That is, Algorithm~\ref{alg: CGSB2} is correct. 
This algorithm can be implemented in a similar way as the Paige-Tarjan algorithm for the relational coarsest partition problem~\cite{PaigeT87} and its complexity can be estimated analogously.
\myend
\end{proof}
} % \newcommand

\VersionA{\ProofThmCGSBt}

%-------------------------------------------------------------------------------

\section{Applications}
\label{section: apps}

As mentioned in the introduction, bisimulations have applications in analyzing expressiveness of DLs, minimizing interpretations and concept learning in DLs.

%-------------------------------------------------------------------------------

\subsection{Comparing the Expressiveness of Description Logics}
\label{section: OIDSJ}

The expressiveness of description logics (DLs) has been studied in a number of works~\cite{Baader96,Borgida96,CadoliPL97,KurtoninaR99,LutzPW11}. 
In~\cite{Baader96} Baader proposed a formal definition of the expressive power of DLs. His definition is liberal in that it allows the compared logics to have different vocabularies. His work provides separation results for some early DLs. 
In~\cite{Borgida96} Borgida showed that certain DLs have the same expressiveness as the two or three variable fragment of first-order logic. The class of DLs considered in~\cite{Borgida96} is large, but the results only concern DLs without the reflexive and transitive closure of roles. 
In~\cite{CadoliPL97} Cadoli et al. considered the expressiveness of hybrid knowledge bases that combine a DL knowledge base with Horn rules. The used DL is $\mathcal{ALCNR}$. 
The work~\cite{KurtoninaR99} by Kurtonina and de~Rijke is a comprehensive work on the expressiveness of DLs that are sublogics of $\mathcal{ALCNR}$. It is based on bisimulation and provides many interesting results. 
In~\cite{LutzPW11} Lutz et al.\ characterized the expressiveness and rewritability of DL TBoxes for the DLs that are sublogics of $\mathcal{ALCQIO}$. They used semantic notions such as bisimulation, equisimulation, disjoint union and direct product. 

In this subsection we compare the expressiveness of the DLs $\mLP$ w.r.t.\ concepts, TBoxes and ABoxes, where $\mL$ stands for \ALCreg and $\Phi \subseteq \{I,O,Q,U,\Self\}$. Our results about separating the expressiveness of DLs are based on bisimulations and naturally extended to the case when instead of \ALCreg we have any sublogic of \ALCreg that extends \ALC. In the workshop paper~\cite{Ali2014}, our results are extended a little further for separating the expressiveness of DLs w.r.t.\ positive concepts.

\BeginDefinition{Equivalence between Concepts, TBoxes or ABoxes}
Two concepts $C$ and $D$ are {\em equivalent} if, for every interpretation~$\mI$, $C^\mI= D^\mI$. 
Two TBoxes $\mathcal{T}_1$ and $\mathcal{T}_2$ are {\em equivalent} if, for every interpretation~$\mI$, $\mI$ is a model of $\mathcal{T}_1$ iff $\mI$ is a model of $\mathcal{T}_2$.
Two ABoxes $\mathcal{A}_1$ and $\mathcal{A}_2$ are {\em equivalent} if, for every interpretation~$\mI$, $\mI$ is a model of $\mathcal{A}_1$ iff $\mI$ is a model of~$\mathcal{A}_2$.
\myend
\EndDefinition

\BeginDefinition{Comparing Description Logics}
We say that a logic $\mL_1$ is {\em at most as expressive as} a logic $\mL_2$ w.r.t. concepts (resp.\ TBoxes, ABoxes), denoted by $\mL_1 \leq_C \mL_2$ (resp.\ $\mL_1 \leq_T \mL_2$, $\mL_1 \leq_A \mL_2$), if every concept (resp.\ TBox, ABox) in $\mL_1$ has an equivalent concept (resp.\ TBox, ABox) in $\mL_2$. 

We say that a logic $\mL_2$ is {\em more expressive} than a logic $\mL_1$ (or $\mL_1$ is {\em less expressive} than $\mL_2$) w.r.t. concepts (resp.\ TBoxes, ABoxes), denoted by $\mL_1 <_C \mL_2$ (resp.\ $\mL_1 <_T \mL_2$, $\mL_1 <_A \mL_2$), if $\mL_1 \leq_C \mL_2$ (resp.\ $\mL_1 \leq_T \mL_2$, $\mL_1 \leq_A \mL_2$) and $\mL_2 \not \leq_C \mL_1$ (resp.\ $\mL_2 \not \leq_T \mL_1$, $\mL_2 \not \leq_A \mL_1$). 
\myend
\EndDefinition

\begin{proposition}\label{Ali:transitivity}
If a logic $\mL_1$ is at most as expressive as a logic $\mL_2$ w.r.t. concepts (resp.\ TBoxes, ABoxes) and a logic $\mL_2$ is at most as expressive as $\mL_3$ w.r.t. concepts (resp.\ TBoxes, ABoxes) then  $\mL_1$ is at most as expressive as $\mL_3$ w.r.t. concepts (resp.\ TBoxes, ABoxes). 
\end{proposition}

The proof of this proposition is straightforward.

\newcommand{\LemmaAliMethod}{Let $\Phi_1$ and $\Phi_2$ be sets of DL-features such that $\Phi_1 \subseteq \Phi_2$. Denote $\mL_1 = \mL_{\Phi_1}$ and $\mL_2 = \mL_{\Phi_2}$. Let $\mI$, $\mI'$ be interpretations and $Z$ an $\mL_1$-bisimulation between $\mI$ and $\mI'$. 
\begin{enumerate}
\item\label{Ali:m1} If $\mL_1 \leq_C \mL_2$, $x\in \Delta^\mI, x'\in \Delta^{\mI'}, Z(x,x')$ holds, and there exists a concept $C$ of $\mL_2$ such that $x\in C^\mI$ but $x'\not \in C^{\mI'}$, then  $\mL_1 <_C \mL_2$.
\item\label{Ali:m2} Suppose that $U \in \Phi_1$ or both $\mI$ and $\mI'$ are unreachable-objects-free. If \mbox{$\mL_1 \leq_T \mL_2$} and there exists a TBox $\mT$ in $\mL_2$ such that $\mI$ is a model of $\mT$ but $\mI'$ is not, then  $\mL_1 <_T \mL_2$.
\item\label{Ali:m3} Suppose $O \in \Phi_1$. If $\mL_1 \leq_A \mL_2$ and there exists an ABox $\mA$ in $\mL_2$ such that $\mI$ is a model of $\mA$ but $\mI'$ is not, then $\mL_1 <_A \mL_2$.
\end{enumerate}}

\begin{lemma}\label{Ali:method}
\LemmaAliMethod
\end{lemma}

\newcommand{\ProofLemmaAliMethod}{\begin{proof}
Consider the first assertion. Suppose $\mL_1 \leq_C \mL_2$, $x \in \Delta^\mI$,  $x' \in \Delta^{\mI'}$, $ Z(x,x') $ holds and there exists a concept $C$ of $\mL_2$ such that $x\in C^\mI$ but $x'\not \in C^{\mI'}$. We prove that $\mL_2 \not \leq_C \mL_1$. For the sake of contradiction, suppose $\mL_2 \leq_C \mL_1$.   
  It follows that there exists a concept $C'$ of $\mL_1$ that is equivalent to $C$. Thus, $x\in C'^\mI$ but $x'\not \in C'^{\mI'}$. Hence, $C'$  is not invariant for $Z$, which contradicts Theorem~\ref{theorem: bs-inv-1}. Therefore, $\mL_1 <_C \mL_2$. 
  
Consider the second assertion. Suppose $\mL_1 \leq_T \mL_2$ and there exists a~TBox $\mT$ in $\mL_2$ such that  $\mI$ is a model of $\mT$ but $\mI'$ is not. We prove that $\mL_2 \not \leq_T \mL_1$. For the sake of contradiction, suppose $\mL_2 \leq_T \mL_1$. It follows that there exists a~TBox $\mT'$ in $\mL_1$  that is equivalent to $\mT$. Thus, $\mI$ is a model of $\mT'$ but $\mI'$ is not, which contradicts Corollary~\ref{cor: TBox-invar} or Theorem~\ref{theorem: TBox-inv-2}. Therefore, $\mL_1 <_T \mL_2$.

Consider the third assertion. Suppose $\mL_1 \leq_A \mL_2$ and there exists an ABox $\mA$ in $\mL_2$ such that  $\mI$ is a model of $\mA$ but $\mI'$ is not. We prove that $\mL_2 \not \leq_A \mL_1$. For the sake of contradiction, suppose $\mL_2 \leq_A \mL_1$. It follows that there exists an ABox $\mA'$ in $\mL_1$  that is equivalent to $\mA$. Thus, $\mI$ is a model of $\mA'$ but $\mI'$ is not, which contradicts Theorem~\ref{theorem: ABox-inv}. Therefore, $\mL_1 <_A \mL_2$.
\myend
\end{proof}
} % \newcommand

%\VersionA{\ProofLemmaAliMethod}
\ProofLemmaAliMethod

In the rest of this subsection, we assume that $\CN$ and $\RN$ are not empty and $\IN$ contains at least two individual names. Let $\{a,b\} \subseteq \IN$, $A \in \CN$ and $r \in \RN$. 

\newcommand{\LemmaAliMainT}{\begin{enumerate}
\item For any pair $\tuple{\mL_1,\mL_2}$ among $\tuple{\mL_I,\mL_{OQU\Self}}, \tuple{\mL_Q,\mL_{IOU\Self}}, \tuple{\mL_{\Self},\mL_{IOQU}}$, we have that: 
$\mL_1 \not\leq_C \mL_2$,\ \ 
$\mL_1 \not\leq_T \mL_2$,\ \ 
$\mL_1 \not\leq_A \mL_2$.
\item $\mL_O \not\leq_C \mL_{IQU\Self}$ and 
$\mL_O \not\leq_T \mL_{IQU\Self}$.
\item $\mL_U \not\leq_C \mL_{IOQ\Self}$ and 
$\mL_U \not\leq_A \mL_{IOQ\Self}$.
\end{enumerate}}

\begin{lemma}\label{Ali:main2}\ 
\LemmaAliMainT
\end{lemma}

\newcommand{\ProofLemmaAliMainT}{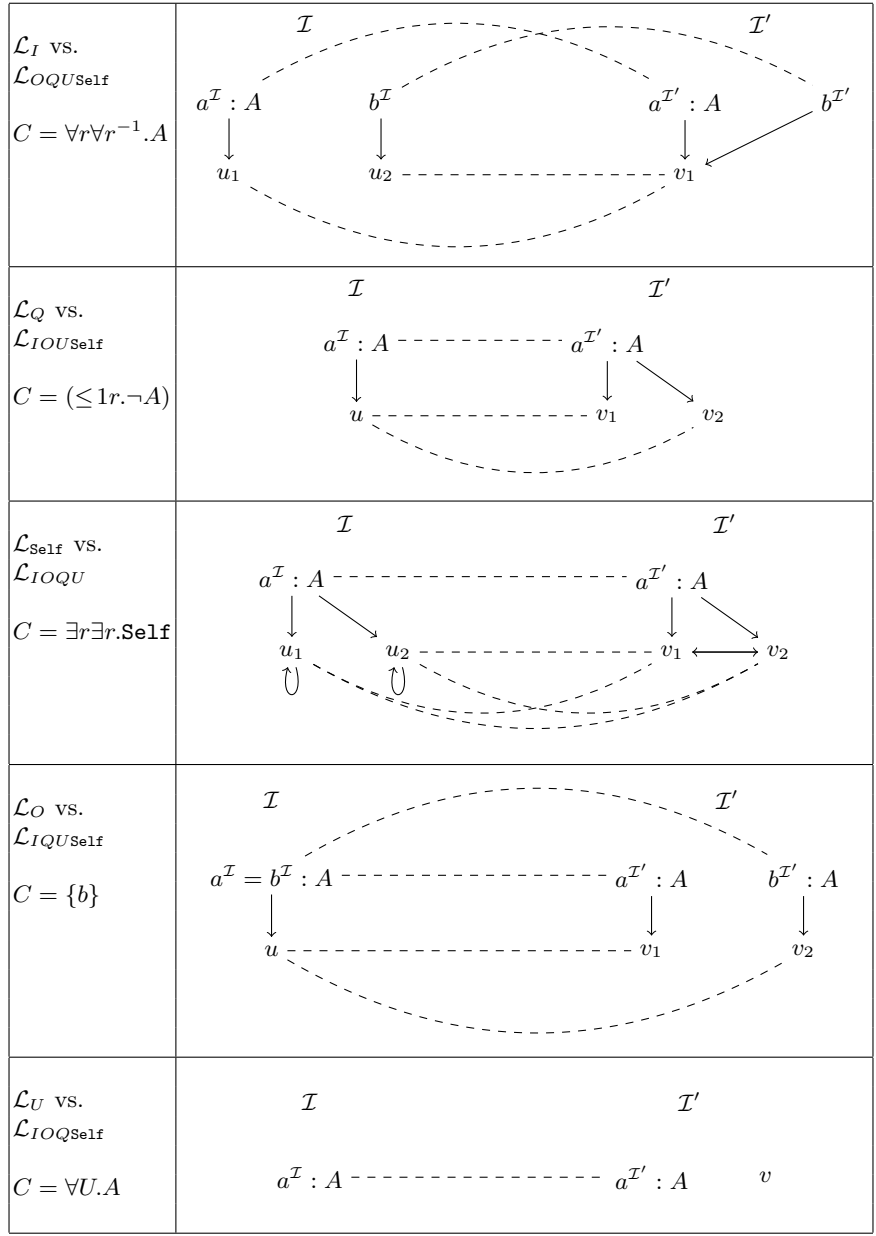
\begin{figure}[t!]
\begin{center}
\begin{tabular}{|l|c|}
\hline
& 
\multirow{9}{*}{
\begin{tikzpicture}
\node (I) {$\mI$};
\node (p1) [node distance=1cm, below of=I, left of=I] {${a^\mI}:A$};
\node (p2) [node distance=2cm, right of=p1]{${b^{\mI}}$};
\node (p3) [node distance=1cm, below of=p1]{${u_1}$};
\node (p4) [node distance=1cm, below of=p2]{${u_2}$};
\node (J) [node distance=6cm, right of=I]{$\mI'$};
\node (q1) [node distance=1cm, below of=J, left of=J] {${a^{\mI'}}:A$};
\node (q2) [node distance=1cm, below of=J, right of=J] {$b^{\mI'}$};
\node (q3) [node distance=1cm, below of=q1]{${v_1}$};
\draw[-, dashed, bend left] (p1) to node [swap] {} (q1);
\draw[-, dashed, bend left] (p2) to node [swap] {} (q2);
\draw[-, dashed, bend right] (p3) to node [swap] {} (q3);
\draw[-, dashed] (p4) to node [swap] {} (q3);
\draw[->] (p1) to node  { } (p3);
\draw[->] (p2) to node  { } (p4);
\draw[->] (q1) to node  {} (q3);
\draw[->] (q2) to node  {} (q3);
\end{tikzpicture}
} \\
$\mL_I$ vs. & \\
$\mL_{OQU\Self}$ & \\
 & \\
$C= \V r\V r^{-1}.A$ & \\
 & \\
 & \\
 & \\
 & \\
\hline
& 
\multirow{8}{*}{
   \begin{tikzpicture}
   \node (I) {$\mI$};
   \node (p1) [node distance=0.7cm, below of=I] {${a^\mI}:A$};
   \node (p3) [node distance=1cm, below of=p1]{${u}$};
   \node (J) [node distance=4cm, right of=I]{$\mI'$};
   \node (q1) [node distance=0.7cm, below of=J, left of=J] {${a^{\mI'}}:A$};
   \node (q3) [node distance=1cm, below of=q1]{${v_1}$};
   \node (q4) [node distance=1.4cm, right of=q3]{${v_2}$};
   \draw[-, dashed] (p1) to node [swap] {} (q1);
   \draw[-, dashed] (p3) to node [swap] {} (q3);
   \draw[-, dashed, bend right] (p3) to node [swap] {} (q4);
   \draw[->] (p1) to node  { } (p3);
   \draw[->] (q1) to node  {} (q3);
   \draw[->] (q1) to node  {} (q4);
   \end{tikzpicture}
} \\
$\mL_Q$ vs. & \\
$\mL_{IOU\Self}$ & \\
& \\
$C=(\leq\!1r.\lnot A)$ & \\
& \\
& \\
& \\
\hline 
& 
\multirow{9}{*}{
   \begin{tikzpicture}
   \node (I) {$\mI$};
   \node (p1) [node distance=0.7cm, below of=I, left of=I] {${a^\mI}:A$};
   \node (p3) [node distance=1cm, below of=p1]{${u_1}$};
   \node (p4) [node distance=1.4cm, right of=p3]{${u_2}$};
   \node (J) [node distance=5cm, right of=I]{$\mI'$};
   \node (q1) [node distance=0.7cm, below of=J, left of=J] {${a^{\mI'}}:A$};
   \node (q3) [node distance=1cm, below of=q1]{${v_1}$};
   \node (q4) [node distance=1.4cm, right of=q3]{${v_2}$};
   \draw[-, dashed] (p1) to node [swap] {} (q1);
   \draw[-, dashed] (p4) to node [swap] {} (q3);
   \draw[-, dashed, bend right] (p3) to node [swap] {} (q4);
   \draw[-, dashed, bend right] (p3) to node [swap] {} (q3);
   \draw[-, dashed, bend right] (p4) to node [swap] {} (q4);
   \draw[->] (p1) to node  { } (p3);
   \draw[->] (p3)[loop below] to node  {} (p3);
   \draw[->] (p1) to node  { } (p4);
   \draw[->] (p4)[loop below] to node  {} (p4);
   \draw[->] (q1) to node  {} (q3);
   \draw[->] (q1) to node  {} (q4);
   \draw[->] (q3) to node  {} (q4);
   \draw[->] (q4) to node  {} (q3);      
   \end{tikzpicture}
} \\
$\mL_\Self$ vs. & \\
$\mL_{IOQU}$ & \\ 
& \\
$C= \E r\E r.\Self$ & \\
& \\
& \\
& \\
& \\
\hline
& 
\multirow{10}{*}{
 \begin{tikzpicture}
 \node (I) {$\mI$};
 \node (p1) [node distance=1cm, below of=I] {${a^\mI=b^\mI}:A$};
 \node (p3) [node distance=1cm, below of=p1]{${u}$};
 \node (J) [node distance=6cm, right of=I]{$\mI'$};
 \node (q1) [node distance=1cm, below of=J, left of=J] {${a^{\mI'}}:A$};
 \node (q2) [node distance=1cm, below of=J, right of=J] {$b^{\mI'}:A$};
 \node (q3) [node distance=1cm, below of=q1]{${v_1}$};
 \node (q4) [node distance=1cm, below of=q2]{${v_2}$};
 \draw[-, dashed] (p1) to node [swap] {} (q1);
 \draw[-, dashed, bend left] (p1) to node [swap] {} (q2);
 \draw[-, dashed] (p3) to node [swap] {} (q3);
 \draw[-, dashed, bend right] (p3) to node [swap] {} (q4);
 \draw[->] (p1) to node  { } (p3);
 \draw[->] (q1) to node  {} (q3);
 \draw[->] (q2) to node  {} (q4);
 \end{tikzpicture}
} \\
$\mL_O$ vs. & \\
$\mL_{IQU\Self}$ & \\ 
 & \\
$C = \{b\}$ & \\
 & \\
 & \\
 & \\
 & \\
 & \\
\hline
& 
\multirow{6}{*}{
\begin{tikzpicture}
\node (I) {$\mI$};
\node (p1) [node distance=1cm, below of=I] {$a^\mI:A$};
\node (J) [node distance=5cm, right of=I]{$\mI'$};
\node (q1) [node distance=4.5cm, right of=p1] {${a^{\mI'}:A}$};
\node (q3) [node distance=1.5cm, right of=q1]{$v$};
\draw[-, dashed] (p1) to node [swap] {} (q1);
\end{tikzpicture}
} \\
$\mL_U$ vs. & \\
$\mL_{IOQ\Self}$ & \\ 
& \\
$C= \V U.A$ & \\
& \\
\hline
\end{tabular} 
\end{center}

\vspace{-1em}

\caption{An illustration for the proof of Lemma~\ref{Ali:main2}.\label{Ali:fig:main2}}
\end{figure}

\begin{proof}
Let us compare $\mL_I$ with $\mL_{OQU\Self}$. Consider the interpretations $\mI$, $\mI'$ and the relation $Z$ shown in the first part of Figure~\ref{Ali:fig:main2}. The arrows denote the instances of $r$ in $\mI$ and $\mI'$. The instances of $A$ in $\mI$ and $\mI'$ are explicitly indicated in the figure. Let $B^\mI = B^{\mI'} = \emptyset$ for all $B \in \CN \setminus \{A\}$, $s^\mI = s^{\mI'} = \emptyset$ for all $s \in \RN \setminus \{r\}$, and $c^\mI = a^\mI$, $c^{\mI'} = a^{\mI'}$ for all $c \in \IN \setminus \{a,b\}$. 
The dotted lines in the figure indicate the instances of a binary relation $Z \subseteq \Delta^\mI \times \Delta^{\mI'}$. 
It can be checked that $Z$ is an $\mL_{OQU\Self}$-bisimulation between $\mI$ and $\mI'$. Consider the concept $C= \V r\V r^{-1}.A$ of $\mL_I$. Clearly, $a^\mI \in C^\mI$ but $a^{\mI'} \not \in C^{\mI'}$. By Theorem~\ref{theorem: bs-inv-1}, $C$ does not have any equivalent concept in $\mL_{OQU\Self}$. Hence, $\mL_I \not \leq_C \mL_{OQU\Self}$. Consider the TBox $\mT = \{A \sqsubseteq C\}$. Since $\mI \models \mT$ but $\mI' \not \models \mT$, by Theorem~\ref{theorem: TBox-inv-2}, $\mT$ does not have any equivalent TBox in $\mL_{OQU\Self}$. Hence $\mL_I \not \leq_T \mL_{OQU\Self}$. Consider the ABox $\mA = \{C(a)\}$. Since $\mI \models \mA$ but $\mI' \not \models \mA$, by Theorem~\ref{theorem: ABox-inv}, $\mA$ does not have any equivalent ABox in $\mL_{OQU\Self}$. Hence $\mL_I \not \leq_A \mL_{OQU\Self}$.   

The proofs for the other pairs of logics can be done similarly, using $\mI$, $\mI'$, $C$ specified in the next parts of Figure~\ref{Ali:fig:main2}. For the parts without the presence of $b$, let $b^\mI = a^\mI$ and $b^{\mI'} = a^{\mI'}$. 
\myend
\end{proof}
} % newcommand

%\VersionA{\ProofLemmaAliMainT}
\ProofLemmaAliMainT

\newcommand{\TheoremAliMain}{Let $\Phi$ and $\Phi'$ be subsets of $\{I,O,Q,U,\Self\}$.
\begin{enumerate}
\item If $\Phi \subset \Phi'$ then $\mLP <_C \mL_{\Phi'}$.
\item If $\Phi \not \subseteq \Phi'$ then $\mLP \not\leq_C \mL_{\Phi'}$.
\item If $\Phi \subset \Phi'$ and $\Phi' \setminus \Phi \neq\{U\}$ then $\mLP <_T \mL_{\Phi'}$.
\item If $\Phi \not \subseteq \Phi'$ and $\Phi \setminus \Phi' \neq \{U\}$ then $\mLP \not\leq_T \mL_{\Phi'}$.
\item If $\Phi \subset \Phi'$ and $\Phi' \setminus \Phi \neq\{O\}$ then $\mLP <_A \mL_{\Phi'}$.
\item If $\Phi \not \subseteq \Phi'$ and $\Phi \setminus \Phi' \neq \{O\}$ then $\mLP \not\leq_A \mL_{\Phi'}$.
\end{enumerate}}

\begin{theorem}\label{Ali:th:main}
\TheoremAliMain
\end{theorem}

\newcommand{\ProofTheoremAliMain}{\begin{proof}
Consider the first assertion and suppose $\Phi \subset \Phi'$. Since every concept of $\mLP$ is also a concept of $\mL_{\Phi'}$, we have that $\mLP \leq_C \mL_{\Phi'}$. Since $\Phi' \setminus \Phi \neq \emptyset$, at least one feature among $I$, $O$, $Q$, $U$, $\Self$ belongs to $\Phi' \setminus \Phi$. Consider the case $I \in \Phi' \setminus \Phi$. The cases of other features are similar and omitted. For the sake of contradiction, suppose $\mL_{\Phi'} \leq_C \mLP$. Since $\mL_I \leq_C \mL_{\Phi'}$ and $\mLP \leq_C \mL_{OQU\Self}$, it follows that  $\mL_I \leq_C \mL_{OQU\Self}$, which contradicts Lemma~\ref{Ali:main2}. Therefore, $\mLP <_C \mL_{\Phi'}$. 

Consider the second assertion and suppose $\Phi \not\subseteq \Phi'$. Since $\Phi \setminus \Phi' \neq \emptyset$, at least one feature among $I$, $O$, $Q$, $U$, $\Self$ belongs to $\Phi \setminus \Phi'$. Consider the case $I \in \Phi \setminus \Phi'$. The cases of other features are similar and omitted. For the sake of contradiction, suppose $\mLP \leq_C \mL_{\Phi'}$. Since $\mL_I \leq_C \mLP$ and $\mL_{\Phi'} \leq_C \mL_{OQU\Self}$, it follows that  $\mL_I \leq_C \mL_{OQU\Self}$, which contradicts Lemma~\ref{Ali:main2}. Therefore, $\mLP \not\leq_C \mL_{\Phi'}$. 

Consider the third assertion and suppose $\Phi \subset \Phi'$ and $\Phi' \setminus \Phi \neq\{U\}$. At least one feature among $I$, $O$, $Q$, $\Self$ belongs to $\Phi' \setminus \Phi$. Consider the case $I \in \Phi' \setminus \Phi$. The cases of other features are similar and omitted. Since $\Phi \subset \Phi'$, \mbox{$\mLP \leq_T \mL_{\Phi'}$}. For the sake of contradiction, suppose $\mL_{\Phi'} \leq_T \mLP$. Since \mbox{$\mL_I \leq_T \mL_{\Phi'}$} and $\mLP \leq_T \mL_{OQU\Self}$, it follows that $\mL_I \leq_T \mL_{OQU\Self}$, which contradicts Lemma~\ref{Ali:main2}. Therefore, \mbox{$\mLP <_T \mL_{\Phi'}$}.

Consider the fourth assertion and suppose $\Phi \not\subseteq \Phi'$ and $\Phi \setminus \Phi' \neq\{U\}$. At least one feature among $I$, $O$, $Q$, $\Self$ belongs to $\Phi \setminus \Phi'$. Consider the case $I \in \Phi \setminus \Phi'$. The cases of other features are similar and omitted. For the sake of contradiction, suppose $\mLP \leq_T \mL_{\Phi'}$. Since $\mL_I \leq_T \mLP$ and $\mL_{\Phi'} \leq_T \mL_{OQU\Self}$, it follows that $\mL_I \leq_T \mL_{OQU\Self}$, which contradicts Lemma~\ref{Ali:main2}. Therefore, $\mLP \not\leq_T \mL_{\Phi'}$.

Consider the fifth assertion and suppose $\Phi \subset \Phi'$ and $\Phi' \setminus \Phi \neq\{O\}$. At least one feature among $I$, $Q$, $U$, $\Self$ belongs to $\Phi' \setminus \Phi$. Consider the case $I \in \Phi' \setminus \Phi$. The cases of other features are similar and omitted. Since $\Phi \subset \Phi'$, \mbox{$\mLP \leq_A \mL_{\Phi'}$}. For the sake of contradiction, suppose $\mL_{\Phi'} \leq_A \mLP$. Since \mbox{$\mL_I \leq_A \mL_{\Phi'}$} and $\mLP \leq_A \mL_{OQU\Self}$, it follows that $\mL_I \leq_A \mL_{OQU\Self}$, which contradicts Lemma~\ref{Ali:main2}. Therefore, \mbox{$\mLP <_A \mL_{\Phi'}$}.

Consider the last assertion and suppose $\Phi \not\subseteq \Phi'$ and $\Phi \setminus \Phi' \neq\{O\}$. At least one feature among $I$, $Q$, $U$, $\Self$ belongs to $\Phi \setminus \Phi'$. Consider the case $I \in \Phi \setminus \Phi'$. The cases of other features are similar and omitted. For the sake of contradiction, suppose $\mLP \leq_A \mL_{\Phi'}$. Since $\mL_I \leq_A \mLP$ and $\mL_{\Phi'} \leq_A \mL_{OQU\Self}$, it follows that $\mL_I \leq_A \mL_{OQU\Self}$, which contradicts Lemma~\ref{Ali:main2}. Therefore, $\mLP \not\leq_A \mL_{\Phi'}$.
\myend
\end{proof}
} % newcommand

%\VersionA{\ProofTheoremAliMain}
\ProofTheoremAliMain

%-------------------------------------------------------------------------------

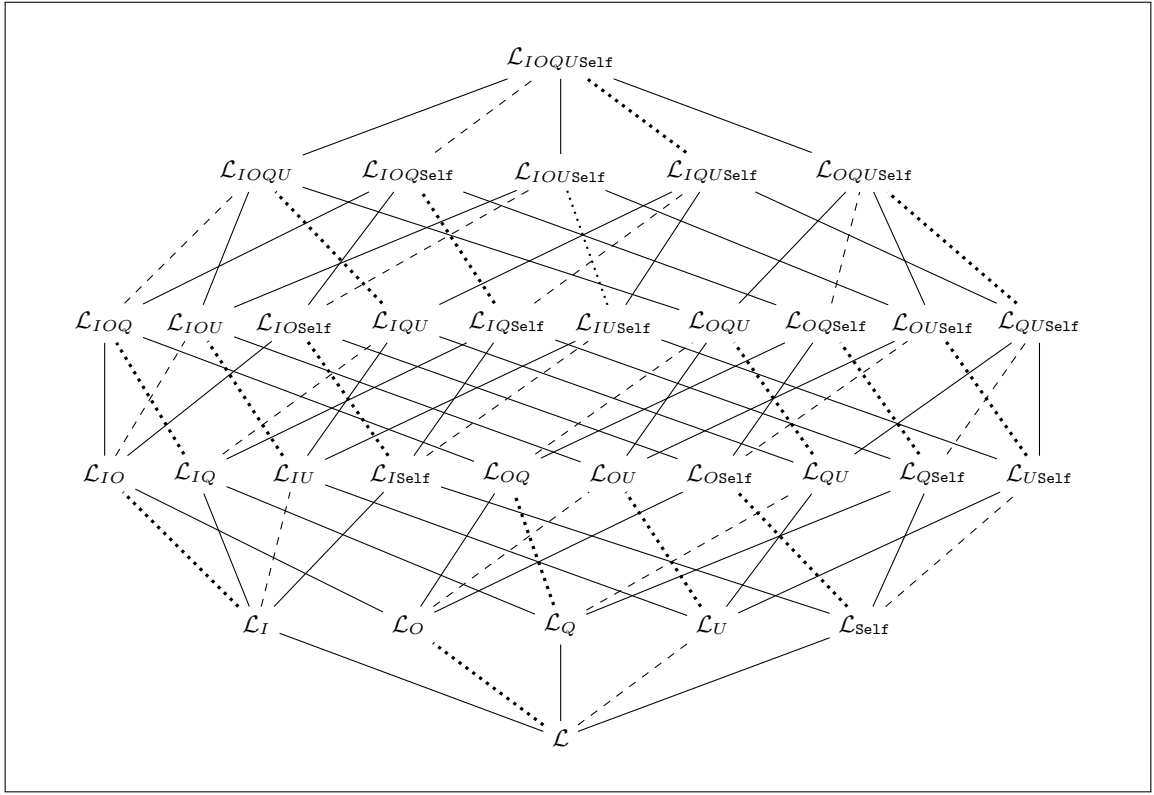
\begin{figure}[t!] 
\mybox{
\begin{center}
\begin{tikzpicture}
\node (I) {$\mL$};
\node (p3) [node distance=1.5cm, above of=I] {$\mL_Q$};
\node (p4) [node distance=2cm, right of=p3] {$\mL_U$};
\node (p2) [node distance=2cm, left of=p3] {$\mL_O$};
\node (p1) [node distance=2cm, left of=p2] {$\mL_I$};
\node (p5) [node distance=2cm, right of=p4] {$\mL_{\Self}$};
\node (x) [node distance=2cm, above of=p3] {};
\node (q5) [node distance=0.7cm, left of=x] {$\mL_{OQ}$};
\node (q6) [node distance=0.7cm, right of=x] {$\mL_{OU}$};
\node (q4) [node distance=1.4cm, left of=q5] {$\mL_{I\Self}$};
\node (q3) [node distance=2.8cm, left of=q5] {$\mL_{IU}$};
\node (q2) [node distance=4.1cm, left of=q5] {$\mL_{IQ}$};
\node (q1) [node distance=5.3cm, left of=q5] {$\mL_{IO}$};
\node (q7) [node distance=1.4cm, right of=q6] {$\mL_{O\Self}$};
\node (q8) [node distance=2.8cm, right of=q6] {$\mL_{QU}$};
\node (q9) [node distance=4.2cm, right of=q6] {$\mL_{Q\Self}$};
\node (q10) [node distance=5.6cm, right of=q6] {$\mL_{U\Self}$};
\node (r1) [node distance=2cm, above of=q1] {$\mL_{IOQ}$};
\node (r2) [node distance=2cm, above of=q2] {$\mL_{IOU}$};
\node (r3) [node distance=2cm, above of=q3] {$\mL_{IO\Self}$};
\node (r4) [node distance=2cm, above of=q4] {$\mL_{IQU}$};
\node (r5) [node distance=2cm, above of=q5] {$\mL_{IQ\Self}$};
\node (r10) [node distance=2cm, above of=q6] {$\mL_{IU\Self}$};
\node (r6) [node distance=2cm, above of=q7] {$\mL_{OQU}$};
\node (r7) [node distance=2cm, above of=q8] {$\mL_{OQ\Self}$};
\node (r8) [node distance=2cm, above of=q9] {$\mL_{OU\Self}$};
\node (r9) [node distance=2cm, above of=q10] {$\mL_{QU\Self}$};
\node (s3) [node distance=6cm, above of=p3] {$\mL_{IOU\Self}$};
\node (s2) [node distance=6cm, above of=p2] {$\mL_{IOQ\Self}$};
\node (s1) [node distance=6cm, above of=p1] {$\mL_{IOQU}$};
\node (s4) [node distance=6cm, above of=p4] {$\mL_{IQU\Self}$};
\node (s5) [node distance=6cm, above of=p5] {$\mL_{OQU\Self}$};
\node (t) [node distance=1.5cm, above of=s3] {$\mL_{IOQU\Self}$};
%
%\draw[-, dashed,bend right] (I) to node [swap] {} (p1);
%
\draw[-] (I) to node  { } (p1);
\draw[-, dotted, very thick] (I) to node  { } (p2);
\draw[-] (I) to node  { } (p3);
\draw[-, dashed] (I) to node  { } (p4);
\draw[-] (I) to node  { } (p5);
\draw[-, dotted, very thick] (p1) to node  { } (q1);
\draw[-] (p1) to node  { } (q2);
\draw[-,dashed] (p1) to node  { } (q3);
\draw[-] (p1) to node  { } (q4);
\draw[-] (p2) to node  { } (q1);
\draw[-] (p3) to node  { } (q2);
\draw[-] (p4) to node  { } (q3);
\draw[-] (p5) to node  { } (q4);
\draw[-] (p2) to node  { } (q5);
\draw[-, dotted, very thick] (p3) to node  { } (q5);
\draw[-, dashed] (p2) to node  { } (q6);
\draw[-, dotted, very thick] (p4) to node  { } (q6);
\draw[-] (p2) to node  { } (q7);
\draw[-, dotted, very thick] (p5) to node  { } (q7);
\draw[-, dashed] (p3) to node  { } (q8);
\draw[-] (p4) to node  { } (q8);
\draw[-] (p3) to node  { } (q9);
\draw[-] (p5) to node  { } (q9);
\draw[-] (p4) to node  { } (q10);
\draw[-, dashed] (p5) to node  { } (q10);
\draw[-] (q1) to node  { } (r1);
\draw[-, dotted, very thick] (q2) to node  { } (r1);
\draw[-] (q5) to node  { } (r1);
\draw[-,dashed] (q1) to node  { } (r2);
\draw[-, dotted, very thick] (q3) to node  { } (r2);
\draw[-] (q6) to node  { } (r2);
\draw[-] (q1) to node  { } (r3);
\draw[-, dotted, very thick] (q4) to node  { } (r3);
\draw[-] (q7) to node  { } (r3);
\draw[-,dashed] (q2) to node  { } (r4);
\draw[-] (q3) to node  { } (r4);
\draw[-] (q8) to node  { } (r4);
\draw[-] (q2) to node  { } (r5);
\draw[-] (q4) to node  { } (r5);
\draw[-] (q9) to node  { } (r5);
\draw[-,dashed] (q5) to node  { } (r6);
\draw[-] (q6) to node  { } (r6);
\draw[-, dotted, very thick] (q8) to node  { } (r6);
\draw[-] (q5) to node  { } (r7);
\draw[-] (q7) to node  { } (r7);
\draw[-, dotted, very thick] (q9) to node  { } (r7);
\draw[-] (q6) to node  { } (r8);
\draw[-, dashed] (q7) to node  { } (r8);
\draw[-, dotted, very thick] (q10) to node  { } (r8);
\draw[-] (q8) to node  { } (r9);
\draw[-,dashed] (q9) to node  { } (r9);
\draw[-] (q10) to node  { } (r9);
\draw[-] (q3) to node  { } (r10);
\draw[-,dashed] (q4) to node  { } (r10);
\draw[-] (q10) to node  { } (r10);
\draw[-,dashed] (r1) to node  { } (s1);
\draw[-] (r2) to node  { } (s1);
\draw[-, dotted, very thick] (r4) to node  { } (s1);
\draw[-] (r6) to node  { } (s1);
\draw[-] (r1) to node  { } (s2);
\draw[-] (r3) to node  { } (s2);
\draw[-] (r4) to node  { } (s4);
\draw[-, dotted, very thick] (r5) to node  { } (s2);
\draw[-] (r7) to node  { } (s2);
\draw[-] (r2) to node  { } (s3);
\draw[-, dotted, thick] (r10) to node  { } (s3);
\draw[-,dashed] (r3) to node  { } (s3);
\draw[-] (r8) to node  { } (s3);
\draw[-] (r10) to node  { } (s4);
\draw[-, dashed] (r5) to node  { } (s4);
\draw[-] (r9) to node  { } (s4);
\draw[-] (r6) to node  { } (s5);
\draw[-,dashed] (r7) to node  { } (s5);
\draw[-] (r8) to node  { } (s5);
\draw[-, dotted, very thick] (r9) to node  { } (s5);
\draw[-] (s1) to node  { } (t);
\draw[-,dashed] (s2) to node  { } (t);
\draw[-] (s3) to node  { } (t);
\draw[-, dotted, very thick] (s4) to node  { } (t);
\draw[-] (s5) to node  { } (t);
\end{tikzpicture}
\end{center}
}
\caption{Comparing the expressiveness of description logics, where $\mathcal{ALC} \leq \mL \leq \mathcal{ALC}_{reg}$. If there is a path from a logic $\mL_2$ down to a logic $\mL_1$ that contains either a normal edge or at least two edges then $\mL_2$ is more expressive than $\mL_1$ w.r.t.\ concepts, TBoxes and ABoxes. If the path is a dotted edge then $\mL_2$ is more expressive than $\mL_1$ w.r.t.\ concepts and TBoxes. If the path is a dashed edge then $\mL_2$ is more expressive than $\mL_1$ w.r.t.\ concepts and ABoxes.\label{Ali:fig:IOUQSelf}}
\end{figure} 

\begin{definition}\em
We define \ALC to be the sublogic of \ALCreg such that the role constructors $\varepsilon$, $R \circ S$, $R \sqcup S$, $R^*$ and $C?$ are disallowed. We say that $\mL$ is a {\em sublogic of \ALCreg that extends \ALC}, denoted by $\mathcal{ALC} \leq \mL \leq \mathcal{ALC}_{reg}$, if it extends \ALC with some of those role constructors. For $\mathcal{ALC} \leq \mL \leq \mathcal{ALC}_{reg}$ and $\Phi \subseteq \{I,O,Q,U,\Self\}$, let $\mLP$ be defined as usual in the spirit of Definition~\ref{def:HGFDA}.
\myend
\end{definition}

\newcommand{\CorAliMain}{Let $\mL$ be any sublogic of \ALCreg that extends \ALC and let $\Phi$ and $\Phi'$ be subsets of $\{I,O,Q,U,\Self\}$.
\begin{enumerate}
\item If $\Phi \subset \Phi'$ then $\mLP <_C \mL_{\Phi'}$.
\item If $\Phi \not \subseteq \Phi'$ then $\mLP \not\leq_C \mL_{\Phi'}$.
\item If $\Phi \subset \Phi'$ and $\Phi' \setminus \Phi \neq\{U\}$ then $\mLP <_T \mL_{\Phi'}$.
\item If $\Phi \not \subseteq \Phi'$ and $\Phi \setminus \Phi' \neq \{U\}$ then $\mLP \not\leq_T \mL_{\Phi'}$.
\item If $\Phi \subset \Phi'$ and $\Phi' \setminus \Phi \neq\{O\}$ then $\mLP <_A \mL_{\Phi'}$.
\item If $\Phi \not \subseteq \Phi'$ and $\Phi \setminus \Phi' \neq \{O\}$ then $\mLP \not\leq_A \mL_{\Phi'}$.
\end{enumerate}}

\begin{corollary}\label{Ali:cor:main}
\CorAliMain
\end{corollary}

\newcommand{\ProofCorAliMain}{\begin{proof}
Observe that the concepts $C$ listed in Figure~\ref{Ali:fig:main2} do not use any of the role constructors $\varepsilon$, $R \circ S$, $R \sqcup S$, $R^*$, $C?$. All the lemmas and theorems given in Section~\ref{section: OIDSJ} hold for the case when $\mL$ is a sublogic of \ALCreg that extends \ALC. Their proofs do not require any change.
\myend
\end{proof}
} % newcommand

%\VersionA{\ProofCorAliMain}
\ProofCorAliMain

%-------------------------------------------------------------------------------

Figure~\ref{Ali:fig:IOUQSelf} illustrates the relationships between the expressiveness of all the DLs that extend $\mL$, where $\mathcal{ALC} \leq \mL \leq \mathcal{ALC}_{reg}$, with any non-empty combination of the features $I$, $O$, $Q$, $U$, $\Self$. 
Note that the problems whether $\mLP <_T \mL_{\Phi'}$ when $\Phi' \setminus \Phi = \{U\}$ and whether $\mLP <_A \mL_{\Phi'}$ when $\Phi' \setminus \Phi = \{O\}$ remain open.

%-------------------------------------------------------------------------------

\subsection{Interpretation Minimization: Applications}
\label{section: IUEKA}

Minimizing an interpretation in a DL is not the same as minimizing an ontology in that DL. From the logical point of view, an ontology is specified by a knowledge base, which may have zero or infinitely many models. It is possible that interpretation minimization may have some effects or may form a starting point for the study on ontology minimization. However, this is a challenging topic of automated reasoning in DLs and is beyond the scope of this paper. In this subsection we only discuss applications of interpretation minimization, which is useful when one is dealing with a~specific interpretation, e.g., with the {\em unique intended model} of a rule-based knowledge base in a DL or with a counterexample of an instance checking problem in a~DL. 

Note that if a knowledge base $\KB$ has the unique intended model $\mI$ then a problem of checking $\KB \models \varphi$, where $\varphi$ is a terminological axiom, a role inclusion axiom or an individual assertion of the form $C(a)$, $R(a,b)$ or $a = b$, is usually defined to be equivalent to the problem of checking whether $\mI \models \varphi$. In this case, it makes sense to reduce $\mI$ to $\mI' = \mIq$ when $\Phi \subseteq \{I,O,U\}$, and to $\mI' = \mIqQS$ in the other case. According to Theorems~\ref{theorem: mz1} and~\ref{theorem: mz2}, $\mI'$ is a minimal version of $\mI$ w.r.t.\ essential aspects. Furthermore, by Theorems~\ref{theorem: quo-x} and~\ref{theorem: quo-y}, when $\varphi$ is a query of one of the mentioned forms, $\mI \models \varphi$ iff $\mI' \models \varphi$, and hence $\KB \models \varphi$ iff $\mI' \models \varphi$. Clearly, the reduction is useful as it can be used for answering many queries. 

We present below exemplary types of rule-based knowledge bases in DLs that have the unique intended model: 
%as well as a counterexample of an instance checking problem in a DL for that interpretation minimization is useful. 
\begin{itemize}
\item{\bf Acyclic knowledge bases}: The notion of acyclic knowledge bases in DLs is widely used (see, e.g., \cite{LbRoughification} for a definition). Under the unique name assumption and the closed world assumption, an acyclic knowledge base $\KB$ has the {\em standard model} (see~\cite{LbRoughification} for details). The {\em unique intended model} of such a $\KB$ can be defined to be its standard model. We refer the reader to~\cite{LbRoughification} for an example.

\item{\bf \RLp}: \RL is a profile of OWL~2~Full recommended by W3C. It hase \PTIME data complexity. Knowledge bases in \RL may be unsatisfiable (i.e., inconsistent), since their translations into Datalog may also need negative clauses as constraints. In~\cite{CaoNS14-OWL2RL} Cao et al.\ introduced \RLz as the logical formalism of \RL that ignores the predefined data types. They then introduced \RLp as a maximal fragment of \RLz with the property that every knowledge base $\KB$ expressed in \RLp can be translated to an equivalent Datalog program $P$ without negative clauses. The {\em unique intended model} of such a $\KB$ is the least Herbrand model of that Datalog program~$P$. 

\item{\bf WORL} and {\bf SWORL}: In~\cite{CaoNS14-WORL} Cao et at.\ introduced a Web ontology rule language called WORL, which combines a~variant of \RL with eDatalog$^\lnot$. Similarly to the work on \RLp~\cite{CaoNS14-OWL2RL}, they disallowed those features of \RL that play the role of constraints\footnote{I.e., the ones that are translated to negative clauses of the form $\varphi \to \bot$.}, allowed unary external checkable predicates, additional features like negation and the constructor $\geq\!n\,R.C$ to occur at the left hand side of $\sqsubseteq$ in concept inclusion axioms. They adopted some restrictions for the additional features to guarantee a~translation of WORL programs into eDatalog$^\lnot$. They also defined the rule language SWORL (stratified WORL) and developed the well-founded semantics for WORL and the standard semantics for SWORL via translation into eDatalog$^\lnot$. Both WORL with respect to the well-founded semantics and SWORL with respect to the standard semantics have \PTIME data complexity.
The {\em unique intended model} of a WORL knowledge base $\KB$ can be defined to be the well-founded model of $\KB$, and the {\em unique intended model} of a SWORL knowledge base $\KB$ can be defined to be the standard model of $\KB$. 
\end{itemize}

%-------------------------------------------------------------------------------

\subsection{Concept Learning in Description Logics}
\label{section: JHPSM}

Concept learning in DLs is useful for making decision rules as in traditional binary classification. It is also useful in ontology engineering, e.g., for suggesting definitions of important concepts. 
The major settings of concept learning in DLs are as follows:
\begin{description}
\item[{\bf Setting~1:}] Given a knowledge base $\KB$ and sets $E^+$, $E^-$ of named individuals, learn a concept $C$ in a DL $L$ such that: (a)~$\KB \models C(a)$ for all $a \in E^+$, and (b)~$\KB \models \lnot C(a)$ for all $a \in E^-$. 
The set $E^+$ (resp.\ $E^-$) contains positive (resp. negative) examples of~$C$.

\item[{\bf Setting~2:}] This setting differs from Setting~1 only in that the condition (b) is replaced by the weaker one: $\KB \not\models C(a)$ for all $a \in E^-$.

\item[{\bf Setting~3:}] Given an interpretation $\mI$ and sets $E^+$, $E^-$ of named individuals, learn a concept $C$ in $L$ such that: (a)~$\mI \models C(a)$ for all $a \in E^+$, and (b)~$\mI \models \lnot C(a)$ for all $a \in E^-$. 
Note that $\mI \not\models C(a)$ is the same as $\mI \models \lnot C(a)$.
\end{description}

In~\cite{CohenH94} Cohen and Hirsh studied PAC-learnability of an early DL formalism called CLASSIC. They proposed a concept learning algorithm based on ``least common subsumers''. In~\cite{LambrixL98} Lambrix and Larocchia proposed a simple concept learning algorithm based on concept normalization. 
Badea and Nienhuys-Cheng~\cite{BadeaN00}, Iannone et al.~\cite{IannonePF07}, Fanizzi et al.~\cite{FanizzidEL08}, Lehmann and Hitzler~\cite{LehmannH10} studied concept learning in DLs by using refinement operators as in inductive logic programming. The works~\cite{BadeaN00,IannonePF07} use Setting~1, while the works~\cite{FanizzidEL08,LehmannH10} use Setting~2. 

Bisimulations in DLs have been used for concept learning in DLs in a number of papers. 
%~\cite{LbRoughification,KSE2012,SoICT2012,ICCCI2012a,TranHHNN2013,TranNH2014}. All of these papers acknowledge the workshop version~\cite{BSDL-CSP} of the current paper as the logical basis. 
In the rest of this subsection, we present a~survey about them. 

\subsubsection{Bisimulation-Based Concept Learning in DLs Using Setting~3}\ 

\medskip

\noindent
In~\cite{LbRoughification} Nguyen and Sza{\l}as generalized our notion of bisimulations in DLs and some of our results~\cite{BSDL-CSP} to model indiscernibility of objects for concept learning in DLs. It also concerns concept approximation by using bisimulation and Pawlak's rough set theory~\cite{Pa91,PawlakS07}. The generalization deals with the following: the main language is $\mLP$ using the signature $\Sigma$, but the concept to be learned may be restricted to a language $\mL_{\Phi^\dag}$ with a signature $\Sigma^\dag$, where $\Phi^\dag \subseteq \Phi$ and $\Sigma^\dag \subseteq \Sigma$. For that they introduced the language $\mLSP$ and $\mLSPD$-bisimulation between two interpretations $\mI$ and $\mI'$ (see~\cite{LbRoughification} for details).  

An $\mLSPD$-bisimulation between $\mI$ and itself is called an {\em $\mLSPD$-auto-bisimulation of~$\mI$}. An $\mLSPD$-auto-bisimulation of $\mI$ is said to be the {\em largest} if it is larger than or equal to ($\supseteq$) any other $\mLSPD$-auto-bisimulation of~$\mI$.

An {\em information system in $\mLSP$} is a finite interpretation in $\mLSP$. 
It can be given explicitly or specified somehow, e.g., by a knowledge base in the Web ontology rule language OWL~2~RL$^+$~\cite{CaoNS14-OWL2RL} (using the standard semantics) or WORL~\cite{CaoNS14-WORL} (using the well-founded semantics) or SWORL~\cite{CaoNS14-WORL} (using the stratified semantics).

Given an interpretation $\mI$ in $\mLSP$, by $\sim_\SdPdI$ we denote the largest $\mLSPD$-auto-bisimulation of $\mI$, and by $\equiv_\SdPdI$ we denote the binary relation on $\Delta^\mI$ with the property that \mbox{$x \equiv_\SdPdI x'$} iff $x$ is $\mLSPD$-equivalent to $x'$.

The following theorem correspond to Propositions~\ref{prop: JFDSP} and~\ref{prop: LABS}.

\begin{theorem}{\cite[Theorem~19.3]{LbRoughification}}\label{theorem: SDFHG}
Let $\Sigma$ and $\SigmaDag$ be DL-signatures such that $\SigmaDag \subseteq \Sigma$, $\Phi$ and $\Phi^\dag$ be sets of DL-features such that $\Phi^\dag \subseteq \Phi$, and $\mI$ be an interpretation in $\mLSP$. Then:
\begin{itemize}
\item the largest $\mLSPD$-auto-bisimulation of $\mI$ exists and is an equivalence relation,
\item if $\mI$ is finitely branching w.r.t. $\mLSPD$ then the relation $\equiv_\SdPdI$ is the largest $\mLSPD$-auto-bisimulation of $\mI$ (i.e.\ the relations $\equiv_\SdPdI$ and $\sim_\SdPdI$ coincide).
\end{itemize}
\end{theorem}

We say that a~set $Y$ is {\em split} by a~set $X$ if $Y \setminus X \neq \emptyset$ and $Y \cap X \neq \emptyset$. Thus, $Y$ is not split by $X$ if either $Y \subseteq X$ or $Y \cap X = \emptyset$.
A partition $P = \{Y_1,\ldots,Y_n\}$ is {\em consistent} with a~set $X$ if, for every $1 \leq i \leq n$, $Y_i$ is not split by $X$.

\begin{theorem}{\cite[Theorem~19.4]{LbRoughification}} \label{theorem: HGSDK}
Let $\mI$ be an interpretation in $\mLSP$, and let $X \subseteq \Delta^\mI$, $\SigmaDag \subseteq \Sigma$ and $\Phi^\dag \subseteq \Phi$. Then:
\begin{enumerate}
\item if there exists a~concept $C$ of $\mLSPD$ such that $X = C^\mI$ then the partition of $\Delta^\mI$ by $\sim_\SdPdI$ is consistent with $X$,
\item if the partition of $\Delta^\mI$ by $\sim_\SdPdI$ is consistent with $X$ then there exists a~concept $C$ of $\mLSPD$ such that $C^\mI = X$.
\end{enumerate}
\end{theorem}

Let $\mI$ be an information system in $\mLSP$ and let $A_d \in \CN$ be a~concept name standing for the ``decision attribute''. Suppose that $A_d$ can be expressed by a~concept $C$ in $\mLSPD$, for some specific $\SigmaDag \subseteq \Sigma \setminus \{A_d\}$ and $\Phi^\dag \subseteq \Phi$. How can we learn that concept $C$ on the basis of $\mI$? That is, how can we learn a~definition of $A_d$ in $\mLSPD$ on the basis of $\mI$?

The idea of~\cite{LbRoughification} for this task is based on the following observation:
\begin{quote}
if $A_d$ is definable in $\mLSPD$ then, by the first assertion of Theorem~\ref{theorem: HGSDK}, $A_d^\mI$~must be the union of some equivalence classes of $\Delta^\mI$ w.r.t. $\sim_\SdPdI$.
\end{quote}
Nguyen and Sza{\l}as~\cite{LbRoughification} proposed the following method:
\begin{enumerate}
\item Starting from the partition $\{\Delta^\mI\}$, make subsequent granulations to reach the partition corresponding to $\sim_\SdPdI$. 
The granulation process can be stopped as soon as the current partition is consistent with $A_d^\mI$ (or when some criteria are met).
In the granulation process, we denote the blocks created so far in all steps by $Y_1, \ldots, Y_n$, where the current partition $\{Y_{i_1},\ldots,Y_{i_k}\}$ may consist of only some of them. We do not use the same subscript to denote blocks of different contents (i.e., we always use new subscripts obtained by increasing $n$ for new blocks). We take care that, for each $1 \leq i \leq n$:
   \begin{itemize}
   \item $Y_i$ is characterized by an appropriate concept $C_i$ (such that $Y_i = C_i^\mI$),

   \item we keep information about whether $Y_i$ is split by $A_d^\mI$,

   \item if $Y_i \subseteq A_d^\mI$ then $\LargestContainer[i] := j$, where $1 \leq j \leq n$ is the subscript of the largest block $Y_j$ such that $Y_i \subseteq Y_j \subseteq A_d^\mI$.
   \end{itemize}

\item At the end, let $j_1,\ldots,j_h$ be all the indices from $\{i_1,\ldots,i_k\}$ such that $Y_{j_t} \subseteq A_d^\mI$ for $1 \leq t \leq h$, and let $\{l_1,\ldots,l_p\} = \{\LargestContainer[j_t] \mid 1 \leq t \leq h\}$.
Let $C$ be a~simplified form of $C_{l_1} \mor \ldots \mor C_{l_p}$.
Return $C$ as the result.
\end{enumerate}

%How to split a block and make a concept $C_i$ that characterizes the block $Y_i$ will be described shortly. 

In~\cite{KSE2012} Tran et al.\ generalized and extended the concept learning method of~\cite{LbRoughification} for DL-based information systems. They took attributes as basic elements of the language. Each attribute may be discrete or numeric. A Boolean attribute is treated as a concept name. They also allowed data roles and the features $F$ (functionality) and $N$ (unqualified number restriction). If $\sigma$ is a data role and $d$ belongs to the range of $\sigma$ then $\E \sigma.\{d\}$ is a concept. Concepts $\geq\!n\,R$ and $\leq\!n\,R$, where $R$ is a basic role, mean $\geq\!n\,R.\top$ and $\leq\!n\,R.\top$, respectively, and can be used when the feature $N$ is allowed. The concept $\leq\!1\,r$ (resp.\ $\leq\!1\,r^-$) is used to express functionality (resp.\ inverse functionality) of $r$. 
The Hennessy-Milner property (Theorem~\ref{theorem: H-M}) is reformulated in a straightforward way for the extended language $\mLSP$~\cite{KSE2012}. 

\begin{figure}[t!]
\mybox{
%\parbox{15.5cm}{
\begin{itemize}
\item $A$, where $A \in \SigmaDagC$ % this is related to~\eqref{bs:eqB}
\item $A=d$, where $A \in \SigmaDagA\setminus\SigmaDagC$ and $d \in \Range(A)$
\item $A \leq d$ and $A < d$, where $A \in \SigmaDagNA$, $d \in \Range(A)$ and $d$ is not a minimal element of $\Range(A)$
\item $A \geq d$ and $A > d$, where $A \in \SigmaDagNA$, $d \in \Range(A)$ and $d$ is not a maximal element of $\Range(A)$
\item $\E \sigma.\{d\}$, where $\sigma \in \SigmaDagDR$ and $d \in \Range(\sigma)$
\item $\E r.C_i$, $\E r.\top$ and $\V r.C_i$, where $r \in \SigmaDagOR$ and $1 \leq i \leq n$
\item $\E r^-.C_i$, $\E r^-.\top$ and $\V r^-.C_i$, if $I \in \Phi^\dag$, $r \in \SigmaDagOR$ and $1 \leq i \leq n$
\item $\{a\}$, if $O \in \Phi^\dag$ and $a \in \SigmaDagI$ % this is related to~\eqref{bs:eqO0}
\item $\leq\!1\,r$, if $F \in \Phi^\dag$ and $r \in \SigmaDagOR$
\item $\leq\!1\,r^-$, if $\{F,I\} \subseteq \Phi^\dag$ and $r \in \SigmaDagOR$
\item $\geq\!l\,r$ and $\leq\!m\,r$, if $N \in \Phi^\dag$, $r \in \SigmaDagOR$, $0 < l \leq \sharp\Delta^\mI$ and $0 \leq m < \sharp\Delta^\mI$
\item $\geq\!l\,r^-$ and $\leq\!m\,r^-$, if $\{N,I\} \subseteq \Phi^\dag$, $r \in \SigmaDagOR$, $0 < l \leq \sharp\Delta^\mI$ and $0 \leq m < \sharp\Delta^\mI$
\item $\geq\!l\,r.C_i$ and $\leq\!m\,r.C_i$, if $Q \in \Phi^\dag$, $r \in \SigmaDagOR$, $1 \leq i \leq n$, $0 < l \leq \sharp C_i$ and $0 \leq m < \sharp C_i$
\item $\geq\!l\,r^-.C_i$ and $\leq\!m\,r^-.C_i$, if $\{Q,I\}\! \subseteq\!\Phi^\dag$, $r\!\in\!\SigmaDagOR$, $1 \leq i \leq n$, $0 < l \leq \sharp C_i$ and $0 \leq m < \sharp C_i$$\!\!$
\item $\E r.\Self$, if $\Self \in \Phi^\dag$ and $r \in \SigmaDagOR$ \end{itemize}
%} %\parbox
} %\mybox

\vspace{-1em}

\caption{Basic selectors. Here, $\SigmaDagA$ denotes the set of attributes of $\SigmaDag$, $\Range(A)$ denotes the range of the attribute~$A$, $\SigmaDagNA$ denotes the set of numeric attributes of $\SigmaDag$, $\SigmaDagDR$ denotes the set of data roles of $\SigmaDag$, $\Range(\sigma)$ denotes the range of the data role $\sigma$, $\SigmaDagOR$ denotes the set of (object) role names of $\SigmaDag$, $n$ is the number of blocks created so far when granulating $\Delta^\mI$, and $C_i$ is the concept characterizing the block~$Y_i$.\label{fig: Selectors}}
\end{figure}

Reconsider the process of granulating $\{\Delta^\mI\}$ for computing the partition corresponding to $\sim_\SdPdI$. The works~\cite{LbRoughification,KSE2012} use the concepts listed in Figure~\ref{fig: Selectors} (on page~\pageref{fig: Selectors}) as {\em basic selectors} for the granulation process. Let the current partition of $\Delta^\mI$ be $\{Y_{i_1},\ldots,Y_{i_k}\}$. If a block $Y_{i_j}$ ($1 \leq j \leq k$) is split by $D^\mI$, where $D$ is a selector, then splitting $Y_{i_j}$ by $D$ is done as follows:
      \begin{itemize}
      \item $s := n+1$,\ \ $t := n+2$,\ \ $n := n+2$,
      \item $Y_s := Y_{i_j} \cap D^\mI$,\ \ $C_s := C_{i_j} \mand D$,
      \item $Y_t := Y_{i_j} \cap (\lnot D)^\mI$,\ \ $C_t := C_{i_j} \mand \lnot D$,
      \item the new partition of $\Delta^\mI$ becomes $\{Y_{i_1},\ldots,Y_{i_k}\} \setminus \{Y_{i_j}\} \cup \{Y_s,Y_t\}$.
      \end{itemize}

It was proved in~\cite{KSE2012} that using the basic selectors listed in Figure~\ref{fig: Selectors} is sufficient to granulate $\Delta^\mI$ to obtain the partition corresponding to~$\sim_\SdPdI$. In practice, we prefer as simple as possible definitions for the learned concept. Therefore, it is worth using also other selectors~\cite{LbRoughification,KSE2012,TranNH2014} (despite that they are expressible by the basic selectors over~$\mI$). 

In~\cite{TranNH2014} Tran et al.\ implemented the bisimulation-based concept learning method of~\cite{LbRoughification,KSE2012} (for most of the DLs considered in~\cite{LbRoughification,KSE2012}). They presented a domain partitioning method that use information gain and both basic selectors and extended selectors. The evaluation results of~\cite{TranNH2014} show that the concept learning method of~\cite{LbRoughification,KSE2012} (for Setting~3) is valuable and extended selectors support it significantly.

We refer the reader to~\cite{LbRoughification,KSE2012,TranNH2014} for examples illustrating concept learning for DL-based information systems. 
 
%-------------------------------------------------------------------------------

In~\cite{ICCCI2012a} Divroodi et al.\ studied the C-learnability (possibility of correct learning) of concepts in DLs using Setting~3. They proved that any concept in any DL that extends \ALC with some features amongst $I$, $\Self$, $Q_k$ (qualified number restrictions with numbers bounded by a constant $k$) can be learned if the training information system (specified by an interpretation) is good enough. That is, there exists a learning algorithm such that, for every concept $C$ of those logics, there exists a training information system consistent with $C$ such that applying the learning algorithm to the system results in a concept equivalent to~$C$. 
This shows a good property of the bisimulation-based concept learning method of~\cite{LbRoughification,KSE2012}.

The work~\cite{ICCCI2012a} uses {\em bounded bisimulation} in DLs and a new version of the algorithms proposed in~\cite{LbRoughification,KSE2012} that minimizes modal depths of resulting concepts. 
The notion of (bounded) $\mLSPd$-bisimulation, where $d$ is a natural number standing for the bound, is defined in~\cite{ICCCI2012a} appropriately so that the Hennessy-Milner property (similar to Theorem~\ref{theorem: H-M}) holds and, for every interpretations $\mI$, $\mI'$ and elements $x \in \Delta^\mI$, $x' \in \Delta^{\mI'}$, we have that $x$ is $\mLSPd$-bisimilar to $x'$ iff $x$ is $\mLSPd$-equivalent to $x'$ (i.e., for every concept $C$ of $\mLSP$ with the modal depth bounded by $d$, $x \in C^\mI$ iff $x' \in C^{\mI'}$). 

\subsubsection{Bisimulation-Based Concept Learning in DLs Using Settings~1 and~2}\ 

\medskip

\noindent
In~\cite{SoICT2012} Ha et al.\ developed bisimulation-based methods, called \BBCLearn and dual-\BBCLearn, for concept learning in DLs using Setting~1. Their method uses models of $\KB$ and bisimulations in those models to guide the search for the concept to be learned. 

In~\cite{TranHHNN2013} Tran et al.\ developed a bisimulation-based method, called \BBCLearnS, for concept learning in DLs using Setting~2. Their method is based on the dual-\BBCLearn method~\cite{SoICT2012}. 
They made appropriate changes for dealing with the condition ``$\KB \not\models C(a)$ for all $a \in E^-$'' instead of ``$\KB \models \lnot C(a)$ for all $a \in E^-$''.  

The concept learning methods \BBCLearn, dual-\BBCLearn and \BBCLearnS are formulated for the DLs $\mLP$ with $\mL = \mathcal{ALC}$, $\Phi \subseteq \{F,I,N,O,Q,U,\Self\}$, (discrete and numeric) attributes and data roles. These DLs disallow the PDL-like role constructors, but it still covers a large class of DLs and well-known DLs like \ALC, \SHIQ, \SHOIQ, \SROIQ. 
We refer the reader to~\cite{SoICT2012,TranHHNN2013} for illustrative examples about \BBCLearn, dual-\BBCLearn and \BBCLearnS.

These methods were not implemented and evaluated. As bisimulation is the notion for characterizing indiscernibility of objects in DLs, one can hope that they are promising. %Implementing, evaluating and improving these methods will be done by Thanh-Luong Tran for his PhD thesis under the supervision of the second author of the current paper and Dr.~Thi-Lan-Giao Hoang.

%-------------------------------------------------------------------------------

\section{Conclusions}
\label{section: conc}

We have studied bisimulations in a uniform way for a large class of DLs with useful ones like the DL \SROIQ of OWL~2. In comparison with~\cite{KurtoninaR99,LutzPW11}, this class allows also the role constructors of PDL, the concept constructor $\E r.\Self$ and the universal role as well as role axioms. Our main contributions are the following:
\begin{itemize}
	\item We proposed to treat named individuals as initial states and gave an appropriate condition for bisimulation. We introduced bisimulation conditions for the universal role and the concept constructor $\E r.\Self$.
	\item We proved that all of the bisimulation conditions \eqref{bs:eqA}-\eqref{bs:eqSelf} can be combined together to guarantee invariance of concepts and the Hennessy-Milner property for the whole class of studied DLs. 
	\item We addressed and gave results on invariance or preservation of ABoxes, RBoxes and knowledge bases in DLs. Independently from~\cite{LutzPW11} we gave results on invariance of TBoxes. By examples, we showed that our results on invariance or preservation of TBoxes, ABoxes, RBoxes and knowledge bases in DLs are strong and cannot be extended in a straightforward way.
	
	\item We introduced a new notion called QS-interpretation, which is needed for dealing with minimizing interpretations in DLs with qualified number restrictions and/or the concept constructor $\E r.\Self$. 
	\item We formulated and proved results on minimality of quotient interpretations w.r.t.\ the largest auto-bisimulations. 
	\item We adapted Hopcroft's automaton minimization algorithm and the Paige-Tarjan algorithm to give efficient algorithms for computing the partition corresponding to the largest auto-bisimulation of a finite interpretation in any DL of the considered family. The adaptation requires special treatments for the allowed constructors of the considered DLs.
	\item We provided results about separating the expressiveness of the DLs that extend~$\mL$, where $\mathcal{ALC} \leq \mL \leq \mathcal{ALC}_{reg}$, with any combination of the features $I$, $O$, $Q$, $U$, $\Self$. Our separation results are w.r.t.\ concepts, TBoxes and ABoxes. 
	Our work on the expressiveness of DLs differs significantly from all of~\cite{Baader96,Borgida96,CadoliPL97,KurtoninaR99,LutzPW11}, as the class of considered DLs is much larger than the ones considered in those works and our separation results are obtained not only w.r.t.\ concepts and TBoxes but also w.r.t.\ ABoxes.
\end{itemize}

We also gave a survey on bisimulation-based concept learning in DLs and discussed applications of interpretation minimization.

This paper is a comprehensive work on bisimulations for DLs. Our results found the logical basis for concept learning in DLs~\cite{LbRoughification,KSE2012,SoICT2012,ICCCI2012a,TranHHNN2013,TranNH2014}. These cited papers are pioneering ones in applying bisimulation to concept learning and approximation in DLs. That is, our results, especially the ones on the largest auto-bisimulations, are very useful for machine learning in the context of DLs.

%-------------------------------------------------------------------------------

\bigskip

\noindent
{\bf Acknowledgements.} 
This work was supported by the Polish National Science Centre (NCN) under Grant No.~2011/01/B/ST6/02759.

%-------------------------------------------------------------------------------

%-------------------------------------------------------------------------------

\VersionB{
\newpage
\appendix

\newcommand{\Proof}[4]{

\smallskip

\noindent{\bf {#1}~\ref{#2}. }{\em #3}{#4}
}	

\section{Proofs}

In this appendix we present proofs for the results of this paper. To increase readability, we recall our lemmas and theorems before presenting their proofs. 

\Proof{Lemma}{lemma: bs-inv-1}{\LemmaBsInvF}{\ProofBsInvF}
\Proof{Corollary}{cor: TBox-invar}{\CorTBoxInvar}{\ProofCorTBoxInvar}
\Proof{Theorem}{theorem: TBox-inv-2}{\TheoremTBoxInvS}{\ProofThmTBoxInvS}
\Proof{Theorem}{theorem: ABox-inv}{\TheoremABoxInv}{\ProofThmABoxInv}
\Proof{Theorem}{theorem: preserving RBox}{\TheoremPreservingRBox}{\ProofThmPreservingRBox}
\Proof{Lemma}{Ali:method}{\LemmaAliMethod}{\ProofLemmaAliMethod}

\smallskip

Recall that, for Lemma~\ref{Ali:main2}, Theorem~\ref{Ali:th:main} and Corollary~\ref{Ali:cor:main}, we assume that $\CN$ and $\RN$ are not empty and $\IN$ contains at least two individual names. Let $\{a,b\} \subseteq \IN$, $A \in \CN$ and $r \in \RN$. 

\Proof{Lemma}{Ali:main2}{\LemmaAliMainT}{\ProofLemmaAliMainT}
\Proof{Theorem}{Ali:th:main}{\TheoremAliMain}{\ProofTheoremAliMain}
\Proof{Corollary}{Ali:cor:main}{\CorAliMain}{\ProofCorAliMain}
\Proof{Theorem}{theorem: H-M}{\TheoremHM}{\ProofThmHM}
\Proof{Theorem}{theorem: LABS}{\PropLABS}{\ProofPropLABS}
\Proof{Theorem}{theorem: quo1}{\TheoremQuoF}{\ProofThmQuoF}
\Proof{Theorem}{theorem: quo-x}{\TheoremQuoX}{\ProofThmQuoX}
\Proof{Theorem}{theorem: mz1}{\TheoremMZF}{\ProofThmMZF}
\Proof{Lemma}{lemma: HGEAO}{\LemmaHGEAO}{\ProofLemmaHGEAO}
\Proof{Theorem}{theorem: quo-y}{\TheoremQuoY}{\ProofThmQuoY}
\Proof{Theorem}{theorem: mz2}{\TheoremMZS}{\ProofThmMZS}
\Proof{Theorem}{theorem: CGSB}{\TheoremCGSB}{\ProofThmCGSB}

} % \VersionB

%-------------------------------------------------------------------------------

\end{document}